\documentclass[twocolumn,letterpaper,pre,showpacs,superscriptaddress]{revtex4} \usepackage[T1]{fontenc}                  \usepackage[latin9]{inputenc} 
\usepackage{float}      
\usepackage{amstext}  
\usepackage{amsmath}   
\usepackage{graphicx} 
\usepackage{esint} 
\usepackage{hyperref}

\providecommand{\tabularnewline}{\\}
\floatstyle{ruled}
\newfloat{algorithm}{tbp}{loa}
\floatname{algorithm}{Algorithm}

\begin{document}

\title{The Geometry of Nonlinear Least Squares, with applications to   Sloppy Models and Optimization}

\author{Mark K. Transtrum}

\affiliation{Laboratory of Atomic and Solid State Physics, Cornell   University, Ithaca, New York 14853, USA}

\author{Benjamin B. Machta}

\affiliation{Laboratory of Atomic and Solid State Physics, Cornell   University, Ithaca, New York 14853, USA}

\author{James P. Sethna}

\affiliation{Laboratory of Atomic and Solid State Physics, Cornell   University, Ithaca, New York 14853, USA}

\begin{abstract}
  Parameter estimation by nonlinear least squares minimization is a   common problem that has an elegant geometric interpretation: the   possible parameter values of a model induce a manifold within the   space of data predictions. The minimization problem is then to find   the point on the manifold closest to the experimental data. We show that the model manifolds of a large class of models, known as \emph{sloppy models}, have many universal features; they are characterized by a geometric series of widths, extrinsic curvatures, and parameter-effects curvatures, which we describe as a hyper-ribbon.  A number of common difficulties in optimizing least squares problems are due to this common geometric structure.  First, algorithms tend to run into the boundaries of the model manifold, causing parameters to diverge or become unphysical before they have been optimized. We introduce the model graph as an extension of the model manifold to remedy this problem. We argue that appropriate priors can remove the boundaries and further improve the convergence rates. We show that typical fits will have many evaporated parameters unless the data are very accurately known.  Second, `bare' model parameters are usually ill-suited to describing model behavior; cost contours in parameter space tend to form hierarchies of plateaus and long narrow canyons.  Geometrically, we understand this inconvenient parameterization as an extremely skewed coordinate basis and show that it induces a large parameter-effects curvature on the manifold.  By constructing alternative coordinates based on geodesic motion, we show that these long narrow canyons are transformed in many cases into a single quadratic, isotropic basin.  We interpret the modified Gauss-Newton and Levenberg-Marquardt fitting algorithms as an Euler approximation to geodesic motion in these natural coordinates on the model manifold and the model graph respectively.  By adding a geodesic acceleration adjustment to these algorithms, we alleviate the difficulties from parameter-effects curvature, improving both efficiency and success rates at finding good fits.
\end{abstract}


\maketitle

\section{Introduction \label{sec:Introduction}}

An ubiquitous problem in mathematical modeling involves estimating parameter values from observational data. One of the most common approaches to the problem is to minimize a sum of squares of the deviations of predictions from observations. A typical problem may be stated as follows: given a regressor variable, $t$, sampled at a set of points $\{t_{m}\}$ with observed behavior $\{y_{m}\}$ and uncertainty $\{\sigma_{m}\}$, what values of the parameters, $\theta$, in some model $f(t,\theta)$, best reproduce or explain the observed behavior? This optimal value of the parameters is known as the best fit.

To quantify how good a fit is, the standard approach is to assume that the data can be reproduced from the model plus a stochastic term that accounts for any discrepancies. That is to say \[ y_{m} = f(t_{m},\theta)+\zeta_{m},\] where $\zeta_{m}$ are random variables assumed to be independently distributed according to $\mathit{N}(0,\sigma_{m})$. Written another way, the residuals given by
\begin{equation} r_{m}(\theta) =   \frac{y_{m}-f(t_{m},\theta)}{\sigma_{m}}, \label{eq:rdefinition}
\end{equation} 
are random variables that are independently, normally distributed with zero mean and unit variance. The probability distribution function of the residuals is then 
\begin{equation}         P(\vec{r},\theta)=\frac{1}{(2\pi)^{M/2}}\exp\left(-\frac{1}{2}\sum_{m=1}^{M}r_{m}(\theta)^{2}\right),
  \label{eq:Prob(r)}
\end{equation} 
where $M$ is the number of residuals. The stochastic part of the residuals is assumed to enter through its dependence on the observed data, while the parameter dependence enters through the model. This distinction implies that while the residuals are random variables, the matrix of derivatives of the residuals with respect to the parameters is not. We represent this Jacobian matrix by $J_{m\mu}$:\[ J_{m\mu}=\partial_{\mu}r_{m}.\]

For a given set of observations $\{y_{m}\}$, the distribution in Eq.~\eqref{eq:Prob(r)} is a likelihood function, with the most likely, or best fit, parameters being those that minimize the cost function, $C$, defined by \begin{equation}   C(\theta)=\frac{1}{2}\sum_{m}r_{m}(\theta)^{2},\label{eq:CostDef}\end{equation} which is a sum of squares. Therefore, if the noise is Gaussian (normally) distributed, minimizing a sum of squares is equivalent to a maximum likelihood estimation.

If the model happens to be linear in the parameters it is a linear least squares problem and the best fit values of the parameters can be expressed analytically in terms of the observed data and the Jacobian. If, however, the model is nonlinear, the best fit cannot be found so easily. In fact, finding the best fit of a nonlinear problem can be a very difficult task, notwithstanding the many algorithms that are designed for this specific purpose.

For example, a nonlinear least squares problem may have many local minima. Any search algorithm that is purely local will at best converge to a local minima and fail to find the global best fit. The natural solution is to employ a search method designed to find a global minima, such as a genetic algorithm or simulated annealing. We will not address such topics in this paper, although the geometric framework that we develop could be applied to such methods. We find, surprisingly, that most fitting problems do not have many local minima. Instead, we find a universality of cost landscapes, as we discuss later in section~\ref{sec:Manifolds-with-Boundaries}, consisting of only one, or perhaps very few, minima.

Instead of difficulties from local minima, the best fit of a nonlinear least squares problem is difficult to find because of \textit{sloppiness}, particularly if the model has many parameters. Sloppiness is the property that the behavior of the model responds very strongly to only a few combinations of parameters, known as stiff parameter combinations, and very weakly to all other combinations of parameters, which are known as sloppy parameter combinations. Although the sloppy model framework has been developed in the context of systems biology~\cite{Brown2003,Brown2004,Casey2007,Daniels2008,Gutenkunst2007,Gutenkunst2007a,Gutenkunst2008}, models from many diverse fields have been shown to lie within the sloppy model universality class~\cite{Waterfall2006}.

In this paper we present the geometric framework for studying nonlinear least squares models. This approach has a long, interesting history, originating with Jeffreys in 1939~\cite{Jeffreys1998}, and later continued by Rao~\cite{Rao1945,Rao1949} and many others~\cite{Amari2007,Murray1993}.  An equivalent, alternative formulation began with Beale in 1960~\cite{Beale1960}, and continued with the work of Bates and Watts~\cite{Bates1980,Bates1981,Bates1983,Bates1988} and others~\cite{Cook1985,Cook1986,Clarke1987}. The authors have used this geometric approach previously to explain the extreme difficulty of the data fitting process~\cite{Transtrum2010}; of which this work is a continuation.

In section~\ref{sec:ModelManifold} we present a review of the phenomenon of sloppiness and describes the \textit{model manifold}, i.e. the geometric interpretation of a least squares model. The geometric picture naturally illustrates two major difficulties that arise when optimizing sloppy models. First, parameters tend to diverge or drift to unphysical values, geometrically corresponding to running off the edge of the manifold, as we describe in section~\ref{sec:Manifolds-with-Boundaries}.  This is a consequence of the model manifold having boundaries that give it the shape of a curving hyper-ribbon in residual space with a geometric hierarchy of widths and curvatures. We show, in section~\ref{sec:The-Model-Graph} that the \textit{model graph}, the surface formed by plotting the residual output versus the parameters, can help to remove the boundaries and improve the fitting process. Generalizing the model graph suggests the use of priors as additional residuals, as we do in section~\ref{sec:Priors}.  We see there that the natural scales of the experiment can be a guide to adding priors to the cost function that can significantly improve the convergence rate.

The second difficulty is that the model's `bare' parameters are often a poor coordinate choice for the manifold. In section~\ref{sec:Extended-Geodesic-Coordinates} we construct new coordinates, which we call \textit{extended geodesic coordinates}. The coordinates remove the effects of the bad coordinates all the way to the edge of the manifold. The degree to which extended geodesic coordinates are effective at facilitating optimization is related to the curvature of the manifold. Section~\ref{sec:Curvature} discusses several measures of curvature and explores curvature of sloppy models. We show that the \textit{parameter-effects} curvature is typically the dominant curvature of a sloppy model, explaining why extended geodesic coordinates can be a huge simplification to the optimization process.  We also show that typical best fits will usually have many evaporated parameters and then define a new measure of curvature, the \textit{optimization curvature}, that is useful for understanding the limitation of iterative algorithms.

We apply geodesic motion to numerical algorithms in section~\ref{sec:Applications-to-Algorithms}, where we show that the modified Gauss-Newton method and Levenberg-Marquardt method are an Euler approximation to geodesic motion. We then add a geodesic acceleration correction to the Levenberg-Marquardt algorithm and achieve much faster convergence rates over standard algorithms and more reliability at finding good fits.

\section{The Model Manifold \label{sec:ModelManifold}}

\begin{figure*}
  \includegraphics[width=7in]{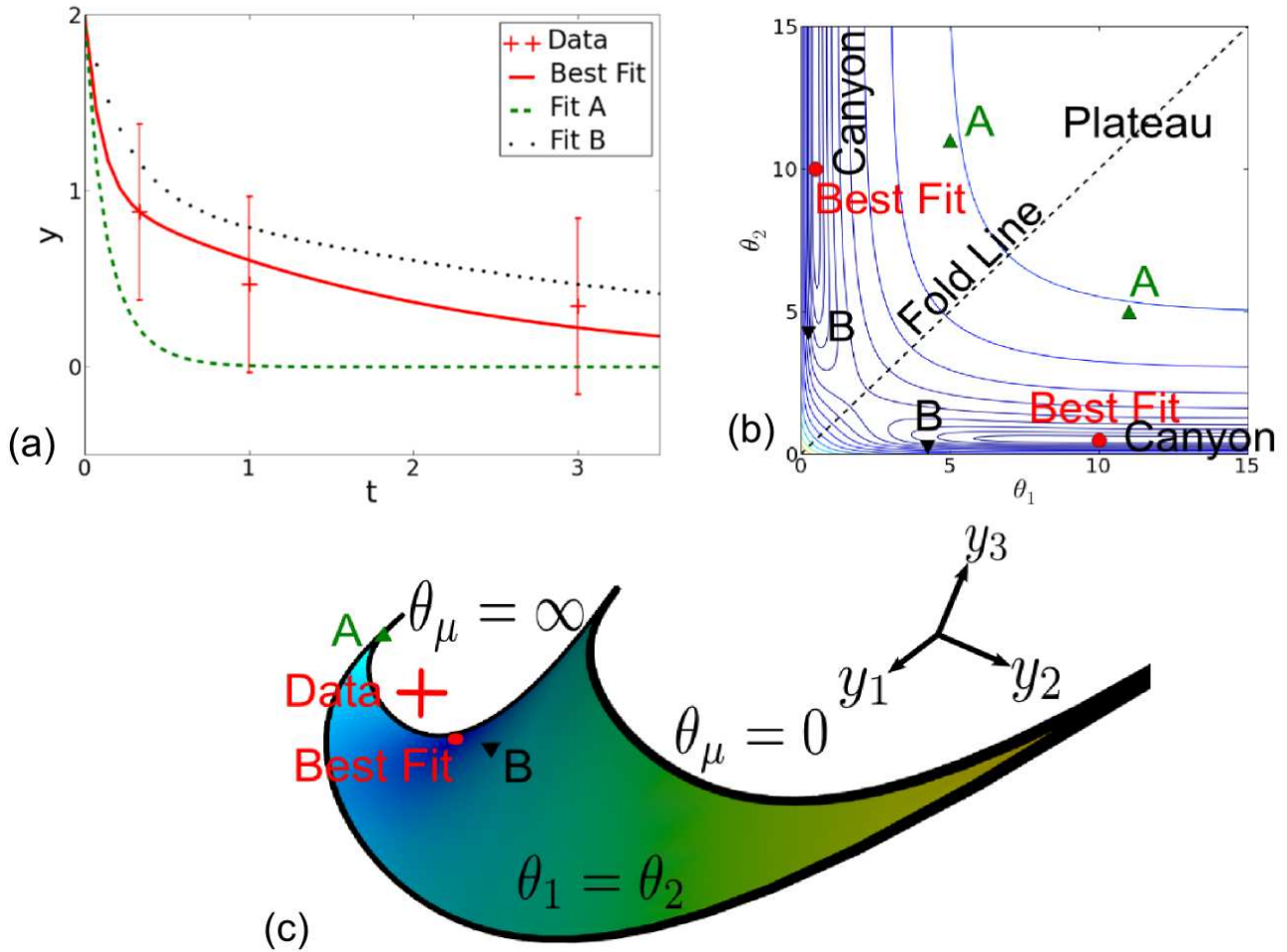}\caption{\label{fig:Fitting}(a)     \textbf{Fitting a nonlinear function to data}, in this case the     sum of two exponentials to three data points. Fit A has rate     constants which decay too quickly, resulting in a poor fit; B is     an improvement over Fit A, although the rates are too slow; the     best fit minimizes the cost (the sum of the squares of the     residuals, which are deviations of model from data points) (b)     \textbf{Contours of constant Cost in parameter space.} Note the     {}``plateau'' in the region of large rates where the model is     essentially independent of parameter changes. Note also the long,     narrow canyon at lower rates, characteristic of a sloppy     model. The sloppy direction is parallel to the canyon and the     stiff direction is against the canyon wall.  (c) \textbf{Model       predictions in data space}. The experimental data is represented     by a single point. The set of all possible fitting parameters     induce a manifold of predictions in data space. The best fit is     the point on the manifold nearest to the data. The plateau in (b)     here is the small region around the short cusp near the corner.}

\end{figure*}

In this section we review the properties of sloppy models and the geometric picture naturally associated with least squares models.  To provide a concrete example of sloppiness to which we can apply the geometric framework, consider the problem of fitting three monotonically decreasing data points to the model \[ y(t,\theta)=e^{-t\theta_{1}}+e^{-t\theta_{2}}.\] Although simple, this model illustrates many of the properties of more complicated models. Figure~\ref{fig:Fitting}a is an illustration of the data and several progressively better fits. Because of the noise, the best fit does not pass exactly through all the data points, although the fit is within the errors.

A common tool to visualize the parameter dependence of the cost is to plot contours of constant cost in parameters space, as is done for our toy model in Figure~\ref{fig:Fitting}b. This view illustrates many properties of sloppy models. This particular model is invariant to a permutation of the parameters, so the plot is symmetric for reflections about the $\theta_{1}=\theta_{2}$ line. We refer to the $\theta_{1}=\theta_{2}$ linear as the {}``fold line'' for geometric reasons that will be apparent in section~\ref{sec:The-Model-Graph}. Around the best fit, cost contours form a long narrow canyon. The direction along the length of the canyon is a sloppy direction, since this parameter combination hardly changes the behavior of the model, and the direction up a canyon wall is the stiff direction. Because this model has few parameters, the sloppiness is not as dramatic as it is for most sloppy models.  It is not uncommon for real-life models to have canyons with an aspect ratios much more extreme than in Fig.~\ref{fig:Fitting}b, typically $1000:1$ or more for models with $10$ or more parameters~\cite{Gutenkunst2007a}.

\begin{figure}
  \includegraphics[width=3.25in]{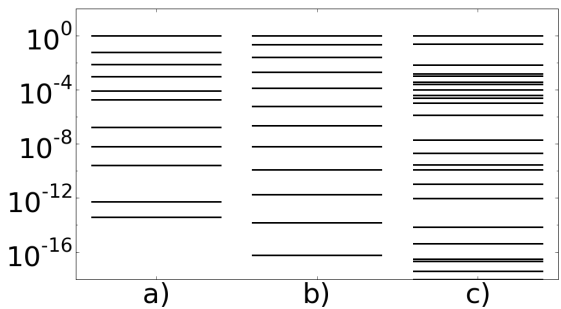}\caption{\label{fig:Eigenvalue-Plot}Hessian     eigenvalues for three sloppy models.  Note the extraordinarily     large range of eigenvalues ($15$-$17$ orders of magnitude,     corresponding to to valley aspect ratios of $10^{7}$-$10^{9}$) in     Fig.~\ref{fig:Fitting}b. Notice also the roughly equal fractional     spacing between eigenvalues--there is no clean separation between     important (stiff) and irrelevant (sloppy) direction in parameter     space.  a) The model formed by summing six exponential terms with     rates and amplitudes. We use this model to investigate curvature     in section~\ref{sec:Curvature} and as a test problem to compare     algorithms in section~\ref{sub:Algorithm-Comparisons}. b) The     linear problem of fitting polynomials is sloppy with the Hessian     given by the Hilbert matrix. c) A more practical model from     systems biology of signaling the epidermal growth factor in rat     pheochromocytoma (PC12) cells~\cite{Brown2004}, which also has a     sloppy eigenvalue spectrum. Many more examples can be found in ~\cite{Gutenkunst2007a,Waterfall2006}.}

\end{figure}

Sloppiness can be quantified by considering the quadratic approximation of the cost around the best fit. The Hessian (second derivative) matrix, $H_{\mu\nu}$, of the cost at the best fit has eigenvalues that span many orders of magnitude and whose logarithms tend to be evenly spaced, as illustrated in Fig.~\ref{fig:Eigenvalue-Plot}. Eigenvectors of the Hessian with small eigenvalues are the sloppy directions, while those with large eigenvalues are the stiff directions. In terms of the residuals, the Hessian is given by

\begin{eqnarray}
  H_{\mu\nu} & = & \partial_{\mu}\partial_{\nu}C\nonumber \\
  & = &   \sum_{m}\partial_{\mu}r_{m}\partial_{\nu}r_{m}+\sum_{m}r_{m}\partial_{\mu}\partial_{\nu}r_{m}\label{eq:Hessian}\\
  & \approx &   \sum_{m}\partial_{\mu}r_{m}\partial_{\nu}r_{m}.\label{eq:HessianApprox}\\
  & = & \left(J^{T}J\right)_{\mu\nu}\end{eqnarray}
In the third and fourth line we have made the approximation that
at the best fit the residuals are negligible. Although the best fit
does not ordinarily corresponds to the residuals being exactly zero,
the Hessian is usually dominated by the term in Eq.~\eqref{eq:HessianApprox}
when evaluated at the best fit. Furthermore, the dominant term, $J^{T}J$,
is a quantity important geometrically which describes the model-parameter
response for all values of the parameters independently of the data.
The approximate Hessian is useful to study the sloppiness of a model
independently of the data at points other than the best fit.  It also shares the sloppy spectrum of the exact Hessian. We call the eigenvectors of $J^TJ$ the local \emph{eigenparameters} as they embody the varying stiff and sloppy combinations of the `bare' parameters.

In addition to the stiff and sloppy parameter combinations near the best fit, Fig.~\ref{fig:Fitting}b also illustrates another property common to sloppy models. Away from the best fit the cost function often depends less and less strongly on the parameters. The contour plot shows a large plateau where the model is insensitive to all parameter combinations. Because the plateau occupies a large region of parameter space, most initial guesses will lie on the plateau. When an initial parameter guess does begin on a plateau such as this, even finding the canyon can be a daunting task.

The process of finding the best fit of a sloppy model, usually consists of two steps. First, one explores the plateau to find the canyon.  Second, one follows the canyon to the best fit. One will search to find a canyon and follow it, only to find a smaller plateau within the canyon that must then be searched to find another canyon. Qualitatively, the initial parameter guess does not fit the data, and the cost gradient does not help much to improve the fit. After adjusting the parameters, one finds a particular parameter combination that can be adjusted to fit some clump of the data. After optimizing this parameter combination (following the canyon), the fit has improved but is still not optimal.  One must then search for another parameter combination that will fit another aspect of the data, i.e. find another canyon within the first.  Neither of these steps, searching the plateau or following the canyon, is trivial.

Although plotting contours of constant cost in parameter space can be an useful and informative tool, it is not the only way to visualize the data. We now turn to describing an alternative geometric picture that helps to explain why the the processes of searching plateaus and following canyons can be so difficult. The geometric picture provides a natural motivation for tools to improve the optimization process.

Since the cost function has the special form of a sum of squares, it has the properties of a Euclidean distance. We can interpret the residuals as components of an $M$-dimensional residual vector. The $M$-dimensional space in which this vector lives is a Euclidean space which we refer to as \textit{data space}. By considering Eq.~\eqref{eq:rdefinition}, we see that the residual vector is the difference between a vector representing the data and vector representing the model (in units of the standard deviation). If the model depends on $N$ parameters, with $N<M,$ then by varying those $N$ parameters, the model vector will sweep out an $N$-dimensional surface embedded within the $M$-dimensional Euclidean space. We call this surface the model manifold, it is sometimes also known as the expectation or regression surface~\cite{Barndorff-Nielsen1986,Bates1988}.  The model manifold of our toy model is shown in Fig.~\ref{fig:Fitting}c.  The problem of minimizing the cost is thus translated into the geometric problem of finding the point on the model manifold that is closest to the the data.

In transitioning from the parameter space picture to the model manifold picture, we are now faced with the problem of minimizing a function on a curved surface. Optimization on manifolds is a problem that has been given much attention in recent decades~\cite{Gabay1982,Mahony1994,Mahony2002,Manton2004,Peeters1993,Smith1993,Smith1994,Udriste1994,Yang2007,Absil2008}. The general problem of minimizing a function on a manifold is much more complicated than our problem; however, because the cost function is linked here to the structure of the manifold the problem at hand is much simpler.

The metric tensor measures distance on the manifold corresponding to infinitesimal changes in the parameters. It is induced from the Euclidean metric of the data space and is found by considering how small changes in parameters correspond to changes in the residuals. The two are related through the Jacobian matrix, \[ dr_{m}=\partial_{\mu}r_{m}d\theta^{\mu}=J_{m\mu}d\theta^{\mu},\] where repeated indices imply summation. (We also employ the convention that Greek letters index parameters, while Latin letters index data points, model points, and residuals.) The square of the distance moved in data space is then \begin{equation}   dr^{2}=(J^{T}J)_{\mu\nu}d\theta^{\mu}d\theta^{\nu}.\label{eq:FirstFundForm}\end{equation} Eq.~\eqref{eq:FirstFundForm} is known as the first fundamental form, and the coefficient of the parameter infinitesimals is the metric tensor, \[ g_{\mu\nu}=(J^{T}J)_{\mu\nu}=\sum_{m}\partial_{\mu}r_{m}\partial_{\nu}r_{m}.\] The metric tensor corresponds to the approximate Hessian matrix in Eq.~\eqref{eq:HessianApprox}; therefore, the metric is the Hessian of the cost at a point assuming that the point exactly reproduced the data.

Qualitatively, the difference between the metric tensor and the Jacobian matrix is that the former describes the local intrinsic properties of the manifold while the latter describes the local embedding.  For nonlinear least squares fits, the embedding is crucial, since it is the embedding that defines the cost function.  To understand how the manifold is locally embedded, consider a singular value decomposition of the Jacobian \[ J = U \Sigma V^T, \] where $V$  is an $N \times N$ unitary matrix satisfying $V^TV=1$ and $\Sigma$ is an $N \times N$ diagonal matrix of singular values.  The matrix $U$ is almost unitary, in the sense that it is an $M \times N$ matrix satisfying $U^T U = 1$; however, $U U^T$ is not the identity~\cite{Press2007}.  In other words, the columns of $U$ contain $N$ residual space vectors that are orthonormal spanning the range of $J$ and not the whole embedding space. In terms of the singular value decomposition, the metric tensor is then given by \[ g = V \Sigma^2 V^T, \] showing us that $V$ is the matrix whose columns are the local eigenparameters of the metric with eigenvalues $\lambda_i = \Sigma_{ii}^2$.  

The singular value decomposition tells us that the Jacobian maps metric eigenvectors onto the data space vector $U_i$ and stretched by an amount $\sqrt{\lambda_i}$.  We  hence denote the columns of $U$ the \emph{eigenpredictions}. The product of singular values describes the mapping of local volume elements of parameter space to data space.  A unit hyper-cube of parameter space is stretched along the eigenpredictions by the appropriate singular values to form a skewed, hyper-parallelepiped of volume $\sqrt{ |g|}$.

The Jacobian and metric contain the first derivative information relating changes in parameters to changes in residuals or model behavior. The second derivative information is contained in the connection coefficient. The connection itself is a technical quantity describing how basis vectors on the tangent space move from point to point. The connection is also closely related to geodesic motion, introduced properly in section~\ref{sec:Extended-Geodesic-Coordinates}. Qualitatively it describes how the metric changes from point to point on the manifold. The relevant connection is the Riemann, or metric, connection; it is calculated from the metric by \[ \Gamma_{\mu\nu}^{\alpha}=\frac{1}{2}g^{\alpha\beta}(\partial_{\mu}g_{\beta\nu}+\partial_{\nu}g_{\beta\mu}-\partial_{\beta}g_{\mu\nu}),\] or in terms of the residuals \begin{equation}   \Gamma_{\mu\nu}^{\alpha}=g^{\alpha\beta}\sum_{m}\partial_{\beta}r_{m}\partial_{\mu}\partial_{\nu}r_{m},\label{eq:Connections-raised}\end{equation} where $g^{\mu\nu}=(g^{-1})^{\mu\nu}$. One could now also calculate the Riemann curvature by application of the standard formulae; however, we postpone a discussion of curvature until section~\ref{sec:Curvature}. For a more thorough discussion of concepts from differential geometry, we refer the reader to any text on the subject~\cite{Misner1973,Spivak1979,Eisenhart1997,Ivancevic2007}.

We have calculated the metric tensor and the connection coefficients from the premise that the cost function, by its special functional form, has a natural interpretation as a Euclidean distance which induces a metric on the model manifold. Our approach is in the spirit of Bates and Watts' treatment of the subject~\cite{Bates1980,Bates1981,Bates1983,Bates1988}.  However, the intrinsic properties of the model manifold can be calculated in an alternative way without reference to the embedding through the methods of Jeffreys, Rao and others~\cite{Jeffreys1998,Rao1945,Rao1949,Murray1993,Amari2007}.  This approach is known as information geometry. We derive these quantities using information geometry in Appendix A.

Given a vector in data space we are often interested in decomposing it into two components; one lying within the tangent space of the model manifold at a point and one perpendicular to the tangent space.  For this purpose, we introduce the projection operators $P^{T}$ and $P^{N}$ which act on data-space vectors and project into the tangent space and its compliment respectively. From the Jacobian at a point on the manifold, these operators are\begin{equation}   P^{T}=\delta-P^{N}=J(g^{-1})J^{T},\label{eq:PTPN}\end{equation} where $\delta$ is the identity operator.  It is numerically more accurate to compute these operators using the singular value decomposition of the Jacobian: \[ P^T=UU^T. \]

Turning to the problem of optimization, the parameter space picture leads one initially to follow the naive, gradient descent direction, $-\nabla_{\mu}C$. An algorithm that moves in the gradient descent direction will decrease the cost most quickly for a given change in the parameters. If the cost contours form long narrow canyons, however, this direction is very inefficient; algorithms tend to zig-zag along the bottom of the canyon and only slowly approach the best fit~\cite{Press2007}.

In contrast, the model manifold defines an alternative direction which we call the Gauss-Newton direction, which decreases the cost most efficiently for a change in the behavior. If one imagines sitting on the surface of the manifold, looking at the point representing the data, then the Gauss-Newton direction in data space is the point directed toward the data but projected onto the manifold. Thus, if $\vec{v}$ is the Gauss-Newton direction in data space, it is given by \begin{eqnarray}
  \vec{v} & = & -P^{T}\vec{r}\nonumber \\
  & = & -J(g^{-1})J^{T}\vec{r}\nonumber \\
  & = & -J(g^{-1})\nabla C\nonumber \\
  & = & -\vec{J}_\mu g^{\mu\nu} \nabla_\nu C, \label{eq:NewtonianVector} 
\end{eqnarray} where we have used the fact that $\nabla C=J^{T}r$. The components of the vector in parameter space, $v^{\mu}$ are related to the vector in data space through the Jacobian 
\begin{equation}       \vec{v}=\vec{J}_{\mu}v^{\mu};\label{eq:NewtonianVector2}
\end{equation} therefore, the direction in parameter space $v^{\mu}$ that decreases the cost most efficiently per unit change in behavior is
\begin{equation}
v^{\mu}=-g^{\mu\nu}\nabla_{\nu}C.\label{eq:Newtonian}\end{equation}

The term 'Gauss-Newton' direction comes from the fact that it is the direction given by the Gauss-Newton algorithm described in section~\ref{sub:Modified-Gauss-Newton-Method}. Because the Gauss-Newton direction multiplies the gradient by the inverse metric, it magnifies motion along the sloppy directions. This is the direction that will move the parameters along the canyon toward the best fit. The Gauss-Newton direction is purely geometric and will be the same in data space regardless of how the model is parametrized. The existence of the canyons are a consequence of bad parameterization on the manifold, which this parameter independent approach can help to remedy. Most sophisticated algorithms, such as conjugate gradient and Levenberg-Marquardt attempt to follow the Gauss-Newton direction as much as possible in order to not get stuck in the canyons.

\begin{figure}
  \includegraphics[width=3.25in]{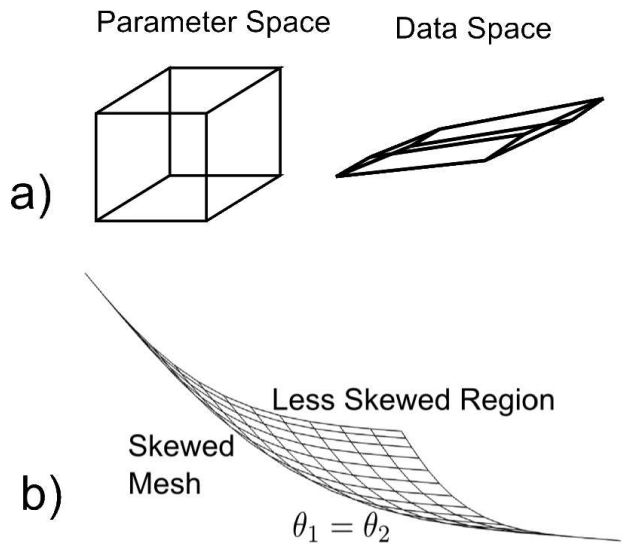}\caption{\label{fig:Skewed}\textbf{Skewed       Coordinates.} A sloppy model is characterized by a skewed     coordinate mesh on the manifold. The volume of the parallel-piped     is given by the determinant of the metric, which is equal to the     product of the eigenvalues. Because sloppy models have many tiny     eigenvalues, these volumes can be very small with extremely skewed     coordinates. Our toy model has extremely skewed coordinates where     the parameters are nearly equal (near the fold line). Most of the     manifold is covered by regions where the coordinates are less     skewed which corresponds to a very small region in parameter     space. }

\end{figure}

The obvious connection between sloppiness and the model manifold is through the metric tensor. For sloppy models, the metric tensor of the model manifold (the approximate Hessian of Eq.~\eqref{eq:HessianApprox}) has eigenvalues spread over many decades. This property is not intrinsic to the manifold however. In fact, one can always reparametrize the manifold to make the metric at a point any symmetric, positive definite matrix. This might naively suggest that sloppiness has no intrinsic geometric meaning, and that it is simply a result of a poor choice of parameters.  The coordinate grid on the model manifold in data space is extremely skewed as in Figure~\ref{fig:Skewed}. By reparametrizing, one can remove the skewedness and construct a more natural coordinate mesh.  We will revisit this idea in section~\ref{sec:Extended-Geodesic-Coordinates}.  We will argue in this manuscript that on the contrary, there is a geometrical component to sloppy nonlinear models that is independent of parameterization and in most cases that the human-picked `bare' parameters naturally illuminate the sloppy intrinsic structure of the model manifold.

In the original parameterization, sections of parameter space are mapped onto very tiny volumes of data space.  We remind the reader that a unit volume of parameter space is mapped into a volume of data space given by $\sqrt{|g|}$.  Because many eigenvalues are nearly zero for sloppy models, the model manifold necessarily occupies a tiny sliver of data space.  In fact, if a region of parameter space has larger eigenvalues by even a small factor, the cumulative effect on the product is that this region of parameter space will occupy most of the model manifold.  We typically find that most of the model manifold is covered by a very small region of parameter space which corresponds to the volumes of (slightly) less skewed meshes.  

We will see when we discuss curvature, that the large range of eigenvalues in the metric tensor usually correspond to a large anisotropy in the extrinsic curvature. Another geometric property of sloppy systems relates to the boundaries that the model imposes on the manifold. The existence of the boundaries for the toy model can be seen clearly in Fig.~\ref{fig:Fitting}c. The surface drawn in the figure corresponds the patch of parameters within $0\leq\theta_{1},\theta_{2}\leq\infty$.  The three boundaries of the surface occur when the parameters reach their respective bounds. The one exception to this is the fold line, which corresponds to when the parameters are equal to one another.  This anomalous boundary ($\theta_{1}=\theta_{2}$) is called the fold line and is discussed further in section~\ref{sec:The-Model-Graph}.  Most nonlinear sloppy models have boundaries.

In the next section we will discuss how boundaries arise on the model manifold and why they pose problems for optimization algorithms. Then, in section~\ref{sec:The-Model-Graph} we describe another surface, the model graph, that removes the boundaries. The surface described by the model graph is equivalent to a model manifold with a linear Bayesian prior added as additional residuals. In section~\ref{sec:Priors} we show that introducing other priors can be even more helpful for keeping algorithms away from the boundaries.

\section{Bounded Manifolds\label{sec:Manifolds-with-Boundaries}}

Sloppiness is closely related to the existence of boundaries on the model manifold. This may seem to be a puzzling claim because sloppiness has previously been understood to be a statement relating to the \textit{local }linearization of model space. Here we will extend this idea and see that it relates to the \textit{global} structure of the manifold and how it produces difficulties for the optimization process.

To understand the origin of the boundaries on model manifolds, consider first the model of summing several exponentials\[ y(t,\theta)=\sum_{\mu}e^{-\theta_{\mu}t}.\] We restrict ourselves to considering only positive arguments in the exponentials, which limits the range of behavior for each term to be between $0$ and $1$. This restriction already imposes boundaries on the model manifold, but those boundaries become much more narrow as we consider the range the model can produce by holding just a few time points fixed.

\begin{figure}

  \includegraphics[width=3.25in]{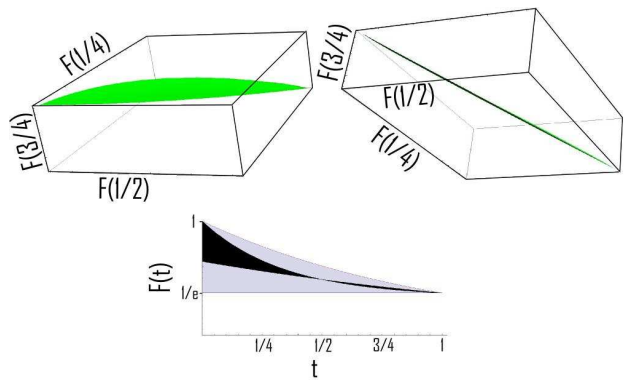}\caption{\label{fig:Range-of-Behavior}Fixing     a few data points greatly restricts the possible range of the     model behavior between those data points (lower). This is a     consequence of interpolation of analytic functions\textbf{.  }In     this case,\textbf{ }$f(t)$ is a sum of three exponentials with six     parameters (amplitudes and rates). Shown above is a three     dimensional slice of possible models plotted in data space, with     the value of $f(0)$ fixed to 1 and the value of $f(1)$ fixed to     $1/e$. With these constraints we are left with a four dimensional     surface, meaning that the manifold of possible data shown here is     indeed a volume.  However, from a carefully chosen perspective     (upper right) this volume can be seen to be extremely thin--in     fact most of its apparent width is curvature of the nearly two     dimensional sheet, evidenced by being able to see both the top     (green) and bottom (black) simultaneously.  Generic aspects of     this picture illustrate the difficulty of fitting nonlinear     problems. Geodesics in this volume are just straight lines in     three dimensions. Although the manifold seems to be only slightly     curved, its extreme thinness means that geodesics travel very     short distances before running into model boundaries,     necessitating the diagonal cutoff in Levenberg-Marquardt     algorithms as well as the priors discussed in section ~\ref{sec:Priors}. }

\end{figure}

Fixing the output of the model at a few time points greatly reduces the values that the model can take on for all the remaining points. Fixing the values that the model takes on at a few data points is equivalent to considering a lower-dimensional cross section of the model manifold, as we have done in Fig.~\ref{fig:Range-of-Behavior}. The boundaries on this cross section are very narrow; the corresponding manifold is long and thin. Clearly, an algorithm that navigates the model manifold will quickly run into the boundaries of this model unless it is actively avoiding them.

\begin{figure}

  \includegraphics[width=3.25in]{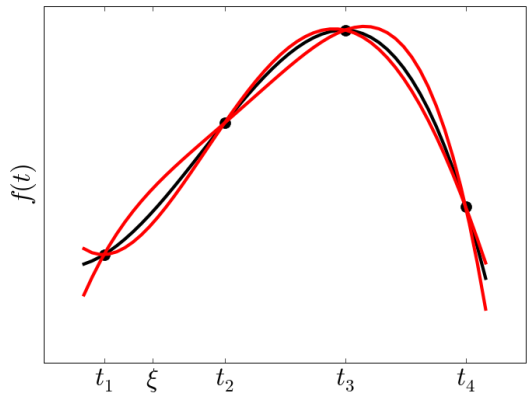}\caption{\label{fig:Interpolation}The     possible values of a model at intermediate time points are     restricted by interpolating theorems. Taking cross sections of the     model manifold corresponds to fixing the model values at a few     time points, restricting the possible values at the remaining     times. Therefore, the model manifold will have a hierarchy of     progressively thinner widths, much like a hyper-ribbon.}

\end{figure}

In general, if a function is analytic, the results presented in Fig.~\ref{fig:Range-of-Behavior} are fairly generic, they come from general theorems governing the interpolation of functions. If a function is sampled at a sufficient number of time points to capture its major features, then the behavior of the function at times between the sampling can be predicted with good accuracy by an interpolating function. For polynomial fits, as considered here, a function, $f(t)$, sampled at $n$ time points, $(t_{1},t_{2},...,t_{n})$, can be fit exactly by a unique polynomial of degree $n-1$, $P_{n-1}(t)$. Then at some interpolating point, $t$, the discrepancy in the interpolation and the function is given by \begin{equation}   f(t)-P_{n-1}(t)=\frac{\omega(t)f^{(n)}(\xi)}{n!},\label{eq:Interpolation-Error}\end{equation} where $f^{(n)}(t)$ is the $n$-th derivative of the function and $\xi$ lies somewhere in the range $t_{1}<\xi<t_{n}$~\cite{Stoer2002}.  The polynomial $\omega(t)$ has roots at each of the interpolating points\[ \omega(t)=(t-t_{1})(t-t_{2})...(t-t_{n}).\] By inspecting Eq.~\eqref{eq:Interpolation-Error}, it is clear that the discrepancy between the interpolation and the actual function will become vanishingly small if higher derivatives of the function do not grow too fast (which is the case for analytic functions) and if the sampling points are not too widely spaced (see Fig.~\ref{fig:Interpolation}).

The possible error of the interpolation function bounds the allowed range of behavior, $\delta f_{n}$, of the model at $t_{0}$ after constraining the nearby $n$ data points, which corresponds to measuring cross sections of the manifold. Consider the ratio of successive cross sections, \[ \frac{\delta f_{n+1}}{\delta   f_{n}}=(t-t_{n+1})(n+1)\frac{f^{n+1}(\xi)}{f^{n}(\xi')},\] if $n$ is sufficiently large, then \[ (n+1)\frac{f^{n+1}(\xi)}{f^{n}(\xi')}\approx\frac{1}{R};\] therefore, we find that \[ \frac{\delta f_{n+1}}{\delta   f_{n}}\approx\frac{t-t_{n+1}}{R}<1\] by the ratio test. Each cross section is thinner than the last by a roughly constant factor $\Delta = \delta t/ R$,  predicting a hierarchy of widths on the model manifold. We describe the shape of a model manifold with such a hierarchy as a hyper-ribbon. We will now measure these widths for a few sloppy models and see that the predicted hierarchy is in fact present.

As a first example, consider the sloppy model of fitting polynomials \begin{equation}   f(t,\theta)=\sum_{m}\theta_{m}t^{m}.\label{eq:Polynomials}\end{equation} If the parameters of the model are allowed to vary over all real values, then one can always fit $M$ data points exactly with an $\left(M-1\right)^{th}$ degree polynomial. However, we wish to artificially restrict the range of the parameters to imitate the limited range of behavior characteristic of nonlinear models. A simple restriction is given by $\sum_{m}\theta_{m}^{2}\leq1$.  This constraint enforces the condition that higher derivatives of the function become small (roughly that the radius of convergence is one) and corresponds to the unit hyper-sphere in parameter space. If this function is sampled at time points $(t_{1},t_{2},...,t_{n})$ then the model vector in data space can be written as \begin{equation}   \vec{f}=\left(\begin{array}{cccc}
      1 & t_{1} & t_{1}^{2} & \cdots\\
      1 & t_{2} & t_{2}^{2} & \cdots\\
      \vdots & \vdots & \vdots & \vdots\\
      1 & t_{n} & t_{n}^{2} &       \cdots\end{array}\right)\left(\begin{array}{c}
      \theta_{0}\\
      \theta_{1}\\
      \theta_{2}\\
      \vdots\end{array}\right).\label{eq:Vandermonde-Polynomials}\end{equation} The matrix multiplying the vector of parameters is an example of a Vandermonde matrix. The Vandermonde matrix is known to be sloppy and, in fact, plays an important role in the sloppy model universality class. The singular values of the Vandermonde matrix are what produce the sloppy eigenvalue spectrum of sloppy models.  Reference~\cite{Waterfall2006} shows that these singular values are indeed broadly spaced in $\log$. For this model, the Vandermonde matrix is exactly the Jacobian.

By limiting our parameter space to a hypersphere for the model in Eq.~\eqref{eq:Polynomials}, the corresponding model manifold is limited to a hyper-ellipse in data space. The principal axes of this hyper-ellipse are the eigenpredictions directions we discussed in section~\ref{sec:ModelManifold}.  The lengths of the principal axes are the singular values. Consequently, there will be a hierarchy of progressively thinner boundaries on the model manifold due to the wide ranging singular values of the Vandermonde matrix. For this model, the purely local property of the metric tensor eigenvalue spectrum is intimately connected to the global property of the boundaries and shape of the model manifold.

\begin{figure}
  \textbf{\includegraphics[width=3.25in]{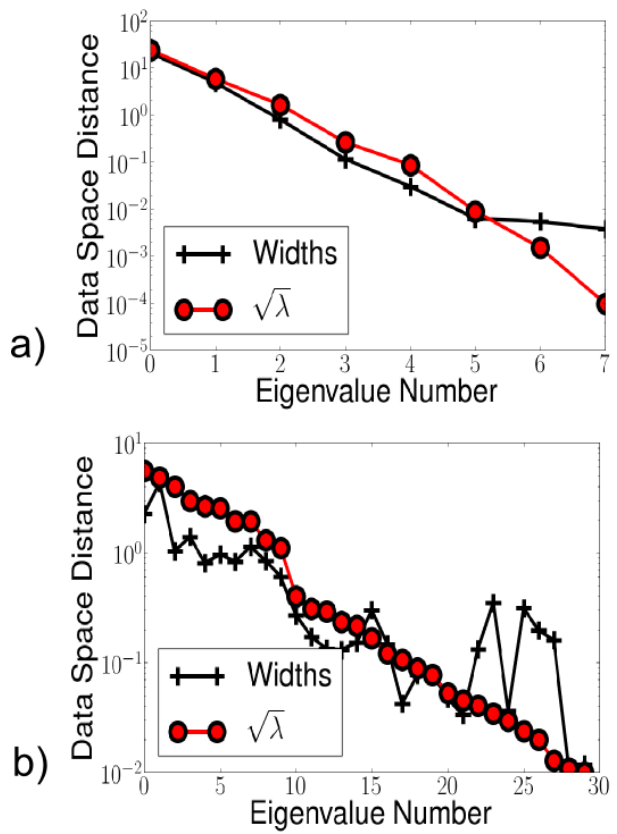}}\caption{\textbf{\label{fig:Cross-sectional-widths}}     a) Geodesic cross-sectional widths of an eight dimensional model     manifold along the eigendirections of the metric from some central     point, together with the square root of the eigenvalues (singular     values of the Jacobian)~\cite{Transtrum2010}. Notice the hierarchy of these data-space     distances -- the widths and singular values each spanning around     four orders of magnitude. To a good approximation, the     cross-sectional widths are given by singular values. In the limit     of infinitely many exponential terms, this model becomes linear.     b) Geodesic cross-sectional widths of a feed-forward artificial     neural network. Once again, the widths nicely track the singular     values.}

\end{figure}

As a second example, consider the model consisting of the sum of eight exponential terms, $y=\sum_{\mu}A_{\mu}e^{-\theta_{\mu}t}$. We use log-parameters, $r_{\theta\mu}=\log\theta_{\mu}$ and $r_{A\mu}=\log A_{\mu}$, to make parameters dimensionless and enforce positivity. We numerically calculate the several widths of the corresponding model manifold in Fig.~\ref{fig:Cross-sectional-widths}a, where we see that they are accurately predicted by the singular values of the Jacobian.  The widths in Fig.~\ref{fig:Cross-sectional-widths} were calculated by considering geodesic motion in each of the eigendirections of the metric from some point located near the center of the model manifold. We follow the geodesic motion until it reaches a boundary; the length in data space of the geodesic is the width.  Alternatively, we can choose $M-N$ orthogonal unit vectors that span the space perpendicular to the tangent plane at a point and a single unit vector given by a eigenprediction of the Jacobian which lies within the tangent plane.  The $M-N+1$ dimensional hyper-plane spanned by these unit vectors intersects the model manifold along a one-dimensional curve.  The width can be taken to be the length of that intersection.  The widths given by these two methods are comparable.

We can show analytically that our exponential fitting problem has model manifold widths proportional to the corresponding singular values of the Jacobian in the limit of a continuous distribution of exponents, $\theta_{\mu}$, using an argument provided to us by Yoav Kallus. In this limit, the sum can be replaced by an integral,\[ y(t)=\int d\theta A(\theta)e^{-t\theta}=\mathcal{L}\left\{   A(\theta)\right\} ,\] where the model is now the Laplace transform of the amplitudes $A(\theta)$.  In this limit the data can be fit without varying the exponential rates, leaving only the linear amplitudes as parameters. If we assume the data has been normalized according to $y(t=0)\leq 1$, then it is natural to consider the hyper-tetrahedron of parameter space given by by $ A_n > 0$ and $\sum A_n \leq 1$.  In parameter space, this tetrahedron has a maximum aspect ratio of $\sqrt{2/M}$, but the mapping to data space distorts the tetrahedron by a constant Jacobian whose singular values we have seen to span many orders of magnitude. The resulting manifold thus must have a hierarchy of widths along the eigenpredictions equal to the corresponding eigenvalues within the relatively small factor $\sqrt{2/M}$.  

As our third example, we consider a feed-forward artificial neural network~\cite{Hertz1991}.  For computational ease, we choose a small network consisting of a layer of four input neurons, a layer of four hidden neurons, and an output layer of two neurons.  We use the hyperbolic tangent function as our sigmoid function and vary the connection weights as parameters.  As this model is not known to reduce to a linear model in any limit, it serves as a test that the agreement for fitting exponentials is not special.  Fig.~\ref{fig:Cross-sectional-widths}b shows indeed that the singular values of the Jacobian agree with geodesic widths again for this model.  

The results in Fig.~\ref{fig:Cross-sectional-widths} is one of our main results and requires some discussion.  Strictly speaking, the singular values of the Jacobian have units of data space distance per unit parameter space distance, while the units of the widths are data space distance independent of parameters. In the case of the exponential model, we have used log-parameters, making the parameters dimensionless.  In the neural network, the parameters are the connection weights whose natural scale is one.  In general, the exact agreement between the singular values and the widths may not agree if the parameters utilize different units or have another natural scale. One must note, however, that the enormous range of singular values implies that the units would have to be radically different from natural values to lead to significant distortions.  

Additionally, the two models presented in Fig.~\ref{fig:Cross-sectional-widths} are particularly easy to fit to data.  The fact that from a centrally located point, geodesics can explore nearly the entire range of model behavior suggests that the boundaries are not a serious impediment to the optimization.  For more difficult models, such as the PC12 model in systems biology~\cite{Brown2004}, we find that the the widths estimated from the singular values and from geodesic motion disagree.  The geodesic widths are much smaller than the singular value estimates.  In this case, although the spacing between geodesic widths is the same as the spacing between the singular values, they are smaller by several orders of magnitude.  We believe that most typical starting points of this model lie near a hyper-corner of the model manifold.  If this is the case, then geodesics will be unable to explore the full range of model behavior without reaching a model boundary.  We argue later in this section that this phenomenon is one of the main difficulties in optimization, and in fact, we find that the PC12 model is a much more difficult fitting problem than either the exponential or neural network problem.

We have seen that sloppiness is the result of skewed coordinates on the model manifold, and we will argue later in section~\ref{sec:Extended-Geodesic-Coordinates} that algorithms are sluggish as a result of this poor parameterization.  Fig.~\ref{fig:Cross-sectional-widths} tells us that the `bare' model parameters are not as perverse as one might naively have thought.  Although the bare-parameter directions are inconvenient for describing the model behavior, the local singular values and eigenpredictions of the Jacobian are useful estimates of the model's global shape.  The fact that the local stiff and sloppy directions coincide with the global long and narrow directions is a nontrivial result that seems to hold for most models.  

To complete our description of a typical sloppy model manifold requires a discussion of curvature, which we postpone until section~\ref{sub:Curvature-in-Sloppy}.  We will see that in addition to a hierarchy of boundaries, the manifold typically has a hierarchy of extrinsic and parameter-effects curvatures whose scales are set by the smallest and widest widths respectively.

We argue elsewhere~\cite{Transtrum2010}, that the ubiquity of sloppy models, appearing everywhere from models in systems biology~\cite{Gutenkunst2007a}, insect flight~\cite{Waterfall2006},  variational quantum wave functions, inter-atomic potentials~\cite{Frederiksen2004}, and a model of the next-generation international linear collider~\cite{Gutenkunst2008}, implies that a large class of models have very narrow boundaries on their model manifolds. The interpretation that multiparameter fits are a type of high-dimensional analytic interpolation scheme, however, also explains why so many models are sloppy. Whenever there are more parameters than effective degrees of freedom among the data points, then there are necessarily directions in parameter space that have a limited effect on the model behavior, implying the metric must have small eigenvalues. Because successive parameter directions have a hierarchy of vanishing effect on model behavior, the metric must have a hierarchy of eigenvalues.

We view most multiparameter fits as a type of multi-dimensional interpolation.  Only a few stiff parameter combinations need to be tuned in order to find a reasonable fit. The remaining sloppy degrees of freedom do not alter the fit much, because they fine tune the interpolated model behavior, which, as we have seen, is very restricted. This has important consequences for interpreting the best fit parameters. One should not expect the best fit parameters to necessarily represent the physical values of the parameters, as each parameter can be varied by many orders of magnitude along the sloppy directions. Although the parameter values at a best fit cannot typically be trusted, one can still make falsifiable predictions about model behavior without knowing the parameter values by considering an ensemble of parameters with reasonable fits~\cite{Brown2003,Brown2004,Casey2007,Gutenkunst2007}.

For our fitting exponential example, part of the model boundary was the `fold lines` where pairs of the exponents are equal (see Fig.~\ref{fig:Fitting}).  No parameters were at extreme values, but the model behavior was nonetheless singular.  Will such internal boundaries arise generically for large nonlinear models?  Model boundaries correspond to points on the manifold where the metric is singular. Typical boundaries occur when parameters are near their extreme values (such as $\pm\infty$ or zero), where the model becomes unresponsive to changes in the parameters. Formally, a singularity will occur if the basis vectors on the model manifold given by $\vec{e}_{\mu}=\partial_{\mu}\vec{r}$ are linearly dependent, which is to say there exist a set of nonzero $\alpha^{\mu}$'s for which \begin{equation}   \alpha^{\mu}\vec{e}_{\mu}=0.\label{eq:singularcondition}\end{equation} In order to satisfy Eq.~\eqref{eq:singularcondition} we may vary $2N$ parameters (the $N$ values of $\alpha^{\mu}$ plus the $N$ parameters of the model) to satisfy $M$ equations. Therefore if $M<2N$ there will exist nontrivial singular points of the metric at non-extreme values of the parameters. 

For models with $M>2N$, we do not expect Eq.~\eqref{eq:singularcondition} to be exactly satisfied generically except at extreme values of the parameters when one or more of the basis vectors vanish, $\vec{e}_{\mu}=0$.  However, many of the data points are interpolating points as we have argued above, and we expect qualitatively to be able to ignore several data points without much information loss. In general, we expect that Eq.~\eqref{eq:singularcondition} could be satisfied to machine precision at nontrivial values of the parameters even for relatively small $N$.

Now that we understand the origin of boundaries on the model manifold, we can discuss why they are problematic for the process of optimization.  It has been observed in the context of training neural networks, that metric singularities (i.e. model boundaries) can have a strong influence on the fitting~\cite{Amari2006}. More generally, the process of fitting a sloppy model to data involves the frustrating experience of applying a black box algorithm to the problem which appears to be converging, but then returns a set of parameters that does not fit the data well and includes parameter values that are far from any reasonable value.  We refer to this drift of the parameters to extreme values as parameter evaporation %
\footnote{The term parameter evaporation was originally used to   describe the drift of parameters to infinite values in the process   of Monte Carlo sampling~\cite{Brown2003a}. In this case the tendency   of parameters to run to unphysical values is a literal evaporation   caused by the finite temperature of the stochastic process. We now   use the term to also describe deterministic drifts in parameters to   extreme values in the optimization process.%
}. This phenomenon is troublesome not just because it causes the algorithm to fail. Often, models are more computationally expensive to evaluate when they are near the extreme values of their parameters. Algorithms will often not just fail to converge, but they will take a long time in the process.

After an algorithm has failed and parameters have evaporated, one may resort to adjusting the parameter values by hand and then reapplying the algorithm. Hopefully, iterating this process will lead to a good fit. Even if one eventually succeeds in finding a good fit, because of the necessity of adjusting parameters by hand, it can be a long and boring process.

\begin{figure}
  \includegraphics[width=3.25in]{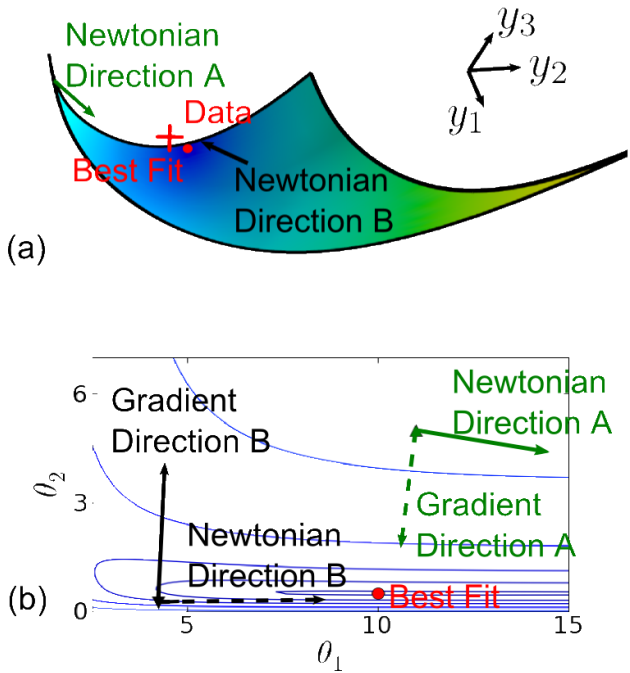}\caption{\label{fig:NaturalDirections}a)     \textbf{Falling off the edge of the model manifold.} The manifold     in data space defines a {}``natural'' direction, known as the     Gauss-Newton direction, in which an algorithm will try to follow     to the best fit. Often this direction will push parameters toward     the edge of the manifold. b) \textbf{Gradient and Gauss-Newton       directions in Parameter space}. The manifold edge corresponds to     infinite values of the parameters. Following the Gauss-Newton     direction to the edge of the manifold will cause parameters to     evaporate while on the plateau. While in a canyon, however, the     Gauss-Newton direction gives the most efficient direction to the     best fit.}

\end{figure}

Parameter evaporation is a direct consequence of the boundaries of the model manifold. To understand this, recall from section~\ref{sec:ModelManifold} that the model manifold defines a natural direction, the Gauss-Newton direction, that most algorithms try to follow. The problem with blindly following the Gauss-Newton direction is that it is purely local and ignores the fact that sloppy models have boundaries. Consider our example model; the model manifold has boundaries when the rates become infinite. If an initial guess has over-estimated or under-estimated the parameters, the Gauss-Newton direction can point toward the boundary of the manifold, as does fit A in Fig.~\ref{fig:NaturalDirections}. If one considers the parameter space picture, the Gauss-Newton direction is clearly nonsensical, pointing away from the best fit. Generally, while on a plateau region, the gradient direction is better at avoiding the manifold boundaries. However, nearer the best fit, the boundary is less important and the Gauss-Newton direction is much more efficient than the downhill direction, as is the case for fit B in Fig.~\ref{fig:NaturalDirections}.

Since the model manifold typically has several narrow widths, it is reasonable to expect that a fit to noisy data will evaporate many parameters to their limiting values (such as $\infty$ or zero), as we explore in section~\ref{sub:parameter-evaporation}. We therefore do not want to prevent the algorithm from evaporating parameters altogether.  Instead, we want to prevent the algorithm from prematurely evaporating parameters and becoming stuck on the boundary (or lost on the plateau).  Using the two natural directions to avoid the manifold boundaries while navigating canyons to the best fit is at the heart of the difficulty in optimizing sloppy models. Fortunately, there exists a natural interpolation between the two pictures which we call the model graph and is the subject of the next section. This natural interpolation is exploited by the Levenberg-Marquardt algorithm, which we discuss in section~\ref{sec:Applications-to-Algorithms}.

\section{The Model Graph\label{sec:The-Model-Graph}}

We saw in Section~\ref{sec:Manifolds-with-Boundaries} that the geometry of sloppiness explains the phenomenon of parameter evaporation as algorithms push parameters toward the boundary of the manifold.  However, as we mentioned in Section~\ref{sec:ModelManifold}, the model manifold picture is a view complementary to the parameter space picture, as illustrated in Fig.~\ref{fig:Fitting}.

The parameter space picture has the advantage that boundaries typically do not exist (i.e. they lie at parameter values equal to $\infty$). If model boundaries occur for parameter values that are not infinite, but are otherwise unphysical, for example, $\theta=0$ for our toy model, it is helpful to change parameters in such a way as to map these boundaries to infinity. For the case of summing exponentials, it is typical to work in $\log\theta$, which puts all boundaries at infinite parameter values and has the added bonus of being dimensionless (avoiding problems of choice of units). In addition to removing boundaries, the parameter space does not have the complications from curvature; it is a flat, Euclidean space.

The disadvantage of the parameter space picture is that motion in parameter space is extremely disconnected from the behavior of the model. This problem arises as an algorithm searches the plateau looking for the canyon and again when it follows the winding canyon toward the best fit.

The model manifold picture and the parameter space picture can be combined to utilize the strengths of both approaches. This combination is called the model graph because it is the surface created by the graph of the model, i.e. the behavior plotted against the parameters.  The model graph is an $N$ dimensional surface embedded in an $M+N$ dimensional Euclidean space. The embedding space is formed by combining the $M$ dimensions of data space with the $N$ dimensions of parameter space. The metric for the model graph can be seen to be \begin{equation} g_{\mu\nu}=g_{\mu\nu}^{0}+\lambda   D_{\mu\nu,}\label{eq:GraphMetric}\end{equation} where $g_{\mu\nu}^{0}=\left(J^{T}T\right)_{\mu\nu}$ is the metric of the model manifold and $D_{\mu\nu}$ is the metric of parameters space.  We discuss common parameter space metrics below. We have introduced the free parameter $\lambda$ in Eq.~\eqref{eq:GraphMetric} which gives the relative weight of the parameter space metric to the data space metric. Most of the work in optimizing an algorithm comes from a suitable choice of $\lambda$, known as the damping parameter or the Levenberg-Marquardt parameter.

If $D_{\mu\nu}$ is the identity, then we call the metric in Eq.~\eqref{eq:GraphMetric} the Levenberg metric because of its role in the Levenberg algorithm~\cite{Levenberg1944}. Another possible choice for $D_{\mu\nu}$ is to populate its diagonal with the diagonal elements of $g_{\mu\nu}^{0}$ while leaving the off-diagonal elements zero. This choice appears in the Levenberg-Marquardt algorithm~\cite{Marquardt1963} and has the advantage that the resulting method is invariant to rescaling the parameters, e.g. it is independent of units.  It has the problem, however, that if a parameter evaporates then its corresponding diagonal element may vanish and the model graph metric becomes singular.  To avoid this dilemma, one often chooses $D$ to have diagonal elements given by the largest diagonal element of $g^0$ yet encountered by the algorithm~\cite{More1977}. This method is scale invariant but guarantees that $D$ is always positive definite. We discuss these algorithms further in section~\ref{sec:Applications-to-Algorithms}.  

It is our experience that the Marquardt metric is much less useful than the Levenberg metric for preventing parameter evaporation. While it may seem counter-intuitive to have a metric (and by extension an algorithm) that is sensitive to whether the parameters are measured in inches or miles, we stress that the purpose of the model graph is to \textit{introduce} parameter dependence to the manifold. Presumably, the modeler is measuring parameters in inches because inches are a more natural unit for the model. By disregarding that information, the Marquardt metric is losing a valuable sense of scale for the parameters and is more sensitive to parameter evaporation. The concept of the natural units will be important in the discussion of priors in section~\ref{sec:Priors}. On the other hand, the Marquardt method is faster at following a narrow canyon and the best choice likely depends on the particular problem.

If the choice of metric for the parameter space is constant, $\partial_{\alpha}D_{\mu\nu}=0$, then the connection coefficients of the model graph (with all lowered indices) are the same as for the model manifold given in Eq.~\eqref{eq:Connections-raised}. The connection with a raised index will include dependence on the parameter space metric:
\[
\Gamma_{\alpha\beta}^{\mu}=(g^{-1})^{\mu\nu}\sum_{m}\partial_{\nu}r_{m}\partial_{\alpha}\partial_{\beta}r_{m},
\]
where $g$ is given by Eq.~\eqref{eq:GraphMetric}.

\begin{figure}
  \includegraphics[width=3.25in]{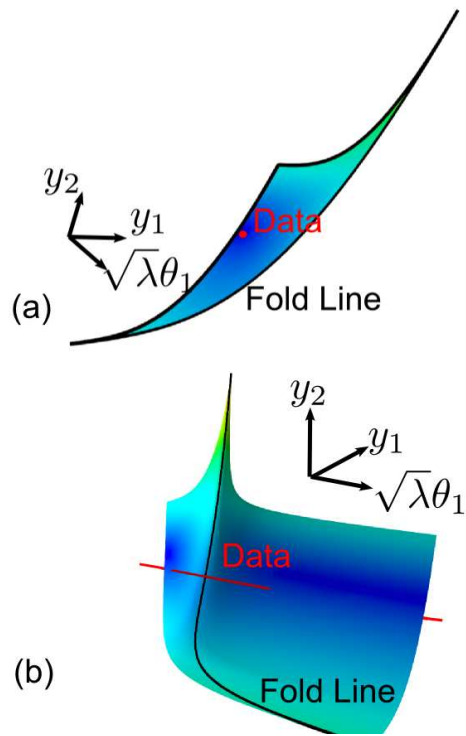}\caption{\label{fig:ModelGraph}The     effect of the damping parameter is to produce a new metric for the     surface induced by the graph of the model versus the input     parameters. (a) \textbf{Model Graph, $\lambda=0$.} If the     parameter is zero, then the resulting graph is simply the original     model manifold, with no extent in the parameter directions. Here     we see a flat two dimensional cross section; the z-axis is a parameter     value multiplied by $\sqrt{\lambda}=0$. (b)\textbf{ Model Graph       $\lambda\neq0$.}  If the parameter is increased, the surface is     \textquotedbl{}stretched\textquotedbl{} into a higher dimensional     embedding space. This is an effective technique for removing the     boundaries, as no such boundary exists     in the model graph. However, this comes at a cost of removing the     geometric connection between the cost function and the structure     of the surface. For very large damping parameters, the model graph     metric becomes a multiple of the parameter space metric, which     rotates the Gauss-Newton direction into the gradient direction.     The damping term therefore interpolates between the parameter     space metric and the data space metric.}

\end{figure}

By considering the model graph instead of the model manifold, we can remove the problems associated with the model boundaries. We return to our example problem to illustrate this point. The embedding space for the model graph is $3+2=5$ dimensional, so we are restricted to viewing $3$ dimensional projections of the embedding space. In Fig.~\ref{fig:ModelGraph} we illustrate the model graph (Levenberg metric) for $\lambda=0$, which is simply the model manifold, and for $\lambda\neq0$, which shows that boundaries of the model manifold are removed in the graph. Since the boundaries occur at $\theta=\infty$, they are infinity far from the origin on the model graph. Even the boundary corresponding to the fold line has been removed, as the fold has opened up like a folded sheet of paper. Since generic boundaries correspond to singular points of the metric, the model graph has no such boundaries as its metric is positive definite for any $\lambda>0$.

\begin{figure*}

  \includegraphics[width=7in]{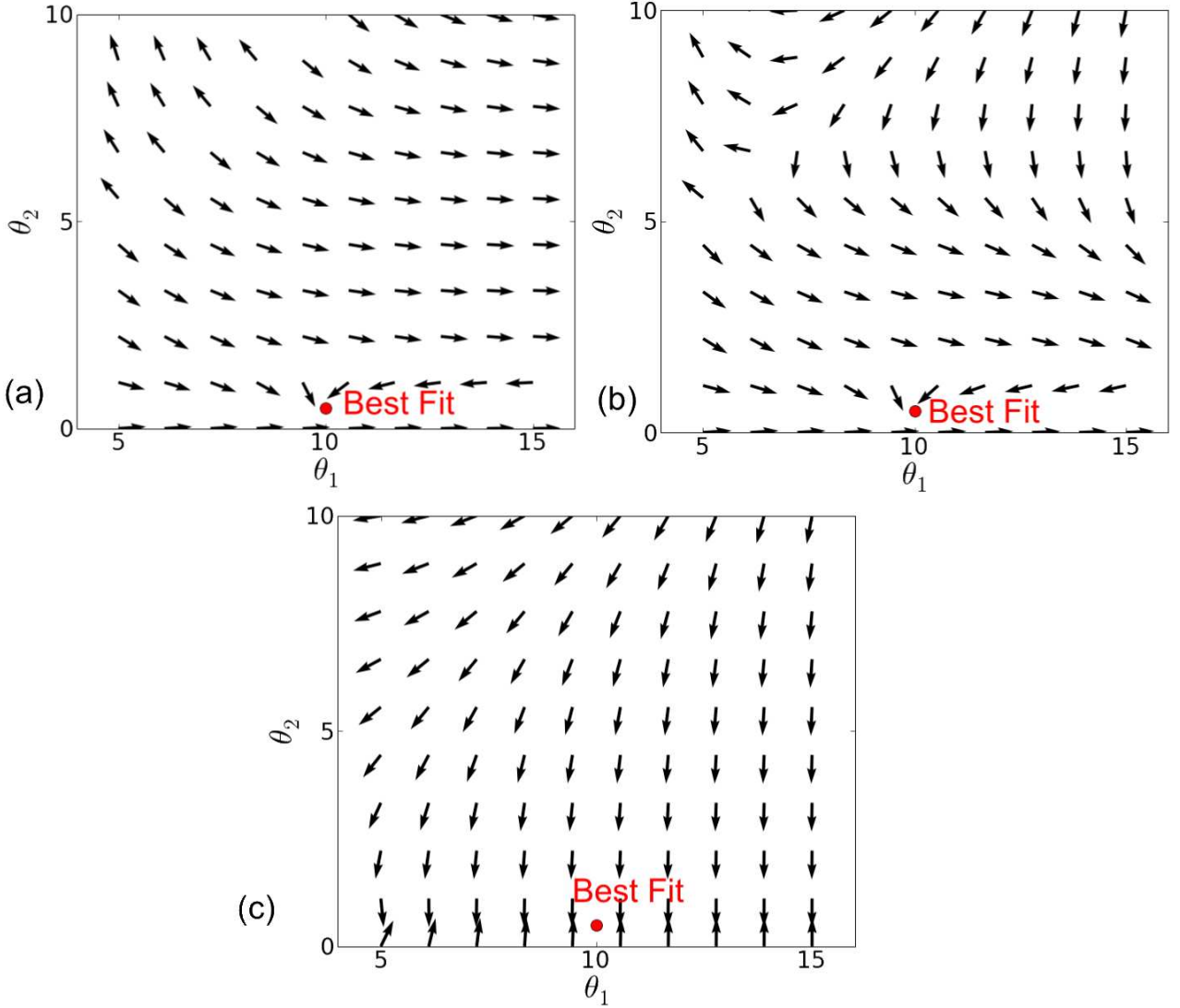}\caption{\label{fig:DirectionRotation}(A)\textbf{Gauss-Newton       Directions.}  The Gauss-Newton direction is prone to pointing     parameters toward infinity, especially in regions where the     metric has very small eigenvalues.  (B) \textbf{Rotated       Gauss-Newton Directions.} By adding a small damping parameter to     the metric, the Gauss-Newton direction is rotated into the     gradient direction. The amount of rotation is determined by the     eigenvalues of the metric at any given point. Here, only a few     points are rotated significantly. (C) \textbf{Gradient       Directions.} For large values of the damping parameter, the     natural direction is rotated everywhere into the gradient     direction.}

\end{figure*}

After removing the boundaries associated with the model manifold, the next advantage of the model graph is to provide a means of seamlessly interpolating between the natural directions of both data space and parameter space. The damping term, $\lambda$, appearing in Eq.~\eqref{eq:GraphMetric} is well suited for this interpolation in sloppy models. If we consider the Levenberg metric, the eigenvectors of the model manifold metric, $g^{0}$, are unchanged by adding a multiple of the identity. However, the corresponding eigenvalues are shifted by the $\lambda$ parameter.  It is the sloppy eigenvalues that are dangerous to the Gauss-Newton direction. Since the eigenvalues of a sloppy model span many orders of magnitude, this means that all the eigenvalues that were originally less than $\lambda$ are cutoff at $\lambda$ in the model graph metric, and the larger eigenvalues are virtually unaffected. By adjusting the damping term, we can essentially wash out the effects of the sloppy directions and preserve the Gauss-Newton direction from the model manifold in the stiff directions. Since the eigenvalues span many orders of magnitude, the parameter does not need to be finely tuned; it can be adjusted very roughly and an algorithm will still converge, as we will see in section~\ref{sec:Applications-to-Algorithms}. We demonstrate how $\lambda$ can interpolate between the two natural directions for our example model in Fig.~\ref{fig:DirectionRotation}.%

\section{Priors\label{sec:Priors}}

In Bayesian statistics, a {\em prior} is an {a-priori} probability
distribution in parameter space, giving information about the relative
probability densities for the model as parameters are varied. For example,
if one has pre-existing measurements of the parameters 
$\theta_m= \theta^0_m \pm \sigma_m$ with normally distributed uncertainties,
then the probability density would be 
$\prod_m 1/\sqrt{2\pi\sigma_m^2} \exp \left[ -(\theta_m-\theta^0_m)^2/(2 \sigma_m^2 \right] $
before fitting to the current data. This corresponds to a 
negative-log-likelihood cost that (apart from an overall constant) is the sum 
of squares, which can be nicely interpreted as the effects of an additional
set of ``prior residuals'' 
\begin{equation}
\label{eq:PriorResiduals}
r_m = (\theta_m - \theta^0_m)/\sigma_m
\end{equation}
(interpreting the pre-existing measurements as extra data points). In this
section, we will explore the more general use of such extra terms, not
to incorporate information about parameter values, but rather to 
incorporate information about the ranges of parameters expected to be useful
in generating good fits.

That is, we want to use priors to prevent parameter combinations which
are not constrained by the data from taking excessively large values --
we want to avoid parameter evaporation. To illustrate again why this
is problematic in sloppy models, consider a linear sloppy model with
true parameters $\theta_0$, but fit to data with added noise $\xi_i$.
The observed best fit is then shifted to 
$\theta = \theta_0 + (J^TJ)^{-1} (J^T) \xi$. The measurement error in
data space $\xi_i$ is thus multiplied by the inverse of the poorly conditioned
matrix $g = J^T J$, so even a small measurement error produces a large
parameter-space error. In section \ref{sub:parameter-evaporation}, we will see in nonlinear
models that such noise will generally shift the best fits to the 
boundary (infinite parameter values) along directions where the noise is large
compared to the width of the model manifold. Thus for example in fitting
exponentials, positive noise in the data point at $t_0=0$ and negative noise
at the data point at the first time $t_1>0$ can lead to one decay rate that
evaporates to infinity, tuned to fit the first data point without affecting
the others. 

In practice, it is not often useful to know that the optimum value of a
parameter is actually infinite -- especially if that divergence is clearly due
to noise. Also, we have seen in Fig.~\ref{fig:NaturalDirections}a that, even if the best fit has
sensible parameters, algorithms searching for the best fits can be led toward
the model manifold boundary. If the parameters are diverging
at finite cost, the model must necessarily become insensitive to the 
diverging parameters, often leading the algorithm to get stuck. Even a very
weak prior whose residuals diverge at the model manifold boundaries can
prevent these problems, holding the parameters in ranges useful for fitting
the data.

In this section, we advocate the use of priors for helping algorithms
navigate the model manifold in finding good fits. These priors are 
pragmatic; they are not introduced to buffer a model with `prior knowledge'
about the system, but to use the data to guess the parameter ranges outside
of which the fits will become insensitive to further parameter changes.
Our priors do not have meaning in the Bayesian sense, and indeed should
probably be relaxed to zero at late stages in the fitting process. 

The first issue is how to guess what ranges of parameter are useful
in fits -- outside of which the model behavior becomes insensitive to 
the parameter values. Consider, for example, the Michaelis-Mentin 
reaction, a saturable reaction rate often arising in systems biology
(for example Reference~\cite{Brown2004}):
\begin{equation}
\frac{d[x^*]}{dt}=\frac{k_x [y^*] [x]}{1+km_x [x]}.
\end{equation}
Here there are two parameters $k_x$ and $km_x$, governing the rate
of production of $[x^*]$ from $[x]$ in terms of the concentration
$[y^*]$, where $[x]+[x^*] = x_{max}$ and $[y]+[y^*] = y_{max}$. 

Several model boundaries can be identified here.  If $k_x$ and $km_x x_{max}$ are both very large, then only their ratio affects the dynamics.  In addition if $km_x$ is very small then it has no effect on the model.  Our prior should enforce our belief that $km_x [x]$ is typically of order $1$.  If it were much larger than one, than we could have modeled the system with one less parameter $k=k_x/km_x$ and if it were much less than one, the second term in the denominator could have been dropped entirely.  Furthermore, if the data is best fit by one of these boundary cases, say $km_x x_{max} \rightarrow \infty$ , it will be fit quite well by taking $km_x x_{max} >> 1$, but otherwise finite.  In a typical model we might expect that $km_x x_{max} =10$ will behave as if it were infinite.

We can also place a prior on $k_x$. Dimensional analysis here involves the time scale at which the model is predictive.  The prior should match the approximate time scale of the model's predictions to the rate of the modeled reaction.  For example, if an experiment takes time series data with precision on the order of seconds with intervals on the order minutes, then a 'fast' reaction is any that takes place faster than a few seconds and a slow reaction is any that happens over a few minutes.  Even if the real reaction happens in microseconds, it makes no sense to extract such information from the model and data.  Similarly, a slow reaction that takes place in years could be well fit by any rate that is longer than a few minutes.  As such we want a prior which prevents $k_x y_{max} x_{max}/\tau$ from being far from $1$, where $\tau$ is the typical timescale of the data, perhaps a minute here.  In summary, we want priors to constrain both $km_x x_{max}$ and $k_x x_{max} y_{max} / \tau$ to be of order one.

We have found that a fairly wide range of priors can be very effective at minimizing the problems associated with parameter evaporation during fitting.  To choose them, we propose starting by changing to the natural units of the problem by dividing by constants, such as time scales or maximum protein concentrations, until all of the parameters are dimensionless.  (Alternatively, priors could be put into the model in the original units, at the expense of more complicated book-keeping.)  In these natural units we expect all parameters to be order 1.

The second issue is to choose a form for the prior. For parameters like these, where both large and near-zero values are to be avoided, we add two priors for every parameter, one which punishes high values, and one which punishes small values:
\begin{equation}
Pr(\theta) = 
\left(\begin{array}{c}
\sqrt{w_h \theta} \\
\sqrt{w_l / \theta} \end{array}\right).
\end{equation}
This prior has minimum contribution to the cost when $\theta^2=\frac{w_l}{w_h}$ so in the proper units we choose $w_h=w_l$.  With these new priors, the metric becomes
\begin{eqnarray}
g_{\mu\nu} & = & \partial_\mu r^{0i} \partial_\nu r^{0i} 
+\partial_{\mu} Pr(\theta) \partial_{\nu} Pr(\theta) \\
& = & g_{\mu\nu}^0 + \delta_{\mu\nu} (\frac{w_l}{\theta^{\mu}} + w_h \theta^{\mu}), \label{eq:PriorMetric}
\end{eqnarray}
which is positive definite for all (positive) values of $\theta$.
As boundaries occur when the metric has an eigenvalue of zero, no boundaries exist for this new model manifold.  This is reminiscent of the metric of the model graph with the difference being that we have permanently added this term to the model.  The best fit has been shifted in this new metric.  

It remains to choose $w_h$ and $w_l$.  Though the choice is likely to be somewhat model specific, we have found that a choice between .001 and $1$ tends to be effective.  That weights of order $1$ can be effective is somewhat surprising.  It implies that good fits can be found while punishing parameters for differing only an order of magnitude from their values given by dimensional analysis.  That this works is a demonstration of the extremely ill-posed nature of these sloppy models, and the large ensemble of potential good fits in parameter space.

A complimentary picture of the benefit of priors takes place in parameter space, where they contribute to the cost:
\begin{equation}
C=C_0+\sum_i w_{h}\theta_i/2+ w_{l}/(2 \theta_i).
\end{equation}
The second derivative of the extra cost contribution with respect to the log of the parameters is given by $\frac{\partial^{2}}{\partial\log\left(\theta\right)^{2}}\left(\frac{Pr(\theta)^{2}}{2}\right)=\frac{w_h \theta}{2}+\frac{w_l}{2 \theta}$.  This is positive definite and concave, making the entire cost surface large when parameters are large.  This in turn makes the cost surface easier to navigate by removing the problems associated with parameter evaporation on plateaus.

\begin{figure}

\includegraphics[width=3in]{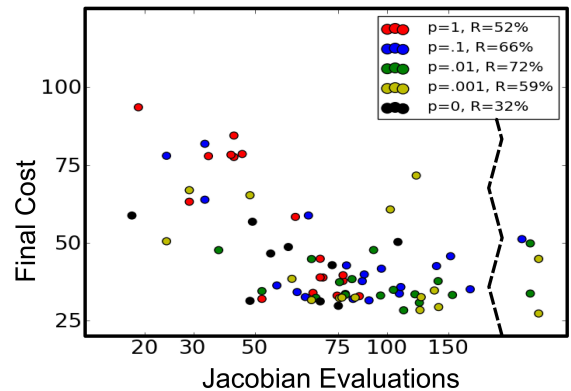}

\caption{\label{fig:PriorsResults} The final cost is plotted against  number of Jacobian evaluations for five strengths of priors.  After  minimizing with priors, the priors are removed and a maximum of 20  further Jacobian evaluations are performed. The prior strength is measured by $p$, with $p=0$ meaning no prior.  The success rate is $R$. The strongest priors converge the fastest, with medium strength priors showing the highest success rate. }

\end{figure}

To demonstrate the effectiveness of this method, we use the PC12 model with 48 parameters described in~\cite{Brown2004}.  We change to dimensionless units as described above.  To create an ensemble, we start from 20 initial conditions, with each parameter taken from a Gaussian distribution in its log centered on 0 (the expected value from dimensional analysis), with a $\sigma=\log 10$ (so that the bare parameters range over roughly two orders of magnitude from .1 to 10).  We put a prior as described above centered on the initial condition, with varying weights.   These correspond to the priors that we would have calculated if we had found those values by dimensional analysis instead.  After minimizing with the priors, we remove them and allow the algorithm to re-minimize.  The results are plotted in Fig.~\ref{fig:PriorsResults}. 

Strikingly, even when a strong prior is centered at parameter values
a factor of $\sim 100$ away from their `true' values, the addition of the prior
in the initial stages of convergence dramatically increases the speed
and success rate of finding the best fit.

In section~\ref{sec:The-Model-Graph}, we introduced the model graph and the Levenberg-Marquardt
algorithm, whose rationale (to avoid parameter evaporation) was similar
to that motivating us here to introduce priors. To conclude this section,
we point out that the model graph metric, Eq.~\eqref{eq:GraphMetric}, and the metric
for our particular choice of prior, Eq.~\eqref{eq:PriorMetric}, both serve to cut off
large steps along sloppy directions. Indeed, the Levenberg-Marquardt
algorithm takes a step identical to that for a model with quadratic priors
(Eq.~\eqref{eq:PriorResiduals}) with $\sigma_m \equiv 1/\sqrt{\lambda}$, except
that the center of the prior is not a fixed set of parameters $\theta_0$, 
but the current parameter set $\theta^*$. (That is, the second derivative of 
the sum of the squares of these residuals, 
$\sum_m [\sqrt{\lambda}(\theta-\theta^*)]^2$ gives $\lambda \delta_{\mu\nu}$,
the Levenberg term in the metric.) This Levenberg term thus acts as a 
`moving prior' -- acting to limit individual algorithmic steps from moving
too far toward the model boundary, but not biasing the algorithm permanently
toward sensible values. Despite the use of a variable $\lambda$ that can 
be used to tune the algorithm toward sensible behavior (Fig.~\ref{fig:DirectionRotation}), we shall
see in section~\ref{sec:Applications-to-Algorithms} that the Levenberg-Marquardt algorithm often fails,
usually because of parameter evaporation. When the useful ranges of 
parameters can be estimated beforehand, adding priors can be a remarkably
effective tool.

\section{Extended Geodesic Coordinates\label{sec:Extended-Geodesic-Coordinates}}

We have seen that the two difficulties of optimizing sloppy models are that algorithms tend to run into the model boundaries and that model parametrization tends to form long, curved canyons around the best fit. We have discussed how the first problem can be improved by the introduction of priors. We now turn our attention to the second problem. In this section we consider the question of whether we can change the parameters of a model in such a way as to remove this difficulty.  We construct coordinates geometrically by considering the motion of geodesics on the manifold.

Given two nearby points on a manifold, one can consider the many paths that connect them. If the points are very far away, there may be complications due to the boundaries of the manifold. For the moment, we assume that the points are sufficiently close that boundaries can be ignored.  The unique path joining the two points whose distance is shortest is known as the geodesic. The parameters corresponding to a geodesic path can be found as the solution of the differential equation

\begin{equation}
  \ddot{x}^{\mu}+\Gamma_{\alpha\beta}^{\mu}\dot{x}^{\alpha}\dot{x}^{\beta}=0,\label{eq:GeodesicODE}
\end{equation}
where $\Gamma_{\alpha\beta}^{\mu}$ are the connection coefficients given by Eq.~\eqref{eq:Connections-raised} and the dot means differentiation with respect to the curve's affine parametrization. Using two points as boundary values, the solution to the differential equation is then the shortest distance between the two points. Alternatively, one can specify a geodesic with an initial point and direction. In this case, the geodesic is interpreted as the path drawn by parallel transporting the tangent vector (also known as the curve's velocity). This second interpretation of geodesics will be the most useful for understanding the coordinates we are about to construct. The coordinates that we consider are polar-like coordinates, with $N-1$ angular coordinates and one radial coordinate.

If we consider all geodesics that pass through the best fit with a normalized velocity, $v^{\mu}v_{\mu}=1$, then each geodesic is identified by $N-1$ free parameters, corresponding to direction of the velocity at the best fit. (The normalization of the velocity does not change the path of the geodesic -- only the time it takes to traverse the path.) These $N-1$ free parameters will be the angular coordinates of the new coordinate system. There is no unique way of defining the angular coordinates. One can choose $N$ orthonormal unit vectors at the best fit, and let the angular coordinates define a linear combination of them. We typically choose eigendirections of the metric (the eigenpredictions of section~\ref{sec:ModelManifold}). Having specified a geodesic with the $N-1$ angular coordinates, the radial coordinate represents the distance moved along the geodesic. Since we have chosen the velocity vector to be normalized to one, the radial component is the parametrization of the geodesic.

We refer to these coordinates as extended geodesic coordinates and denote their Cartesian analog by $\gamma^{\mu}$. These coordinates have the special property that those geodesics that pass through the best fit appears as straight lines in parameter space. (It is impossible for all geodesics to be straight lines if the space is curved.)

In general, one cannot express this coordinate change in an analytic form. The quadratic approximation to this transformation is given by \begin{equation}   \gamma^{\nu}\approx\theta_{bf}^{\nu}+v_{\mu}^{\nu}\delta\theta^{\mu}+\frac{1}{2}\Gamma_{\alpha\beta}^{\nu}\delta\theta^{\alpha}\delta\theta^{\beta}.\label{eq:RNC}\end{equation} The coordinates given in Eq.~\eqref{eq:RNC} are known as Riemann normal coordinates or geodesic coordinates. Within the general relativity community, these coordinates are known as locally inertial reference frames because they have the property that $\Gamma_{\mu\nu}^{\alpha}(x=0)=0$, that is, the Christoffel symbols vanish at the special point around which the coordinates are constructed~\cite{Misner1973}.

\begin{figure}

  \includegraphics[width=3.25in]{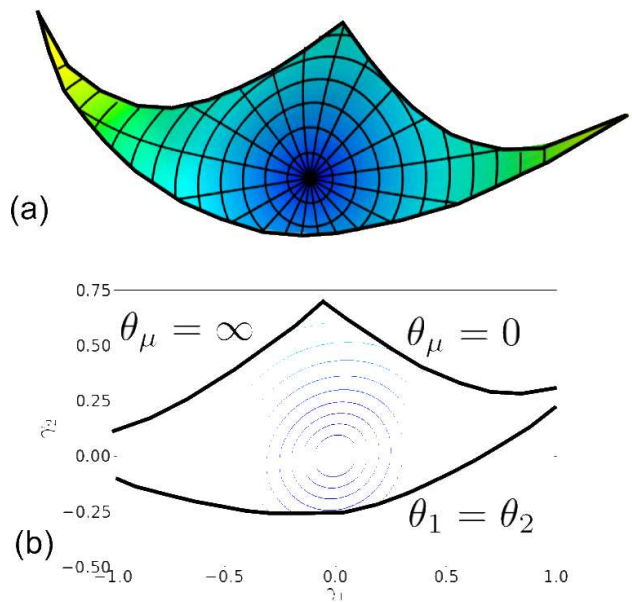}\caption{\label{fig:GeodesicCoords}a)     \textbf{Extended Geodesic Coordinates.}  The parameters of a model     are not usually well suited to describing the behavior of a     model. By considering the manifold induced in data space, one can     construct more natural coordinates based on geodesic motion that     are more well-suited to describing the behavior of a model (black     grid). These coordinates remove all parameter-effects curvature     and are known as extended geodesic coordinates. Note that we have     moved the data point so that the best fit is not so near a     boundary in this picture. b) \textbf{Cost Contours in Extended       Geodesic Coordinates.}  Although the summing exponential model     is nonlinear, that non-linearity does not translate into large     extrinsic curvature. This type of non-linearity is known as     parameter-effects curvature, which the geodesic coordinates     remove. This is most dramatically illustrated by considering the     contours of constant cost in geodesic coordinates. The contours     are nearly circular all the way out to the fold line and the     boundary where the rates are infinite.}

\end{figure}

Let us now consider the shape of cost contours for our example model using extended geodesic coordinates. We can consider both the shape of the coordinate mesh on the manifold in data space, as well as the shape of the cost contours in parameter space. To illustrate the dramatic effect that these coordinates can have, we have adjusted the data so that the best fit does not lie so near the boundary. The results are in Fig.~\ref{fig:GeodesicCoords}.

The extended geodesic coordinates were constructed to make the elongated ellipse that is characteristic of sloppy models become circular. It was hoped that by making the transformation nonlinear, it would straighten out the an-harmonic {}``banana'' shape, rather than magnify it.  It appears that this wish has been granted spectacularly. Not only has the banana been straightened out within the region of the long narrow canyon, but the entire region of parameter space, including the plateau, has been transformed into one manageable, isotropic basin.  Indeed, the cost contours of Fig.~\ref{fig:GeodesicCoords}b are near-perfect circles, all the way to the boundary where the rates go to zero, infinity, or are equal.

To better understand how this elegant result comes about, let's consider how the cost changes as we move along a geodesic that passes through the best fit. The cost then becomes parametrized by the same parameter describing the geodesic, which we call $\tau$. The chain rule gives us, \[ \frac{d}{d\tau}=\frac{d\theta^{\mu}}{d\tau}\frac{\partial}{\partial\theta^{\mu}}=v^{\mu}\partial_{\mu},\] where $v^{\mu}=\dot{\theta}^{\mu}$. Applying this twice to the cost gives: \begin{equation}   \frac{d^{2}C}{d\tau^{2}}=v^{\mu}v^{\nu}g_{\mu\nu}+r_{m}P_{mn}^{N}\partial_{\mu}\partial_{\nu}r_{n}\frac{d\theta^{\mu}}{d\tau}\frac{d\theta^{\nu}}{d\tau}.\label{eq:d2C}\end{equation} The term $v^{\mu}v^{\nu}g_{\mu\nu}$ in Eq.~\eqref{eq:d2C} is the arbitrarily chosen normalization of the velocity vector and is the same at all points along the geodesic. The interesting piece in Eq.~\eqref{eq:d2C} is the expression \[ P^{N}=\delta-J\left(J^{T}J\right)^{-1}J^{T},\] which we recognize as the projection operator that projects out of the tangent space (or into the normal bundle).

Recognizing $P^{N}$ in Eq.~\eqref{eq:d2C}, we see that any deviation of the quadratic behavior of the cost will be when the non-linearity forces the geodesic out of the tangent plane, which is to say that there is an extrinsic curvature. When there is no such curvature, then the cost will be isotropic and quadratic in the extended geodesic coordinates.

If the model happens to have as many parameters as residuals, then the tangent space is exactly the embedding space and the model will be flat. This can be seen explicitly in the expression for $P^{N}$, since $J$ will be a square matrix if $M=N$, with a well-defined inverse:\begin{eqnarray*}
  P^{N} & = & \delta-J\left(J^{T}J\right)^{-1}J^{T}\\
  & = & \delta-JJ^{-1}\left(J^{T}\right)^{-1}J^{T}\\
  & = & 0.\end{eqnarray*} Furthermore, when there are as many parameters as residuals, the extended geodesic coordinates can be chosen to be the residuals themselves, and hence the cost contours will be concentric circles.

In general, there will be more residuals than parameters; however, we have seen in section~\ref{sec:Manifolds-with-Boundaries} that many of those residuals are interpolating points that do not supply much new information. Assuming that we can simply discard a few residuals, then we can {}``force'' the model to be flat by restricting the embedding space. It is, therefore, likely that for most sloppy models, the manifold will naturally be much more flat than one would have expected. We will see when we discuss curvature in section~\ref{sec:Curvature} that most of the non-linearities of a sloppy model do not produce extrinsic curvature, meaning the manifold is typically much more flat that one would have guessed.

Non-linearities that do not produce extrinsic curvature are described as parameter-effects curvature~\cite{Bates1980}. As the name suggests these are {}``curvatures'' that can be removed through a different choice of parameters. By using geodesics, we have found a coordinate system on the manifold that removes all parameter-effects curvature at a point. It has been noted previously that geodesics are linked to zero parameter-effects curvature~\cite{Kass1984}.

We believe it to be generally true for sloppy models that non-linearities are manifested primarily as parameter-effects curvature as we argue in~\cite{Transtrum2010} and in section~\ref{sec:Curvature}. We find similar results when we consider geodesic coordinates in the PC12 model, neural networks, and many other models. Just as for the summing exponential problem that produced Fig.~\ref{fig:GeodesicCoords}b, cost contours for this real-life model are nearly circular all the way to the model's boundary.

\begin{figure*}
  \includegraphics[scale=0.33]{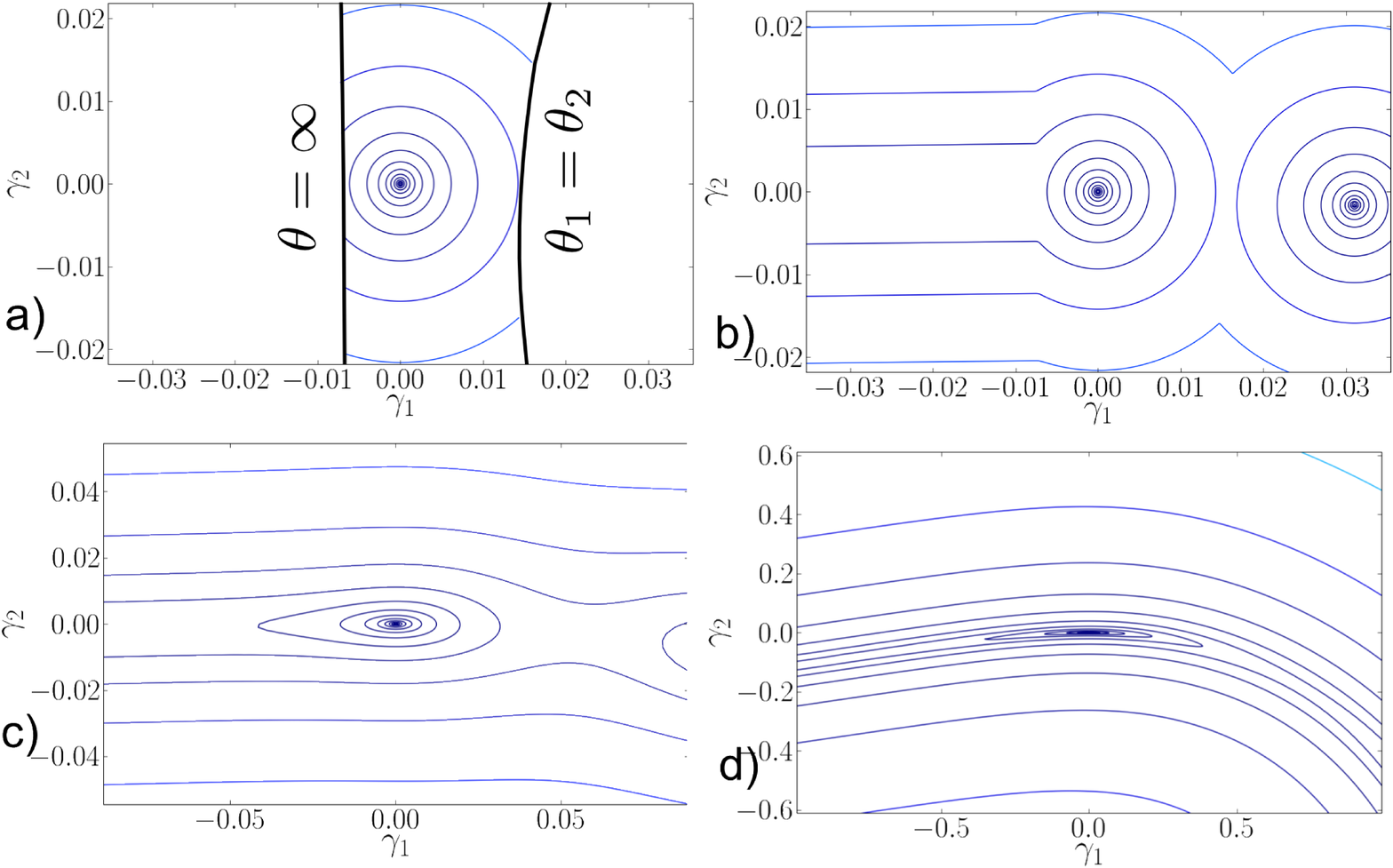}\caption{\label{fig:DeformedContours}By     changing the value of the Levenberg-Marquardt parameter, the     course of the geodesics on the corresponding model graph are     deformed, in turn distorting the shape of the cost contours in the     geodesic coordinates. a) $\lambda=0$ is equivalent to the model     manifold. The cost contours for a relatively flat manifold, such     as that produced by the sum of two exponentials, are nearly     perfect, concentric circles. The geodesics can be evaluated up to     the boundary of the manifold, at which point the coordinates are     no longer defined.  Here we can clearly see the stiff, long     manifold direction (vertical) and the sloppy, thin manifold     direction (horizontal) b) \textbf{Small $\lambda$}, ($\lambda$     much smaller than any of the eigenvalues of the metric) will     produce cost contours that are still circular, but the manifold     boundaries have been removed. In this case the fold line has     disappeared, and cost contours that ended where parameters     evaporated now stretch to infinity. c) \textbf{Moderate $\lambda$     }creates cost contours that begin to stretch in regions where the     damping parameter significantly affects the eigenvalue structure     of the metric. The deformed cost contours begin to take the     plateau and canyon structures of the contours in parameter space.     d) \textbf{Large $\lambda$ }effectively washes out the information     from the model manifold metric, leaving just a multiple of the     parameter space metric.  In this case, the contours are those of     parameter space -- a long narrow curved canyon around the best     fit. This figure analogous to Fig.~\ref{fig:Fitting}b, although     the model here is a more sloppy (and more realistic) example.}
\end{figure*}

Although the model manifold is much more flat than one would have guessed, how does that result compare for the model graph? We observed in section~\ref{sec:The-Model-Graph}, that the model graph interpolates between the model manifold and the parameter space picture. If we find the cost contours for the model graph at various values of $\lambda$, we can watch the cost contours interpolate between the circles in Fig.~\ref{fig:GeodesicCoords}b and the long canyon that is characteristic of parameter space. This can be seen clearly in Fig.~\ref{fig:DeformedContours}.

With any set of coordinates, it is important to know what portion of the manifold they cover. Extended geodesic coordinates will only be defined in some region around the best fit. It is clear from Fig.~\ref{fig:GeodesicCoords} that for our example problem the region for which the coordinates are valid extends to the manifold boundaries. Certainly there are regions of the manifold that are inaccessible to the geodesic coordinates.  Usually, extended geodesic coordinates will be limited by geodesics reaching the boundaries, just as algorithms are similarly hindered in finding the best fit.

\section{Curvature\label{sec:Curvature}}

In this section, we discuss the various types of curvature that one might expect to encounter in a least-squares problem and the measures that could be used to quantify those curvatures. Curvature of the model manifold has had many interesting applications. It has been illustrated by Bates and Watts that the curvature is a convenient measure of the non-linearity of a model~\cite{Bates1980,Bates1981,Bates1988}.  When we discuss the implications of geometry on numerical algorithms this will be critical, since it is the non-linearity that makes these problems difficult.

Curvature has also been used to study confidence regions~\cite{Bates1981,Hamilton1982,Cook1986,Donaldson1987,Wei1994}, kurtosis (deviations from normality) in parameter estimation~\cite{Haines2004}, and criteria for determining if a minimum is the global minimizer~\cite{Demidenko2006}. We will see below that the large anisotropy in the metric produces a similar anisotropy in the curvature of sloppy models. Furthermore, we use curvature as a measure of how far an algorithm can accurately step (section~\ref{sub:kappa}) and to estimate how many parameters a best fit will typically evaporate (section~\ref{sub:parameter-evaporation}).

In our discussion of geodesic coordinates in section~\ref{sec:Extended-Geodesic-Coordinates}, we saw how some of the non-linearity of a model could be removed by a clever choice of coordinates. We also argued that the non-linearity that could not be removed by a coordinate change would be expressed as an extrinsic curvature on the expectation surface. Non-linearity that does not produce an extrinsic curvature is not irrelevant; it can still have strong influence on the model and can still limit the effectiveness of optimization algorithms. Specifically, this type of non-linearity changes the way that distances are measured on the tangent space. They may cause the basis vectors on the tangent space to expand, shrink, or rotate. We follow the nomenclature of Bates and Watts and refer to this type of non-linearity as parameter-effects curvature~\cite{Bates1980,Bates1988}. We emphasize that this is not a {}``real'' curvature in the sense that it does not cause the shape of the expectation surface to vary from a flat surface, but its effects on the behavior of the model is similar to the effect of real curvature. This {}``curvature'' could be removed through a more convenient choice of coordinates, which is precisely what we have done by constructing geodesic coordinates in section~\ref{sec:Extended-Geodesic-Coordinates}.  A functional definition of parameter-effects curvature would be the non-linearities that are annihilated by operating with $P^{N}$.  Alternatively, one can think of the parameter-effects curvature as the curvatures of the coordinate mesh.  We discuss parameter-effects curvature in section~\ref{sub:Parameter-effects-Curvature}.

Bates and Watts refer to all non-linearity that cannot be removed by changes of coordinates as intrinsic curvature~\cite{Bates1988}.  We will not follow this convention; instead, we follow the differential geometry community and further distinguish between intrinsic or Riemann curvature (section~\ref{sub:Intrinsic-(Riemann)-Curvature}) and extrinsic or embedding curvature~\cite{Spivak1979} (section~\ref{sub:Extrinsic-Curvature}).  The former refers to the curvature that could be measured on a surface without reference to the embedding. The latter refers to the curvature that arises due to the manner in which the model has been embedded.  From a complete knowledge of the extrinsic curvature, one could also calculate the intrinsic curvature. Based on our discussion to this point, one would expect that both the intrinsic and the extrinsic curvature should be expressible in terms of some combination of $P^{N}$ and $\partial_{\mu}\partial_{\nu}r_{m}$. This turns out to be the case, as we will shortly see.

All types of curvature appear in least squares models, and we will now discuss each of them.

\subsection{Intrinsic (Riemann) Curvature \label{sub:Intrinsic-(Riemann)-Curvature}}

The embedding plays a crucial role in nonlinear least squares fits -- the residuals embed the model manifold explicitly in data space -- we will be primarily interested in the extrinsic curvature.  However, because most studies of differential geometry focus on the intrinsic curvature, we discuss it.

The Riemann curvature tensor, $R_{\beta\gamma\delta}^{\alpha}$ is one measure of intrinsic curvature. Since intrinsic curvature makes no reference to the embedding space, curvature is measured by moving a vector, $V^{\mu}$, around infinitesimal closed loops and observing the change the curvature induces on the vector, which is expressed mathematically by \[ R_{\beta\gamma\delta}^{\alpha}V^{\beta}=\nabla_{\gamma}\nabla_{\delta}V^{\alpha}-\nabla_{\delta}\nabla_{\gamma}V^{\alpha}.\] This expression in turn can be written independently of $V^{\mu}$ in terms of the Christoffel symbols and their derivatives by the standard formula \[ R_{\beta\gamma\delta}^{\alpha}=\partial_{\gamma}\Gamma_{\beta\delta}^{\alpha}-\partial_{\delta}\Gamma_{\beta\gamma}^{\alpha}+\Gamma_{\beta\delta}^{\epsilon}\Gamma_{\epsilon\gamma}^{\alpha}-\Gamma_{\beta\gamma}^{\epsilon}\Gamma_{\epsilon\delta}^{\alpha}.\] From this we can express $R_{\beta\gamma\delta}^{\alpha}$ in terms of derivatives of the residuals. Even though $R_{\beta\gamma\delta}^{\alpha}$ depends on derivatives of $\Gamma$, suggesting that it would require a third derivative of the residuals, one can in fact represent it in terms of second derivatives and $P^{N}$,\[ R_{\alpha\beta\gamma\delta}=\partial_{\alpha}\partial_{\gamma}r_{m}P_{mn}^{N}\partial_{\beta}\partial_{\delta}r_{n}-\partial_{\alpha}\partial_{\delta}r_{m}P_{mn}^{N}\partial_{\beta}\partial_{\gamma}r_{n},\] which the Gauss-Codazzi equation extended to the case of more than one independent normal direction~\cite{Eisenhart1997}.

\begin{figure}

  \includegraphics[width=3.25in]{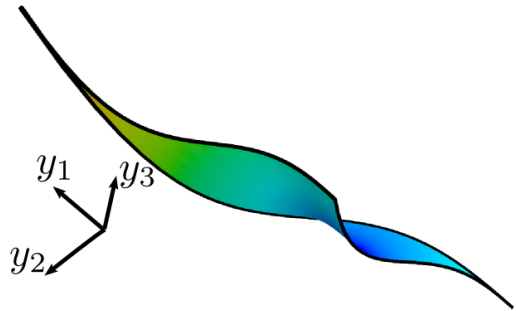}\caption{\label{fig:Intrinsic-Extrinsic-Curvature}\textbf{Intrinsic       and Extrinsic Curvature.} Intrinsic Curvature is inherent to the     manifold and cannot be removed by an alternative embedding. A     model that is the sum of two exponential terms has all types of     curvature. This is the same model manifold as in     Fig.~\ref{fig:Fitting}c, viewed from an alternative angle to     highlight the curvature. From this viewing angle, the extrinsic     curvature becomes apparent. This is also an example of intrinsic     curvature.}

\end{figure}

The toy model that we have used throughout this work to illustrate concepts has intrinsic curvature. The curvature becomes most apparent when viewed from another angle, as in Fig.~\ref{fig:Intrinsic-Extrinsic-Curvature}.

Intrinsic or Riemann curvature is an important mathematical quantity that is described by a single, four-index tensor; however, we do not use intrinsic curvature to study optimization algorithms. Extrinsic and parameter-effects curvature in contrast not be simple tensors but will depend on a chosen direction.  These curvatures are the key to understanding nonlinear least squares fitting.

\subsection{Extrinsic Curvature\label{sub:Extrinsic-Curvature}}

\begin{figure}
  \textbf{\includegraphics[width=3.25in]{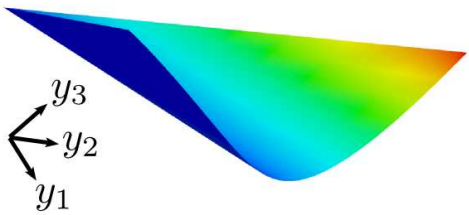}}\caption{\textbf{\label{fig:Ruled-Surface}A       ruled surface} has no intrinsic curvature; however, it may have     extrinsic curvature. The model manifold formed from a single     exponential rate and amplitude is an example of a ruled     surface. This model could be isometrically embedded in another     space to remove the curvature.}

\end{figure}

Extrinsic curvature is easier to visualize than intrinsic curvature since it makes reference to the embedding space, which is where one naturally imagines curved surfaces. It is important to understand that extrinsic and intrinsic curvature are fundamentally different and are not merely different ways of describing the same concept.  In differentiating between intrinsic and extrinsic curvature, the simplest illustrative example is a cylinder, which has no intrinsic curvature but does have extrinsic curvature. One could imagine taking a piece of paper, clearly a flat, two dimensional surface embedded in three dimensional space, and roll it into a cylinder. Rolling the paper does not affect distances on the surface, preserving its intrinsic properties, but changes the way that it is embedded in three dimensional space. The rolled paper remains intrinsically flat, but it now has an extrinsic curvature. A surface whose extrinsic curvature can be removed by an alternative, isometric embedding is known as a ruled surface~\cite{Hilbert1999}. While an extrinsic curvature does not always imply the existence of an intrinsic curvature, an intrinsic curvature requires that there also be extrinsic curvature. Our toy model, therefore, also exhibits extrinsic curvature as in Fig.~\ref{fig:Intrinsic-Extrinsic-Curvature}.  One model whose manifold is a ruled surface is given by a two parameter model which varies an exponential rate and an amplitude: \[ y=Ae^{-\theta t}.\] The manifold for this model with three data points is drawn in Fig.~\ref{fig:Ruled-Surface} \footnote{This example is also a separable nonlinear least squares problem.  Separable problems containing a mixture of linear and nonlinear parameters are amenable to the method known as variable projection~\cite{Golub1973,Kaufman1975,Golub2003}. Variable projection consists of first performing a linear least squares optimization on the linear parameters, making them implicit functions of the nonlinear parameters. The geometric effect of this procedure is to reduce the dimensionality of the model manifold, effectively selecting a sub-manifold which now depends upon the location of the data. We will not discuss this method further in this paper, but we note that it is likely to have interesting geometric properties.}.

There are two measures of extrinsic curvature that we discuss. The first is known as geodesic curvature as it measures the deviation of a geodesic from a straight line in the embedding space. The second measure is known as the shape operator. These two measures are complimentary, and should be used together to understand the way a space is curved.  Both geodesic curvature and the shape operator have analogous measures of parameter-effects curvature that will allow us to compare the relative importance of the two types of curvature.

Measures of extrinsic and parameter effects curvature to quantify non-linearities have been proposed previously by Bates and Watts~\cite{Bates1980,Bates1983,Bates1988}.  Although the measure they use is equivalent to the presentation of the next few sections, their approach is different. The goal of this section is to express curvature measures of non-linearity in a more standard way using the language of differential geometry. By so doing, we hope to make the results accessible to a larger audience.

\subsubsection{Geodesic Curvature \label{sub:Geodesic-Curvature}}

Consider a geodesic parametrized by $\tau$, tracing a path through parameter space, $\theta^{\mu}(\tau)$, which in turn defines a path through residual space, $\vec{r}(\theta(\tau))$. The parametrization allows us to discuss the velocity, $\vec{v}=\frac{d\vec{r}}{d\tau}$, and the acceleration, $\vec{a}=\frac{d\vec{v}}{d\tau}$. A little calculus puts these expressions in a more practical form: \[ \vec{v}=\dot{\theta}^{\mu}\partial_{\mu}\vec{r},\]
 \[
 \vec{a}=\dot{\theta}^{\mu}\dot{\theta}^{\nu}P^{N}\partial_{\mu}\partial_{\nu}\vec{r}.\]  Notice that the normal projection operator emerges naturally in the  expression for $\vec{a}$.

\begin{figure}
  \includegraphics[width=3.25in]{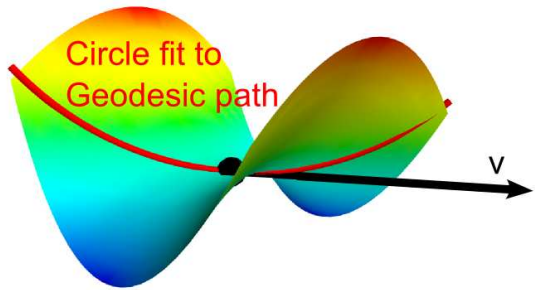}\caption{\label{fig:GeodesicCurvature}\textbf{Geodesic       Curvature. }A direction on a curved surface define a     geodesic. The deviation of the geodesic from a straight line in     the embedding space is measured by the geodesic curvature. It is     the inverse radius of the circle fit to the geodesic path at the     point. }

\end{figure}

For any curve that has instantaneous velocity and acceleration vectors, one can find a circle that local approximates the path. The circle has radius \[ R=\frac{v^{2}}{|\vec{a}|},\] and a corresponding curvature \[ K=R^{-1}=\frac{|\vec{a}|}{v^{2}}.\] Because the path that we are considering is a geodesic, it will be as near a straight line in data space as possible without leaving the expectation surface. That is to say, the curvature of the geodesic path will be a measure of how the surface is curving within the embedding space, i.e.~an extrinsic curvature. The curvature associated with a geodesic path is illustrated in Fig.~\ref{fig:GeodesicCurvature}.

In our previous discussion of geodesics, we saw that a geodesic is fully specified by a point and a direction. Therefore we can define the geodesic curvature of any point on the surface, corresponding to a direction, $v^{\mu}$, by \begin{equation}   K(v)=\frac{|v^{\mu}v^{\nu}P^{N}\partial_{\mu}\partial_{\nu}\vec{r}|}{v^{\alpha}v_{\alpha}}.\label{eq:GeodesicCurvature}\end{equation} At each point an the surface, there is a different value of the geodesic curvature for each direction on the surface.

\subsubsection{Shape Operator\label{sub:Shape-Operator}}

Another measure of extrinsic curvature, complimentary to the geodesic curvature, is the shape operator, $S_{\mu\nu}$. While the geodesic curvature requires us to choose an arbitrary direction on the surface, the shape operator requires us to choose an arbitrary direction normal to the surface.

To understand the shape operator, let us first consider the special case of an $N$-dimensional surface embedded in an $N+1$-dimensional space. If this is the case, then at any point on the surface there is a unique (up to a sign) unit vector normal to the surface, $\hat{n}$. If this is the case, $S_{\mu\nu}$ is given by \begin{equation}   S_{\mu\nu}=\hat{n}\cdot\left(\partial_{\mu}\partial_{\nu}\vec{r}\right).\label{eq:ShapeOperatorn}\end{equation}

\begin{figure}
  \includegraphics[width=3.25in]{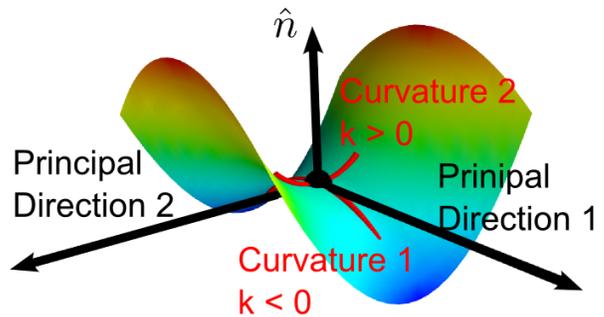}\caption{\label{fig:ShapeOperator}\textbf{Shape       Operator. }Specifying a direction normal to a curved surface,     $\hat{n}$, defines a shape operator.  The eigenvalues of the shape     operator are the principle curvatures and the corresponding     eigenvectors are the directions of principle curvature.}

\end{figure}

$S_{\mu\nu}$ is known as the shape operator because it describes how the surface is shaped around the unit normal, $\hat{n}$. It is a symmetric, covariant rank-2 tensor. We are usually interested in finding the eigenvalues of the shape operator with a single raised index: \[ S_{\nu}^{\mu}=g^{\mu\alpha}S_{\alpha\nu}.\] The eigenvectors of $S^{\mu}_{\nu}$ are known as the principal curvature directions, and the eigenvalues are the extrinsic curvatures in those directions.  In the case that there is only one direction normal to the surface, then the (absolute value of the) eigenvalues of $S_{\nu}^{\mu}$, are equal to the geodesic curvatures in the respective eigendirections.  The eigenvalues, $\left\{ k_{\mu}\right\} $, may be either positive or negative. Positive values indicate that the curvature is toward the direction of the normal, while negative values indicate that it is curving away, as illustrated in Fig.~\ref{fig:ShapeOperator}.

In general, there will not be an unique normal vector. If an $N$-dimensional surface is embedded in an $M$-dimensional space, then there will $M-N$ independent shape operators, and one is left to perform an eigenvalue analysis for each as described above~\cite{Spivak1979}.  Fortunately, for the case of a least squares problem, there is a natural direction to choose: the normal component of the unfit data, $-P^{N}\vec{r}$, making the shape operator \begin{equation}   S_{\mu\nu}=-\frac{\vec{r}P^{N}\partial_{\mu}\partial_{\nu}\vec{r}}{|P^{N}\vec{r}|},\label{eq:ShapeOperator}\end{equation} where we introduce the minus as convention. In general, around an arbitrary vector $\vec{V}$, the shape operator becomes \begin{equation}
  S(\vec{V})_{\mu\nu}=\frac{\vec{V}P^{N}\partial_{\mu}\partial_{\nu}\vec{r}}{|P^{N}\vec{V}|}.\label{eq:ShapeOperatorV}\end{equation}

It should now be clear why these two measures of extrinsic curvature (geodesic curvature and the shape operator) are complimentary. The geodesic curvature is limited by having to choose a direction tangent to the surface, but gives complete information about how that direction is curving into the space normal to the surface. In contrast, the shape operator gives information about all the directions on the surface, but only tells how those directions curve relative to a single normal direction.

\subsection{Parameter-effects Curvature\label{sub:Parameter-effects-Curvature}}

We are now prepared to discuss parameter-effects curvature. We repeat that parameter-effects curvature is not a curvature of the manifold. Instead, it is a measure of the curvatures of the coordinate mesh on the surface. In our experience, parameter-effects curvature is typically the largest of the three types we have discussed. By its very nature, this curvature depends on the choice of the parametrization.  By constructing extended geodesic coordinates in section~\ref{sec:Extended-Geodesic-Coordinates}, we were able to remove the parameter-effects curvature from the model (at a point). In this section we will discuss how to measure the parameter-effects curvature and compare it to the other curvatures that we discussed above.

\begin{figure*}
  \includegraphics[width=7in]{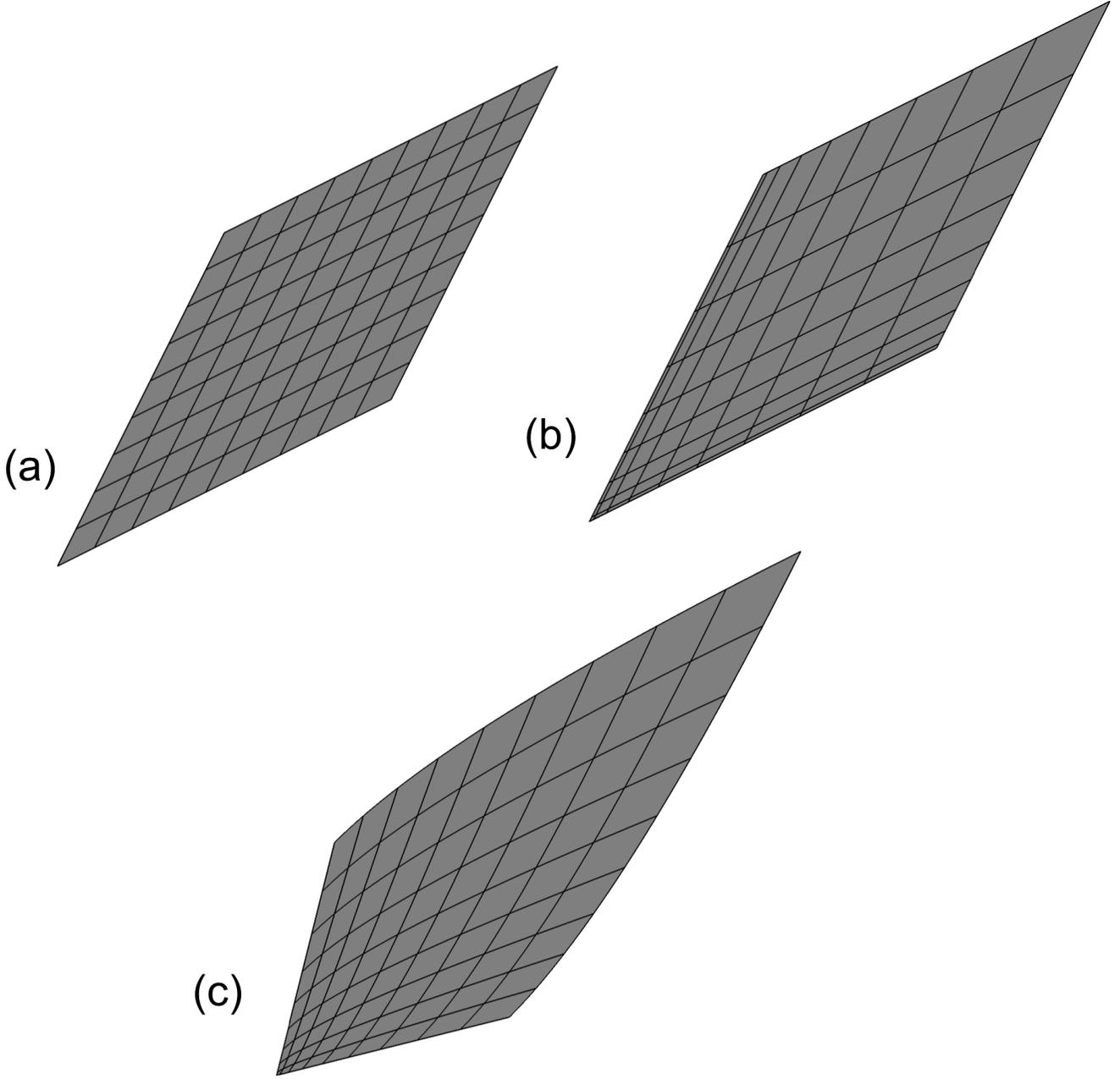}\caption{\label{fig:Parameter-effects-curvature}a)\textbf{       Linear Grid. }A sloppy linear model may have a skewed coordinate     grid, but the shape of the grid is constant, having no parameter     effects curvature.  b)\textbf{ Compressed Grid. }By     reparametrizing the model, the grid may become stretched or     compressed in regions of the manifold. c) \textbf{Rotating,       Compressed Grid. }Another parametrization may not only stretch     the grid, but also cause the coordinates to     rotate. Parameter-effects curvature describes the degree to which     the coordinates are stretching and rotating on the manifold.  With more than two parameters, there is also a torsion parameter-effects curvature (twisting).}

\end{figure*}

To understand the meaning of parameter-effects curvature, let us begin by considering a linear model with no curvature of any type. For simplicity, we consider the parametrization of the xy-plane given by\begin{eqnarray*}
  x & = & \epsilon\theta_{1}+\theta_{2}\\
  y & = & \theta_{1}+\epsilon\theta_{2}.\end{eqnarray*} This parametrization will produce a skewed grid as $\epsilon\rightarrow1$, characteristic of linear sloppy models, such as fitting polynomials. This grid is illustrated in Fig.~\ref{fig:Parameter-effects-curvature}a for $\epsilon=1/2$. By reparametrizing the linear model, we can introduce parameter-effects curvature. For example, if we replace the parameters with their squares (which may be useful if we wish to enforce the positivity of the parameters' effects) \begin{eqnarray*}
  x & = & \epsilon\theta_{1}^{2}+\theta_{2}^{2}\\
  y & = & \theta_{1}^{2}+\epsilon\theta_{2}^{2},\end{eqnarray*} then the corresponding coordinate mesh will become compressed and stretched, as seen in Fig.~\ref{fig:Parameter-effects-curvature}b. Alternatively, if we reparametrize the model as \begin{eqnarray*}
  x & = & \left(\epsilon\theta_{1}+\theta_{2}\right)^{2}\\
  y & = &   \left(\theta_{1}^{2}+\epsilon\theta_{2}^{2}\right)^{2},\end{eqnarray*} in order to limit the region of consideration to the upper-right quarter plane, then the coordinate mesh will stretch and rotate into itself, depicted in Fig.~\ref{fig:Parameter-effects-curvature}c.  With more than two parameters, there is additionally a torsion parameter-effects curvature in which the lines twist around one another.  None of these reparametrization change the intrinsic or extrinsic properties of the model manifold; they merely change how the coordinates describe the manifold. The extent to which coordinate mesh is nonlinear is measured by the parameter-effects curvature.

We now consider how to quantify parameter-effects curvature. We have discussed the normal and tangential projection operators, $P^{N}$ and $P^{T}$, and argued that the normal projection operator would extract the extrinsic and intrinsic curvature from the matrix of second derivatives. Looking back on our expressions for curvature up to this point, we see that each involves $P^{N}$. The complimentary parameter-effects curvature can be found by replacing $P^{N}$ with $P^{T}$ in each expression. Thus, in analogy with Eq.~\eqref{eq:GeodesicCurvature}, we can define the parameter-effects geodesic curvature by \begin{equation} K^{p}(v)=\frac{|v^{\mu}v^{\nu}P^{T}\partial_{\mu}\partial_{\nu}\vec{r}|}{v^{\alpha}v_{\alpha}}. \label{eq:ParameterEffectsCurvature} \end{equation} Likewise, we can define a parameter-effects shape operator by comparison with Eq.~\eqref{eq:ShapeOperator}, \[ S_{\mu\nu}^{p}=-\frac{\vec{r}P^{T}\partial_{\mu}\partial_{\nu}\vec{r}}{|P^{T}\vec{r}|}.\]

Recall that for an $N$-dimensional space embedded in an $M$-dimensional space, there are $M-N$ independent shape operators. This is because the space normal to the tangent space (into which we are projecting the non-linearity) is of dimension $M-N$. The parameter-effects analog must therefore have $N$ independent shape operators, since the projection space (the tangent space) is $N$-dimensional.  Therefore, we are naturally led to define a parameter-effects shape-operator with an additional index to distinguish among the $N$ possible tangent directions,
\[ S^P_{m\mu \nu} = P^T_{mn} \partial_{\mu} \partial_{\nu} r_n. \]
If we resolve these shape operators into the natural basis on the tangent space, $S^P_{m\mu \nu} = S^{p\alpha}_{\mu \nu} \partial_{\alpha}r_m$, we find
\[S^{P\alpha}_{\mu \nu} = g^{\alpha \beta} \partial_\beta \vec{r} \cdot \partial_{\mu} \partial_{\nu} \vec{r} = \Gamma^{\alpha}_{\mu\nu}. \]
Therefore, the parameter-effects curvature is correctly interpreted as the connection coefficients. With this understanding, it is clear that geodesic coordinates remove parameter-effects curvature, since they are the coordinates constructed to give $\Gamma=0$.

Finally, we note that from a complete knowledge of all the curvatures (for all directions) one can determine the matrix of second derivatives completely. Although we do not demonstrate this here, we note it is a consequence of having a flat embedding space.

\subsection{Curvature in Sloppy Models\label{sub:Curvature-in-Sloppy}}

Based on our analysis thus far, we should have two expectations regarding the curvature of sloppy models. First, because of the large spread of eigenvalues of the metric tensor, unit distances measured in parameter space correspond to large ranges of distances in data space. Conversely, one has to move the parameters by large amounts in a sloppy direction in order to change the residuals by a significant amount. Because of this, we expect that the anharmonicities in the sloppy directions will become magnified when we consider the curvature in those directions.  We expect strong anisotropies in the curvatures of sloppy models, with the largest curvatures corresponding to the sloppiest directions.

Secondly, as we saw in section~\ref{sec:Extended-Geodesic-Coordinates}, by changing coordinates to extended geodesic coordinates, we discovered that the manifold generated by our sloppy model was surprisingly flat, i.e.\ had low intrinsic curvature. We have seen that if the model happens to have equal number of data points as parameters, then the model will always be flat. Since many of the data points in a typical sloppy model are just interpolation points, we believe that in general sloppy models have lower extrinsic curvature than one would have naively guessed just by considering the magnitude of the non-linearities.  This explains perhaps why we will find that the dominant curvature of sloppy models is the parameter-effects one.

We can better understand the size of the various curvatures by considering the interpretation presented in section~\ref{sec:Manifolds-with-Boundaries} that sloppy models are a generalized interpolation scheme. If we choose $N$ independent data points as our parametrization, then the interpolating polynomial, $P_{N-1}(t)$ in Eq.~\eqref{eq:Interpolation-Error} is a linear function of the parameters. As discussed below that equation, the manifold in each additional direction will be constrained to within $\epsilon=\delta f_{N+1}$ of $P_{N-1}(t)$. Presuming that this deviation from flatness smoothly varies along the $j$th largest width $W_{j}\sim\delta f_{j}$ of the manifold (i.e., there is no complex or sensitive dependence on parameters), the geodesic extrinsic curvature is 
\begin{equation}
 K=\epsilon/W_{j}^{2}, \label{eq:DeviationFromFlatnessAssume}
\end{equation}
predicting a range of extrinsic curvatures comparable to the range of inverse eigenvalues of the metric. Furthermore, the ratio of the curvature to the inverse width should then be $\epsilon/W_{j}\sim\delta f_{N+1}/\delta f_{j}\sim(\delta t/R)^{N+1-j}$, where $\delta t$ is the spacing of time points at which the model is sampled and $R$ is the time scale over which the model changes appreciably (see the argument in section~\ref{sec:Manifolds-with-Boundaries} following Eq.~\eqref{eq:Interpolation-Error}).

Since we estimate $\epsilon = \delta f_{N+1}$ to be the most narrow width if the model had an additional parameter, we can find the overall scale of the extrinsic curvature to be given by the narrowest width
\[ K_N \approx \frac{1}{W_N}. \]  Additionally, we can find the scale set by the parameter effects curvature by recalling that parameter effects curvature is the curvature of the coordinate mesh.  If we ignore all parameter combinations except the stiffest, then motion in this direction traces out a one-dimensional model manifold.  The parameter-effects curvature of the full model manifold in the stiffest direction now corresponds to the extrinsic curvature of this one-dimensional manifold  %
\footnote{This is strictly only true if the parameter-effects curvature has no compression component.  Bates and Watts observe that typically, the compression is a large part of the parameter-effects curvature~\cite{Bates1980}.  As long as the compression is not significantly larger than the rotation (i.e. is within an order of magnitude), the parameter-effects curvature will be the same order of magnitude as the extrinsic curvature of the one-dimensional model.}%
, and as such is set by the smallest width (which in this case in the only width), i.e.\ the longest width of the full model manifold. The similar structure of parameter-effects curvature and extrinsic curvature, Eqs.~\eqref{eq:GeodesicCurvature} and \eqref{eq:ParameterEffectsCurvature}, suggest that the parameter-effects curvature also be proportional to the inverse eigenvalues (squares of the widths) along the several cross sections. Combining these result, we see that in general the ratio of extrinsic to parameter-effects curvature to be given by ratio of the widest to the most narrow width, 
\begin{equation}
 \frac{K}{K^P} \approx \frac{W_N}{W_1} \approx \sqrt{\frac{\lambda_N}{\lambda_1}}. \label{eq:KKpratio}
\end{equation}

In our experience the ratio of extrinsic to parameter-effects curvature in Eq.~\eqref{eq:KKpratio} is always very small.  When Bates and Watts introduced parameter-effects curvature, they considered its magnitude on twenty four models and found it universally larger than the extrinsic curvature, often much larger~\cite{Bates1980}.  We have here offered an explanation of this effect based on the assumption that the deviation from flatness is given by Eq.~\eqref{eq:DeviationFromFlatnessAssume}.

\begin{figure}
  \includegraphics[width=3.25in]{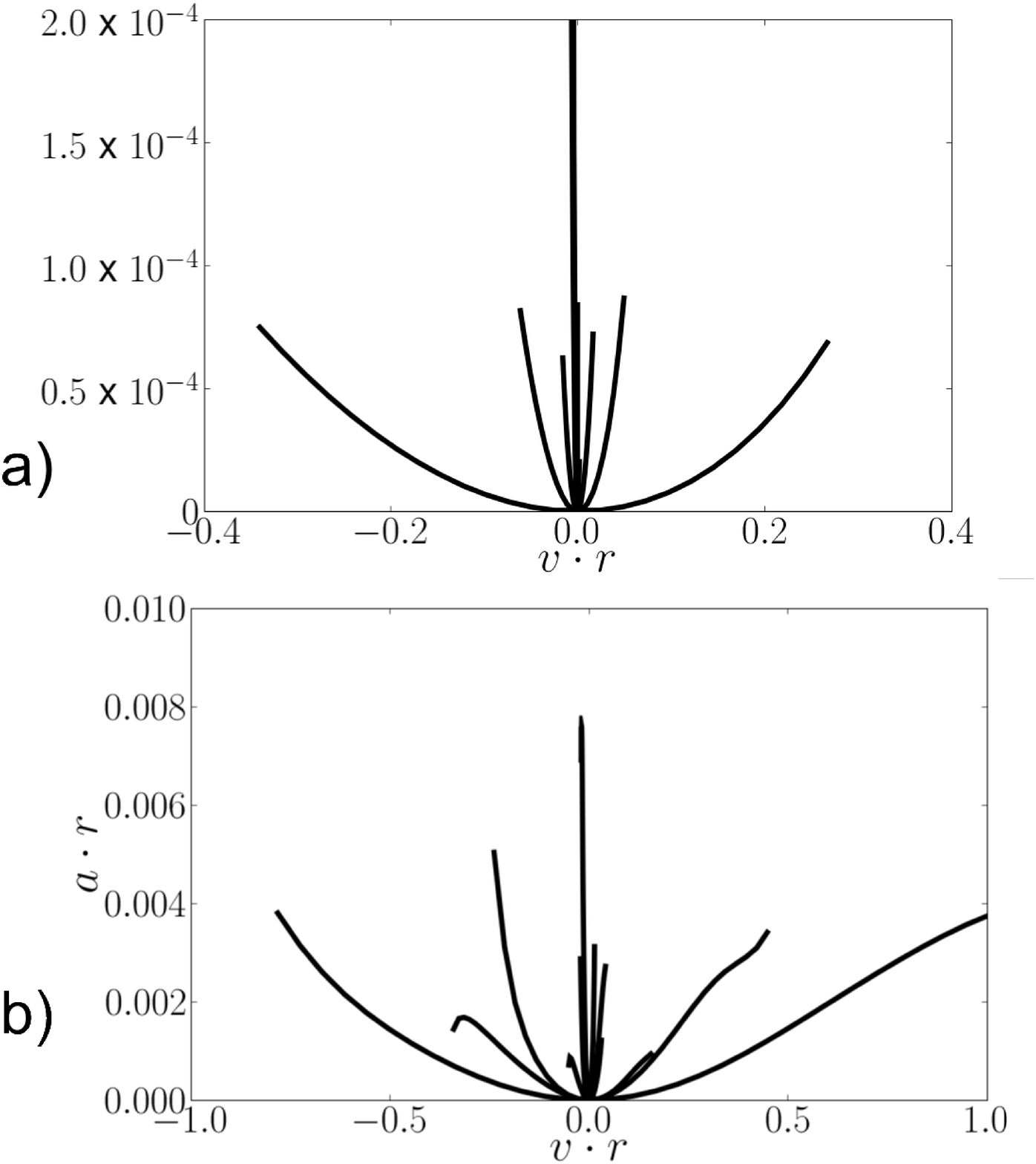}\caption{\label{fig:CrossSectionCurvature}a)     \textbf{Cross sections of a summing exponential model} projected     into the plane spanned by the velocity and acceleration vectors in     data space at an arbitrary point near the center. Notice the     widths of successive cross sections are progressively more narrow,     while the deviations from flatness are uniformly spread across the     width. The magnitude of the deviation from flatness is     approximately the same for each width, giving rise to the     hierarchy of curvatures. b) \textbf{Cross sections of a feed       forward neural network} has many of the same properties as the     exponential model.  In both cases, the curvature is much smaller     than it appears due to the relative scale of the two axes.  In fact, the sloppiest directions (narrowest widths) have an aspect ratio of about one.}

\end{figure}

\begin{figure}
  \textbf{\includegraphics[width=3.25in]{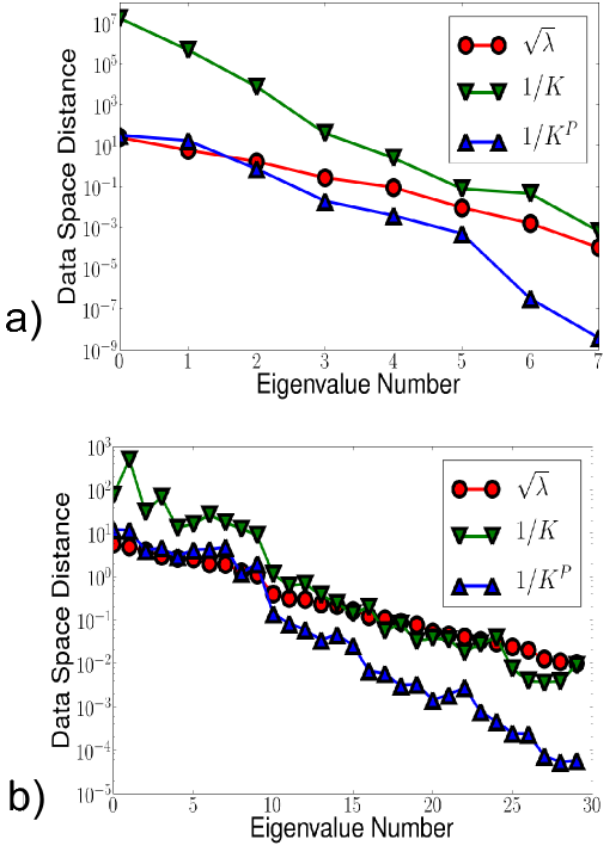}}
  \caption{\label{fig:WidthsCurvatures} The extrinsic and     parameter-effects curvature on the model manifold are strongly     anisotropic, with the largest curvatures along the shortest     widths (see Figs.~\ref{fig:Cross-sectional-widths},~\ref{fig:CrossSectionCurvature}).  The slopes of the (inverse) curvature vs. eigenvalue     lines are roughly twice that of the singular values (which are     equivalent to the widths).  The magnitude of the extrinsic     curvature is set by the most narrow cross sections, while the     magnitude of the parameter-effects curvature is set by the widest     cross-section.  Consequently the parameter-effect curvature is     much larger than the extrinsic curvature.  Here we plot the widths and curvatures for a model of four exponentials (above) from reference~\cite{Transtrum2010} and a feed forward artificial neural network (below) }

\end{figure}

We explicitly check the assumption of Eq.~\eqref{eq:DeviationFromFlatnessAssume} by calculating cross sections for a model of several exponentials and for an artificial neural network.  We have already seen in section~\ref{sec:Manifolds-with-Boundaries} in figure~\ref{fig:Cross-sectional-widths} that these widths span several orders of magnitude as predicted by the singular values of the Jacobian.  In Fig.~\ref{fig:CrossSectionCurvature} we view the data space image of these widths (projected into the plane spanned by the local velocity and acceleration vectors), where we see explicitly that the deviation from flatness is similar for all the cross sections.  In Fig.~\ref{fig:WidthsCurvatures} we see that that the extrinsic curvature is comparable to the narrowest cross section and the parameter-effects curvature is comparable to the widest cross section as we argued above, both for fitting exponentials and for the neural network model.

\begin{figure}
  \includegraphics[width=3.25in]{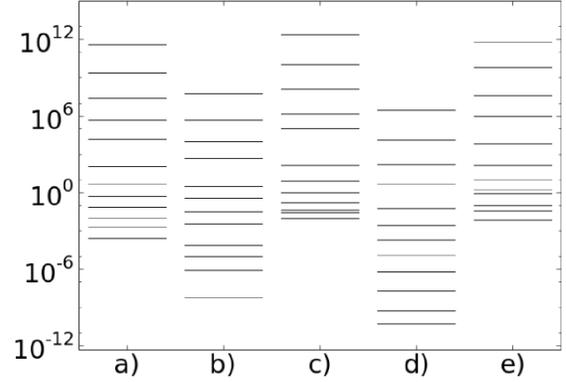}\caption{\label{fig:CurvatureAnisotropy}\textbf{Curvature       Anisotropy. }a) \textbf{Inverse Metric eigenvalues.} The     (inverse) metric has eigenvalues spread over several orders of     magnitude, producing a strong anisotropy in the way distances are     measured on the model manifold. b) \textbf{Geodesic Curvature in       eigendirections of the metric. }The geodesic curvatures also     cover many decades. The shortened distance measurements from the     metric eigenvalues magnify the anharmonicities in the sloppy     directions.  c) \textbf{Parameter-Effects Geodesic Curvature. }The     parameter-effects curvature is much larger than the extrinsic     curvature, but shares the anisotropy. d) \textbf{The eigenvalues       of the Shape Operator.}  The strong curvature anisotropy     described by the geodesic curvature is also illustrated in the     eigenvalue spectrum of the shape operator.  e)     \textbf{Parameter-Effects Shape Operator eigenvalues. }  Two measures (geodesic and shape operator curvatures) span similar ranges, but in both cases  the parameter-effects curvature is a factor of about $10^{5}$ larger     than the extrinsic curvature equivalent.}

\end{figure}

We further illustrate the above analysis by explicitly calculating the curvatures for the sloppy model formed by summing several exponential terms with amplitudes. Fig.~\ref{fig:CurvatureAnisotropy} is a $\log$-plot illustrating the eigenvalues of the inverse metric, the geodesic curvatures in each of those eigendirections, as well as the parameter-effects geodesic curvature in each of those directions. We see the same picture whether we consider the eigenvalues of the shape operator or the geodesic curvature. Both measures of curvature are strongly anisotropic with both extrinsic curvature and parameter-effects curvature covering as many orders of magnitude as the eigenvalues of the (inverse) metric.  However, the extrinsic curvature is smaller by a factor roughly given by Eq.~\eqref{eq:KKpratio}. We will use this large discrepancy between extrinsic and parameter-effects curvature when we improve the standard algorithms in section~\ref{sec:Applications-to-Algorithms}.

\begin{figure}
  \includegraphics[width=3.25in]{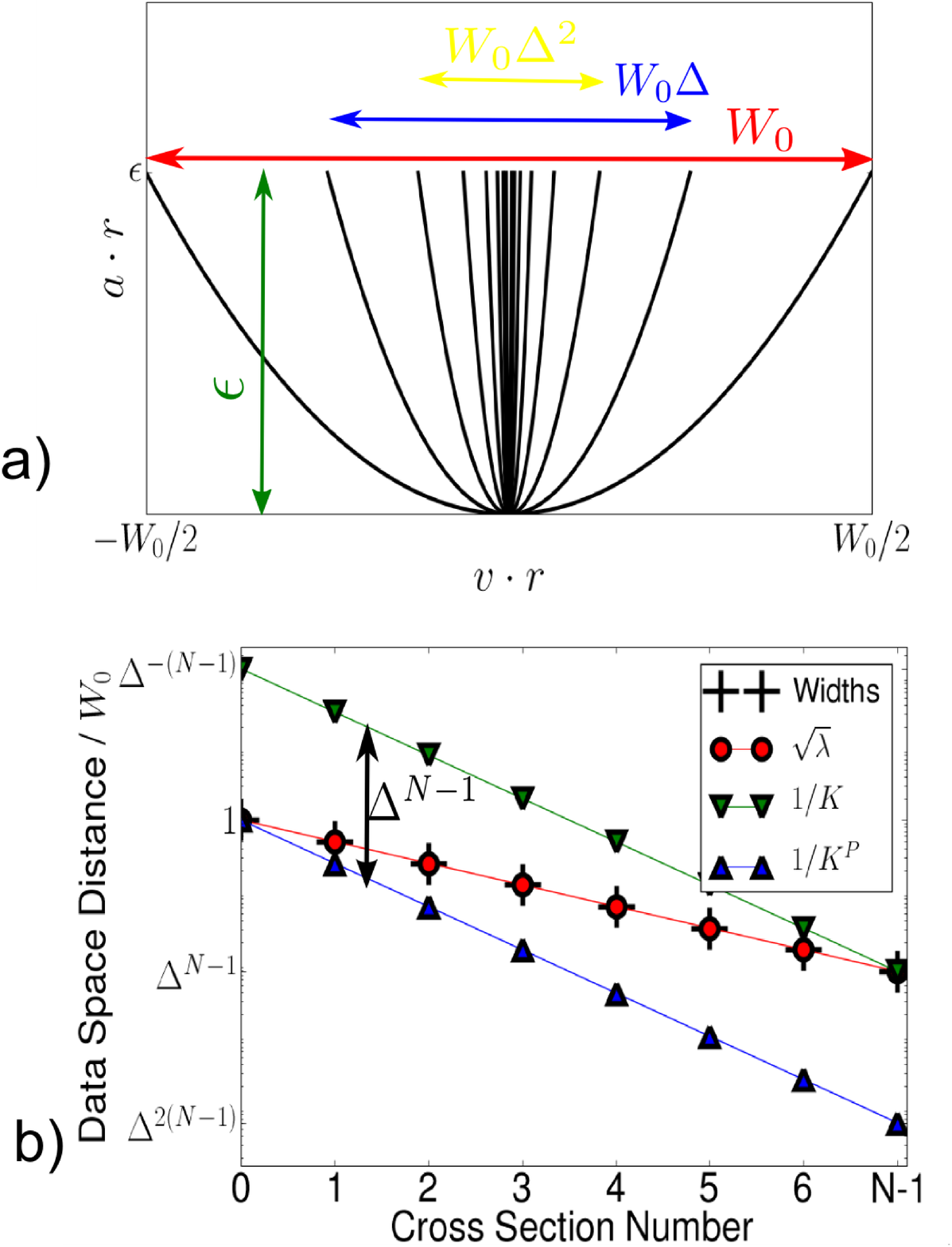}\caption{\label{fig:Caricature} A caricature of the widths and curvatures of a typical sloppy model.  a) The manifold deviates by an amount $\Delta^N$ from a linear model for each width.  As each width is smaller than the last by a factor of $\Delta$ the curvature is largest along the narrow widths.  This summary agrees well with the two real models in Fig.~\ref{fig:CrossSectionCurvature}.  b) The scales of the extrinsic and parameter-effects curvature are set by the narrowest and widest widths respectively.  The parameter-effects curvature is therefore smaller than the extrinsic curvature by a factor of $\Delta^N$.  Both are strongly anisotropic.  Compare this figure to with the corresponding result for the two real models in Fig.~\ref{fig:WidthsCurvatures}. }

\end{figure}

We have seen that manifolds of sloppy models possess a number of universal characteristics.  We saw in section~\ref{sec:Manifolds-with-Boundaries} that they are bounded with a hierarchy of widths which we describe as a hyper-ribbon.  In this section we have seen that the extrinsic and parameter-effects curvature also possess a universal structure summarized in Figs.~\ref{fig:CrossSectionCurvature}-\ref{fig:Caricature}.  A remarkable thing about the parameter-invariant, global structure of a sloppy model manifold is that is typically well-described by the singular values of the parameter-dependent, local Jacobian matrix.  We saw in section~\ref{sec:Manifolds-with-Boundaries} that the singular values correspond to the widths. We have now argued that the largest and smallest singular values set the scale of the parameter-effects and extrinsic curvatures respectively.  This entire structure is a consequence of the observation that most models are a multi-dimensional interpolation scheme.

Let us summarize our conclusions about the geometry of sloppy models. 
We argued in section \ref{sec:Manifolds-with-Boundaries} using interpolation theorems that multiparameter nonlinear
least-squares models should have model manifolds with a hierarchy of widths,
forming a hyper-ribbon with the $n^{th}$ width of order $W_n \sim W_0 \Delta^n$ with
$\Delta$ given by the spacing between data points divided by a radius
of convergence (in some multidimensional sense) and $W_0$ the widest cross section. We discovered in some
cases that the eigenvalues of the Hessian about the best fit agreed
well with the squares of these widths (so $\lambda_n \sim \Delta^{2n}$,
see Fig.~\ref{fig:Cross-sectional-widths}). 
This depends on the choice of parameters and 
the placement of the best fit;
we conjecture that this will usually occur if the `bare' parameters 
are physically or biologically natural descriptions of the model and
have natural units (i.e., dimensionless), and if the best fit is not near
the boundary of the model manifold. 
The parameter $\Delta$ will depend on the model and the
data being fit; it varies (for example) from $0.1$ to $0.9$ among seventeen
systems biology models~\cite{Gutenkunst2007a}.
We argued using interpolation theory
that the extrinsic curvatures should scale as $K_n \sim \epsilon/W_n^2$, where
the total variation $\epsilon \sim W_N$, implying 
$K_n \sim \Delta^N/(W_0 \Delta^{2n} )$ (Fig.~18c). We find this hierarchy both
measured along the eigenvectors of the (parameter-independent) shape operator
(Fig.~\ref{fig:CurvatureAnisotropy}) or the geodesic curvatures measured along the (parameter-dependent)
eigenpredictions at the best fit. Finally, we note that the parameter effects curvature also scales as $1/\Delta^{2n}$ by inspecting the similarity in the two formulae, Eqs.~\eqref{eq:GeodesicCurvature} and \eqref{eq:ParameterEffectsCurvature}.  We argue that the parameter-effects curvature should be roughly given by the extrinsic curvature of a one-dimensional model moving in a stiff direction, which sets the scale of the parameter effects as
$K^P_n \sim W_0/W_n^2 \sim 1/(W_0 \Delta^{2n})$, again either measured along the 
eigendirections of the parameter-effects shape operator or along
eigenpredictions. Thus the entire structure of the manifold can be summarized by three numbers, $W_0$ the stiffest width, $\Delta$ the typical spacing between widths, and $N$ the number of parameters.  We summarize our conclusions in Fig.~\ref{fig:Caricature}.

\subsection{Curvature on the Model Graph\label{sub:Curvature-Model-Graph}}

Most of the non-linearities of sloppy models appear as parameter-effects curvature on the model manifold. On the model graph, however, these non-linearities become extrinsic curvature because the model graph emphasizes the parameter dependence. An extreme version of this effect can be seen explicitly in Fig.~\ref{fig:ModelGraph}, where the model manifold, which had been folded in half, is unfolded in the model graph, producing a region of high curvature around the fold line.

If the Levenberg-Marquardt parameter is sufficiently large, the graph can be made arbitrarily flat (assuming the metric chosen for parameter space is flat, such as for the Levenberg metric). This effect is also visible in Fig.~\ref{fig:ModelGraph} in the regions that stretch toward the boundaries. In these regions, the Levenberg-Marquardt parameter is much larger than the eigenvalues of the metric, making the parameter space metric the dominant contribution, and creating an extrinsically flat region on the model graph.

\begin{figure}
  \includegraphics[width=3.25in]{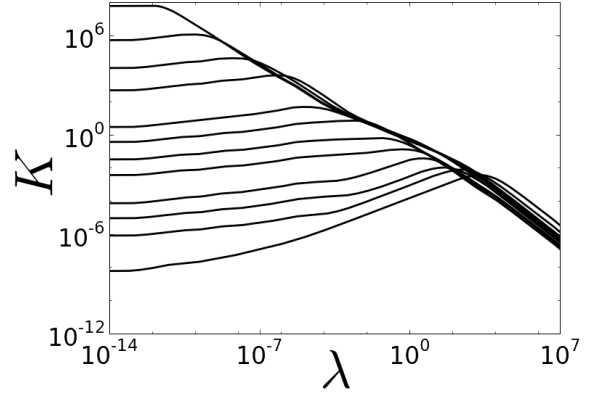}\caption{\label{fig:Model-Graph-Curvature}\textbf{Model       Graph Curvature.}  As the Levenberg-Marquardt parameter,     $\lambda$, is increased, directions with highest curvature become     less curved. For stiff directions with less extrinsic curvature,     the parameter effects curvature may be transformed into extrinsic     curvature. The damping term reduces the large anisotropy in the     curvature. For sufficiently large values of the     Levenberg-Marquardt parameters, all curvatures vanish. }

\end{figure}

To illustrate how the curvature on the model graph is affected by the Levenberg-Marquardt parameter, we consider how the geodesic curvatures in the eigendirections of the metric change as the parameter is increased for a model involving several exponentials with amplitudes and rates.  The results are plotted in Fig.~\ref{fig:Model-Graph-Curvature}.  As the Levenberg-Marquardt parameter is raised, the widely ranging values of the geodesic curvatures may either increase or decrease.  The largest curvature directions (the sloppy directions) tend to flatten, but the directions with the lowest curvature (the stiff directions) direction become more curved. The main effect of the the Levenberg-Marquardt parameter is to decrease the anisotropy in the curvature.

The behavior of the extrinsic curvature as the Levenberg-Marquardt parameter is varied can best be understood in terms of the interplay between parameter-effects curvature and extrinsic curvature. Curvatures decrease as more weight is given to the Euclidean, parameter-space metric. However, as long as the parameter-space metric is not completely dominant, the graph will inherit curvatures from the model manifold.  Since the graph considers model output versus the parameters, curvature that had previously been parameter-effects become extrinsic curvature.  Therefore, directions that had previously been extrinsically flat will be more curved, while the directions with the most curvature will become less curved.

The largest curvatures typically correspond to the sloppy directions. Most algorithms will try to step in sloppy directions in order to follow the canyon. The benefit of the model graph is that it reduces the curvature in the sloppy directions, which allows algorithms to take larger steps. The fact that previously flat directions become extrinsically curved on the model graph does not hinder an algorithm that does not step in these extrinsically flat directions anyway.  The role that curvatures play in determining an algorithm's maximal step size is looked at more closely in the next section.

\subsection{Optimization Curvature\label{sub:kappa}}

The distinction between extrinsic and parameter-effects curvature is not particularly useful in understanding the limitations of an algorithm. An iterative algorithm taking steps based on a local linearization will ultimately be limited by all non-linearities, both extrinsic and parameter-effects. We would like a measure of non-linearity, analogous to curvature, that explains the limitations of stepping in a given direction.

Suppose an algorithm proposes a step in some direction, $v^{\mu}$, then the natural measure of non-linearity should include the directional second derivative, $v^{\mu}v^{\nu}\partial_{\mu}\partial_{\nu}\vec{r}/v^{\alpha}v_{\alpha}$, where we included the normalization in order to remove the scale dependence of $v$. This expression is very similar to the geodesic curvature without the projection operator.

Simply using the magnitude of this expression is not particularly useful because it doesn't indicate whether curvature of the path is improving or hindering the convergence of the algorithm. This crucial bit of information is given by the (negative) dot product with the unit residual vector, \begin{equation}   \kappa(v)=-\frac{v^{\mu}v^{\nu}\partial_{\mu}\partial_{\nu}\vec{r}}{v^{\alpha}v_{\alpha}}\cdot\frac{\vec{r}}{|\vec{r}|},\label{eq:kappa}\end{equation} which we refer to as the \textit{Optimization Curvature}. Since the goal is to reduce the size of the current residual, the negative sign is to produce the convention that for $\kappa>0$, the curvature is helping the algorithm while when $\kappa<0$ the curvature is slowing the algorithm's convergence.

This expression for $\kappa$ has many of the properties of the curvatures discussed in this section. It has the same units as the curvatures we have discussed. It requires the specification of both a direction on the manifold (the proposed step direction, $v$) and a direction in data space (the desired destination, $\vec{r}$), making it a combination of both the geodesic and shape operator measures of curvature.  Furthermore, without the projection operators, it combines both extrinsic and parameter effects curvature into a single measure of non-linearity, although in practice, it is dominated by the parameter-effects curvature. We now consider how $\kappa$ is related to the allowed step size of an iterative algorithm.

Consider the scaled Levenberg step given by \[ \delta\theta^{\mu}=-\left(g^{0}+\lambda   D\right)^{\mu\nu}\partial_{\nu}C\ \delta\tau.\] Each $\lambda$ specifies a direction for a proposed step. For a given $\lambda$, we vary $\delta\tau$ to find how far an algorithm could step in the proposed direction. We determine $\delta\tau$ by performing a line search to minimize the cost in the given direction.  While minimizing the cost at each step may seem like a natural stepping criterion, it is actually a poor choice, as we discuss in section~\ref{sub:Delayed-Gratification}; however, this simple criteria is useful for illustrating the limitations on step size.

\begin{figure*}
  \includegraphics[width=7in]{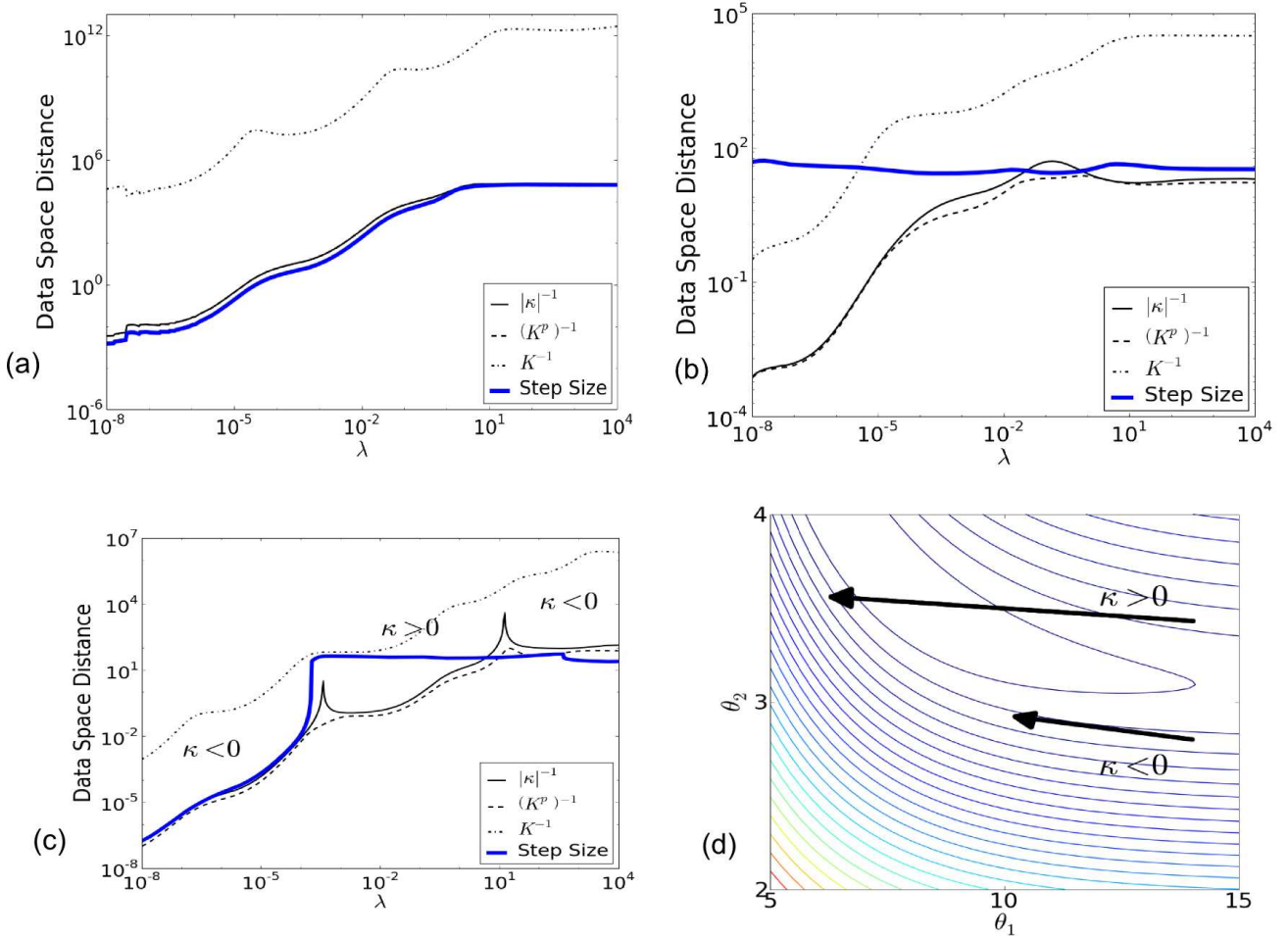}\caption{\label{fig:Curvature-StepSize}a)\textbf{       Curvature and Step Size for $\kappa<0$. }If $\kappa<0$, then the     non-linearities in the proposed direction are diverting the     algorithm away from the desired path. Distances are limited by the     size of the curvature. b) \textbf{Curvature and Step Size for       $\kappa>0$.} The non-linearities may be helpful to an algorithm,     allowing for larger than expected step sizes when $\kappa>0$. c)     \textbf{Curvature and Step Size for $\kappa$ with alternating       sign. }For small $\lambda$, $\kappa<0$ and the non-linearities     are restricting the step size. However, if $\kappa$ becomes     positive (the cusp indicates the change of sign), the possible     step size suddenly increases. d) \textbf{Cost contours for       positive and negative values of $\kappa$. }One can understand     the two different signs of $\kappa$ in terms of which side of the     canyon the given point resides. The upper point has positive     $\kappa$ and can step much larger distances in the Gauss-Newton     direction than can the lower point with negative $\kappa$, which     quickly runs up the canyon wall.}

\end{figure*}

We measure the step size by the motion it causes in the residuals,
$\left\Vert \delta\vec{r}\right\Vert $. This is a convenient choice
because each direction also determines a value for the geodesic curvature
($K$), parameter-effects curvature ($K^{p}$), and an optimization
curvature ($\kappa$), each of which are measured in units of inverse
distance in data space. We compare the step size with the inverse
curvature in each direction in Fig.~\ref{fig:Curvature-StepSize}.

One might assume that the size of the non-linearities always limits the step size, since the direction was determined based on a linearization of the residuals. This is clearly the case for the summing exponentials model in Fig.~\ref{fig:Curvature-StepSize}a, where $\kappa<0$; the step size closely follows the largest of the curvatures, the parameter effects curvature $K^{P}\approx|\kappa|$.

However, the non-linearities on occasion may inadvertently be helpful to an algorithm, as in Fig.~\ref{fig:Curvature-StepSize}b where $\kappa>0$. If the value of $\kappa$ changes sign as we vary $\lambda$, then the distinction becomes clear: steps can be several orders of magnitude larger than expected if $\kappa>0$, otherwise they are limited by the magnitude of $\kappa$. The sign of the parameter $\kappa$ is illustrating something that can be easily understood by considering the cost contours in parameter space, as in Fig.~\ref{fig:Curvature-StepSize}d.  If the canyon is curving {}``into'' the proposed step direction, then the step runs up the canyon wall and must be shortened. However, if the canyon is curving {}``away'' from the proposed step direction, then step runs down the canyon and eventually up the opposite wall, resulting in a much larger step size.

\subsection{Curvature and parameter evaporation\label{sub:parameter-evaporation}}

We have stressed the the boundaries of the model manifold are the major obstacle to optimization algorithms. Because a typical sloppy model has many very narrow widths, it is reasonable to expect the best fit parameters to have several evaporated parameter values when fit to noisy data. In order estimate the expected number of evaporated parameters, however, it is necessary to account for the extrinsic curvature of a model.

\begin{figure}
  \includegraphics[width=3.25in]{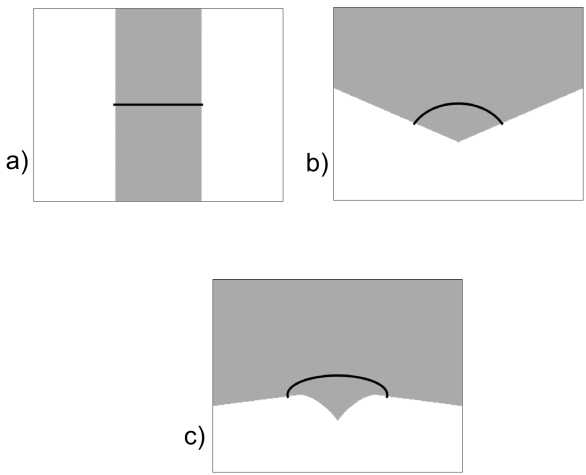}

  \caption{\label{fig:BoundaryProb}The curvature along the width of a     manifold effects if the best fit lies on the boundary or on the     interior. For a cross-sectional width (thick black line), consider     three possibilities: a) extrinsically flat, b) constant curvature     along width, and c) curvature proportional to distance from the     boundary. Grey regions correspond to data points with best fits on     the interior of the manifold, while white regions correspond to     data with evaporated parameters. If the curvature is larger near     the boundaries, there is less data space available for evaporated     best fit parameters.}

\end{figure}

In Fig.~\ref{fig:BoundaryProb} we illustrate how the curvature effects which regions of data space correspond to a best fit with either evaporated or finite parameters. A first approximation is a cross-sectional width with no extrinsic curvature, as in Fig.~\ref{fig:BoundaryProb}a.  If the component of the data parallel to the cross-section does not lie outside the range of the width, the parameter will not evaporate.  If the cross-section has curvature, however, the situation is more complicated, with the best fit depending on the component of the data perpendicular to the cross-section as well.  Figs.~\ref{fig:BoundaryProb}(b) and (c) highlight the regions of data space for which the best fit will not evaporate parameters (gray).

Knowing both the regions of data space corresponding to non-evaporated parameters and the relative probabilities of the possible data (Eq.~\eqref{eq:Prob(r)}), we can estimate the expected number of evaporated parameters for a given a model at the best fit. Using Gaussian data of width $\sigma$ centered on the middle of a cross-section for a problem of fitting exponentials, we find the best fit and count the number zero-eigenvalues of the metric, corresponding to the number of non-evaporated parameters at the fit.

We can derive analytic estimates for the number of evaporated parameters using the approximation that the cross section is either flat or has constant curvature as in Fig.~\ref{fig:BoundaryProb}a and b.  If the cross-section is extrinsically flat then the probability of the corresponding parameter combination not evaporating is given in terms of the error function
\begin{equation}
P^{\textrm{flat}}_n=2\ \textrm{erf} \left( \frac{W_n}{2\sigma} \right), \label{eq:ProbFlat}
\end{equation}
where $W_n$ is the $n^{th}$ width given by $W_n=W_0 \Delta^n$.

A similar formula for the constant curvature approximation is a little more complicated.  It involves integrating the Gaussian centered on the cross section in Fig.~\ref{fig:BoundaryProb} over the gray region.  Since the apex of the gray cone is offset from the center of the Gaussian, we evaluate the integral treating the offset as a perturbation.  We recognize that there are several cases to be considered.  If the noise is smaller than any of the widths, then the probability is approximately one.  However, if the noise is larger than the width but smaller than inverse curvature, the probability is given by $W_n/\sigma$.  Finally, if the noise is larger than any of the widths the probability is $W_n K_n$.  Recall that we characterize a sloppy model manifold by three numbers, $W_0$, $\Delta$, and $N$, the largest width, the average spacing between widths and the number of parameters respectively.  The final result in each of the three cases in terms of these three numbers is given by
\begin{equation}
\label{eq:ProbCurved}
P^{\textrm{curved}}_n  =  \begin{cases}
1 & \text{if $\sigma<W_n$, } \\
\frac{W_0\Delta^n}{\sigma} & \text{if $W_n < \sigma < 1/K_n$, } \\
\Delta^{N-n} & \text{if $1/K_n<\sigma$. } 
\end{cases}
\end{equation}
From our caricature of a typical sloppy model summarized in Fig.~\ref{fig:Caricature}, we estimate how many widths should belong in each category for a given $\sigma$.  Summing the probabilities for the several widths in Eq.~\eqref{eq:ProbCurved} we find the expected number of non-evaporated parameters to be given by
\begin{equation}
  \label{eq:NonEvap}
  \langle N_{\textrm{approx}} \rangle = \frac{2}{1-\Delta} + \frac{\log \sigma/W_0}{\log \Delta} - 1.
\end{equation}

In Table~\ref{tab:NonevaporatedParams} we compare the fraction of non-evaporated parameters with the estimates from Eqs.~\eqref{eq:ProbFlat} and \eqref{eq:ProbCurved}.  We find a large discrepancy when the noise in the data is very large.  In this case there is often a large fraction of non-evaporated parameters even if the noise is much larger than any cross-sectional width.  We attribute this discrepancy to larger curvatures near the corners of the manifold that increase the fraction of data space that can be fit without evaporating parameters.  Since the metric is nearly singular close to a boundary, we expect the extrinsic curvature to become singular also by inspecting Eq.~\eqref{eq:GeodesicCurvature}.  We explicitly calculate the curvature near the boundary and we find that this is in fact the case.

\begin{table}
  \begin{tabular}{|c|c|c|c|c|}
    \hline 
    $\sigma$ & $\langle N\rangle/N$ & $\langle N_{\textrm{flat}}\rangle/N$ &     $\langle N_{\textrm{integral}}\rangle/N$ & $\langle N_{\textrm{approx}} \rangle/N $  \tabularnewline
    \hline
    \hline 
    $10 W_0$ & 0.61 & .0006 & 0.028 & 0.025 \tabularnewline
    \hline 
    $W_0$ & 0.73 & 0.05 & 0.076 & 0.16 \tabularnewline
    \hline 
    $\sqrt{W_0 W_N}$ & 0.87 & 0.50 & 0.52 & 0.60\tabularnewline
    \hline 
    $W_N$ & 0.95 & 0.92 & 0.93 & 1.00 \tabularnewline
    \hline 
    $W_N/10$ & 0.98 & 1.00 & 1.00 & 1.00\tabularnewline
    \hline
  \end{tabular}\caption{\label{tab:NonevaporatedParams}The number of         non-evaporated parameters $\langle N\rangle$ per total number of parameters $N$ at the best fit, for an 8 parameter model of         exponentials and amplitudes. As the noise of the data ensemble grows, the number of non-evaporated parameters at the best fit decreases (i.e.\ more parameters are evaporated by a good fit). Even if the noise is much larger than any of the widths, there are still several     non-evaporated parameters, due to the curvature (see Fig.~\ref{fig:BoundaryProb}). We estimate the expected number of non-evaporated parameters from both a flat manifold approximation (Eq.~\eqref{eq:ProbFlat}) and a constant curvature approximation.  For the constant curvature approximation we show the result of the exact integral of the gaussian over the grey region of Fig.~\ref{fig:BoundaryProb}b as well as our perturbative approximation, Eq.~\eqref{eq:NonEvap}, using the parameters $W_0=6.1$, $\Delta = .11$ and $N=8$. These approximations agree with the numerical results when the noise is small, but for very noisy data there are still several non-evaporated parameters even if the noise is much larger than any of the widths.  Therefore, although our general caricature of the model manifold as a hyper-cylinder of constant curvatures and widths seems to describe the geometry of the sloppy directions, it does not capture the features of the stiff directions.  This discrepancy could be due, for example, by an increase in the curvature near the boundary as in Fig.~\ref{fig:BoundaryProb}c.}

\end{table}

The calculation in Table~\ref{tab:NonevaporatedParams} can be interpreted in several ways.  If one is developing a model to describe some data with known error bars, the calculation can be used to estimate the number of parameters the model could reasonably have without evaporating any at the best fit.  Alternatively for a fixed model, the calculation indicates what level of accuracy is necessary in the data to confidently predict which parameters are not infinite.  Qualitatively, for a given model, the errors must be smaller than the narrowest width for there to be no evaporated parameters.  

Similarly, for experimental data with noise less than any of the (inverse) parameter-effects curvatures the parameter uncertainties estimated by the inverse Fisher information matrix will be accurate since the parameterization is constant over the range of uncertainty. It is important to note, that for models with large numbers of parameters either of these conditions require extremely small, often unrealistically small, error bars.  In general, it is more practical to focus on predictions made by ensembles of parameters with good fits rather than parameter values at the best fit as the latter will depend strongly the noise in the data.

\section{Applications to Algorithmics} \label{sec:Applications-to-Algorithms}

We now consider how the results derived in previous sections can be applied to algorithms. We have stressed that fitting sloppy models to data consist of two difficult steps. The first step is to explore the large, flat plateau to find the canyon. The second step is to follow the canyon to the best fit.

We begin by deriving two common algorithms, the modified Gauss-Newton method and the Levenberg-Marquardt algorithm from the geometric picture in sections~\ref{sub:Modified-Gauss-Newton-Method} and~\ref{sub:Levenberg-Marquardt-Algorithm}.  We then suggest how it may be improved by applying what we call delayed gratification and an acceleration term in sections~\ref{sub:Delayed-Gratification} and~\ref{sub:Using-acceleration}.

We demonstrate that the suggested modifications can offer improvements to the algorithm by applying them to a few test problems in section~\ref{sub:Algorithm-Comparisons}. In comparing the effectiveness of algorithms we make an important observation, that the majority of the computer time for most problems with many parameters is occupied by Jacobian evaluations. As the number of parameters grows, this becomes increasingly the case. Models with many parameters are more likely to be sloppy, so this assumption does not greatly reduce the applicability of the algorithms discussed.

If an algorithm estimates the Jacobian from finite differences of the residuals, then most of the function (residual) evaluations will be spent estimating the Jacobian. (Our function evaluation counts in Table~\ref{tab:Exponential-Results} do not include function evaluations used to estimate Jacobians.) If this is the case, then for any given problem, comparing function evaluations automatically integrates the relative expense of calculating residuals and Jacobians. However, many of the problems we use for comparison are designed to have only a few parameters for quick evaluation, while capturing the essence of larger problems.  We then extrapolate results from small problems to similar, but larger problems. Our primary objective is to reduce the number of Jacobian evaluations necessary for an algorithm to converge. We do not ignore the number of function evaluations, but we but consider reducing the number of function calls to be a lower priority. As we consider possible improvements to algorithms, we will usually be willing to accept a few more function calls if it can significantly reduce the number of Jacobian evaluations that an algorithm requires.

In the next few sections, we discuss the geometric meaning of the Gauss-Newton method (section~\ref{sub:Modified-Gauss-Newton-Method}) and other similar algorithms, such as the Levenberg-Marquardt algorithm (section~\ref{sub:Levenberg-Marquardt-Algorithm}). We then discuss how ideas from differential geometry can lead to ways of improving convergence rates. First, we suggest a method of updating the Levenberg-Marquardt parameter, which we call delayed gratification, in section~\ref{sub:Delayed-Gratification}.  Second, we suggest the inclusion of a geodesic acceleration term in section~\ref{sub:Using-acceleration}. We end the discussion by comparing the efficiency of standard versions of algorithms to those with the suggested improvements in section~\ref{sub:Algorithm-Comparisons}.

\subsection{Modified Gauss-Newton Method\label{sub:Modified-Gauss-Newton-Method}}

The result presented in this paper that appears to be the most likely to lead to a useful algorithm is that cost contours are nearly perfect circles in extended geodesic coordinates as described in section~\ref{sec:Extended-Geodesic-Coordinates}.  The coordinates illustrated in Fig.~\ref{fig:GeodesicCoords} transformed a long, narrow, curved valley into concentric circles. Searching for the best fit in these coordinates would be a straightforward task!  This suggests that an algorithm that begins at an unoptimized point need only follow a geodesic to the best fit. We have thus transformed an optimization problem into a differential equation integration problem.

The initial direction of the geodesic tangent vector (velocity vector) should be the Gauss-Newton direction \begin{equation}   \frac{d\theta^{\mu}}{d\tau}(\tau=0)=-g^{\mu\nu}\partial_{\nu}C.\label{eq:v0}\end{equation} If we assume that the manifold is extrinsically flat (the necessary and sufficient condition to produce concentric circles in extended geodesic coordinates), then Eq.~\eqref{eq:d2C} tells us that the cost will be purely quadratic, \begin{equation}   \frac{d^{2}C}{d\tau^{2}}=g^{\mu\nu}\frac{d\theta^{\mu}}{d\tau}\frac{d\theta^{\nu}}{d\tau}=\text{constant},\label{eq:Cdotdotconstant}\end{equation} which implies that the first derivative of the cost will be linear in $\tau$: \begin{equation}   \frac{dC}{d\tau}=\left(g^{\mu\nu}\frac{d\theta^{\mu}}{d\tau}\frac{d\theta^{\nu}}{d\tau}\right)\tau+\dot{C}(\tau=0).\label{eq:Cdotlinear}\end{equation} A knowledge of $\dot{C}(\tau=0)$ will then tell us how far the geodesic needs to be integrated:\begin{equation}   \tau_{max}=-\frac{\dot{C}(\tau=0)}{g^{\mu\nu}\frac{d\theta^{\mu}}{d\tau}\frac{d\theta^{\nu}}{d\tau}}.\label{eq:taumax}\end{equation} We can calculate the missing piece of Eq.~\eqref{eq:taumax} from the chain rule and Eq.~\eqref{eq:v0},\begin{eqnarray*}
  \dot{C} & = & \frac{d\theta^{\mu}}{d\tau}\partial_{\mu}C\\
  & = & -g^{\mu\nu}\partial_{\nu}C\ \partial_{\mu}C,\end{eqnarray*} which gives us \[ \tau_{max}=1.\]

The simplest method one could apply to solve the geodesic equation would be to apply a single Euler step, which moves the initial parameter guess by \begin{eqnarray}
  \delta\theta^{\mu} & = & \dot{\theta}^{\mu}\delta\tau\nonumber \\
  & = & -g^{\mu\nu}\partial_{\nu}C,\label{eq:MGNStep}\end{eqnarray} since $\delta\tau=1$. Iteratively updating the parameters according to Eq.~\eqref{eq:MGNStep} is known as the Gauss-Newton method. It can be derived without geometric considerations by simply assuming a linear approximation to the residuals. Unless the initial guess is very good, however, the appearance of the inverse Hessian in Eq.~\eqref{eq:MGNStep} (with its enormous eigenvalues along sloppy directions) will result in large, unreliable steps and prevent the algorithm from converging.

The Gauss-Newton method needs some way to shorten its steps. Motivated by the idea of integrating a differential equation, one could imagine taking several Euler steps instead of one. If one chooses a time step to minimize the cost along the line given by the local Gauss-Newton direction, then the algorithm is known as the modified Gauss-Newton method, which is a much more stable algorithm than the simple Gauss-Newton method~\cite{Hartley1961}.

One could also imagine performing some more sophisticated method, such as a Runge-Kutta method. The problem with these approaches is that the sloppy eigenvalues of the inverse metric require the Euler or Runge-Kutta steps to be far too small be competitive with other algorithms. In practice, these techniques are not as effective as the Levenberg-Marquardt algorithm, discussed in the next section.

\subsection{Levenberg-Marquardt Algorithm\label{sub:Levenberg-Marquardt-Algorithm}}

The algorithm that steps according to Eq.~\eqref{eq:MGNStep} using the metric of the model graph, Eq.~\eqref{eq:GraphMetric}, is known as the Levenberg-Marquardt step: \[ \delta\theta^{\mu}=-\left(g^{0}+\lambda   D\right)^{\mu\nu}\partial_{\nu}C.\] If $D$ is chosen to be the identity, then the algorithm is the Levenberg algorithm~\cite{Levenberg1944}. The Levenberg algorithm is simply the Gauss-Newton method on the model graph instead of the model manifold.

If $D$ is chosen to be a diagonal matrix with entries equal to the diagonal elements of $g^{0}$, then the algorithm is the Levenberg-Marquardt algorithm~\cite{Marquardt1963}. As we mentioned in section~\ref{sec:The-Model-Graph}, the Levenberg-Marquardt algorithm, using the Marquardt metric, is invariant to rescaling the parameters. We find this property to often be counterproductive to the optimization process since it prevents the modeler from imposing the proper scale for the parameter values.  In addition we observe that the resulting algorithm is more prone to parameter evaporation.  The purpose for adding $D$ to the metric is to \textit{introduce} parameter dependence to the step direction.

The Levenberg-Marquardt algorithm adjusts $\lambda$ at each step. Typically, when the algorithm has just begun, the Levenberg-Marquardt term will be very large, which will force the algorithm to take small steps in the gradient direction. Later, once the algorithm has descended into a canyon, $\lambda$ will be lowered, allowing the algorithm to step in the Gauss-Newton direction and follow the length of the canyon. The Levenberg-Marquardt parameter, therefore, serves the dual function of rotating the step direction from the Gauss-Newton direction to the gradient direction, as well as shortening the step.

As we mentioned in section~\ref{sec:The-Model-Graph}, when using the Levenberg metric, $\lambda$ will essentially wash out all the sloppy eigenvalues of the original metric and leave the large ones unaffected. The relatively large multiplicative factor separating eigenvalues means that $\lambda$ does not need to be finely tuned in order to achieve convergence. Nevertheless, an efficient method for choosing $\lambda$ is the primary way that the Levenberg-Marquardt algorithm can be optimized. We discuss two common updating schemes here.

A typical method of choosing $\lambda$ at each step is described in Numerical Recipes~\cite{Press2007}. One picks an initial value, say $\lambda=.001$, and tries the proposed step. If the step moves to a point of larger cost, by default, the step is rejected and $\lambda$ is increased by some factor, $10$. If the step has decreased the cost, the step is accepted and $\lambda$ is decreased by a factor of $10$. This method is guaranteed to eventually produce an acceptable step, since for extremely large values of $\lambda$, the method will take an arbitrarily small step in the gradient direction. We refer to this as the traditional scheme for updating $\lambda$.

A more complicated method of choosing $\lambda$ is based on a trust region approach and is described in~\cite{More1977}. As in the previous updating scheme, at each step $\lambda$ is increased until the step goes downhill (all uphill steps are rejected). However, after an accepted step, the algorithm compares the decrease in cost at the new position with the decrease predicted by the linear approximation of the residuals\[ \frac{\left\Vert     \vec{r}\left(\theta_{old}\right)\right\Vert -\left\Vert     \vec{r}\left(\theta_{new}\right)\right\Vert }{\left\Vert     \vec{r}\left(\theta_{old}\right)\right\Vert -\left\Vert     \vec{r}(\theta_{old})+\vec{J}_{\mu}\delta\theta^{\mu}\right\Vert }.\] If this value is very far from unity, then the algorithm has stepped beyond the region for which it trusts the linear approximation and will increase $\lambda$ by some factor even though the cost has decreased; otherwise, $\lambda$ is decreased. This method tunes $\lambda$ so that most steps are accepted, reducing the number of extra function evaluations. As a result, it often needs a few more steps, and therefore, a few more Jacobian evaluations. This algorithm works well for small problems where the computational complexity of the function and the Jacobian are comparable. It is not as competitive using the number of Jacobian evaluations as a measure of success.

These are certainly not the only update schemes available. Both of these criteria reject any move that increases the cost, which is a natural method to ensure that the algorithm does not drift to large costs and never converges. One could imagine devising an update scheme that allows some uphill steps in a controlled way such that the algorithm remains well-behaved. We consider such a scheme elsewhere~\cite{Transtrum2010c} and note that it was a key inspiration for the Delayed Gratification update scheme that we describe below in section~\ref{sub:Delayed-Gratification}.

As we observed in section~\ref{sec:Priors}, the metric formed by the model graph acts similarly to the effect of adding linear Bayesian priors as residuals. The Levenberg-Marquardt algorithm therefore chooses a Gauss-Newton step as though there were such a prior, but then ignores the prior in calculating the cost at the new point. A similar algorithm, known as the iteratively updated Gauss-Newton algorithm, includes the contribution from the prior when calculating the new cost, although the strength of the prior may be updated at each step~\cite{Bakushinskii1992}.

\subsection{Delayed Gratification\label{sub:Delayed-Gratification}}

We have seen that parameter-effects curvatures are typically several orders of magnitude larger than extrinsic curvatures for sloppy models, which means that the model manifold is much more flat than the non-linearities alone suggest and produce the concentric circles in Fig.~\ref{fig:GeodesicCoords}.  When considering only a single step on even a highly curved manifold, if the parameter-effects curvature dominates, the step size will be less than the (inverse) extrinsic curvature and approximating the manifold by a flat surface is a good approximation. Furthermore, we have seen that when the manifold is flat, geodesics are the paths that we hope to follow.

\begin{figure}

  \includegraphics[width=3.25in]{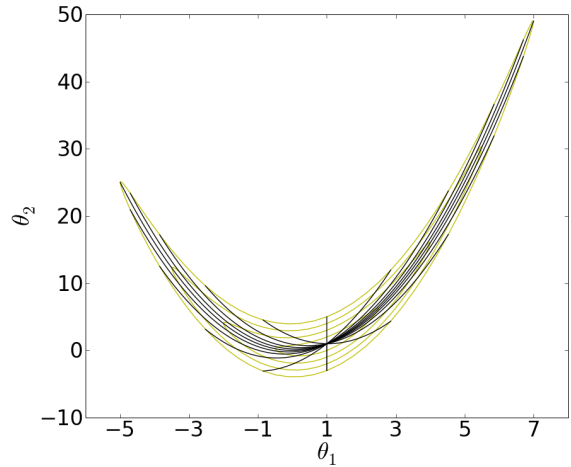}\caption{\label{fig:RosenbrockGeodesics}\textbf{Extended       Geodesic Coordinates for Rosenbrock Function.} The residuals are     one choice of extended geodesic coordinates if the number of     parameters equal the number of data points, as is the case for the     Rosenbrock function. Because the Rosenbrock function is a simple     quadratic, the coordinate transformation can be expressed     analytically. Lines of constant $\rho$ are equi-cost lines, while     lines of constant $\phi$ are the paths a geodesic algorithm should     follow to the best fit. Because the geodesics follow the path of     the narrow canyon, the radial geodesics are nearly parallel to the     equi-cost lines in parameter space. This effect is actually much     more extreme than it appears in this figure because of the     relative scales of the two axes. }

\end{figure}

The Rosenbrock function is a well known test function for which the extended geodesic coordinates can be expressed analytically. It has a long, parabolic shaped canyon and is given by\begin{eqnarray*}
  r_{1} & = & 1-\theta_{1}\\
  r_{2} & = & A\left(\theta_{2}-\theta_{1}^{2}\right),\end{eqnarray*} where $A$ is a parameter that controls the narrowness of the canyon. The Rosenbrock function has a single minimum at $\left(\theta_{1},\theta_{2}\right)=\left(1,1\right)$.  Since there are two residuals and two parameters, the model manifold is flat and the extended geodesic coordinates are the residuals. It is straightforward to solve \begin{eqnarray*}
  \theta_{1} & = & 1-r_{1}\\
  \theta_{2} & = &   \frac{r_{2}}{A}+\left(1-r_{1}\right)^{2}.\end{eqnarray*} If we change to polar coordinates, \begin{eqnarray*}
  r_{1} & = & \rho\sin\phi\\
  r_{2} & = & \rho\cos\phi,\end{eqnarray*} then lines of constant $\phi$ are the geodesic paths that we would like an algorithm to follow toward the best fit, and lines of constant $\rho$ are cost contours. We plot both sets of curves in Fig.~\ref{fig:RosenbrockGeodesics}.

Inspecting the geodesic paths that lead to the best fit in Fig.~\ref{fig:RosenbrockGeodesics} reveals that most of the path is spent following the canyon while decreasing the cost only slightly. This behavior is common to all geodesics in canyons such as this. We would like to devise an update scheme for $\lambda$ in the Levenberg-Marquardt algorithm that will imitate this behavior. The results of section~\ref{sub:kappa} suggest that we will often be able to step further than a trust region would allow, so we start from the traditional update scheme.

The primary feature of the geodesic path that we wish to imitate is that radial geodesics are nearly parallel to cost contours. In the usual update scheme, if a proposed step moves uphill, then $\lambda$ is increased. In the spirit of following a cost contour, one could slowly increase the Levenberg-Marquardt parameter just until the cost no longer increases. If $\lambda$ is fine-tuned until the cost is the same, we call this the equi-cost update scheme. Such a scheme would naturally require many function evaluations for each step, but as we said before, we are primarily interested in problems for which function calls are cheap compared to Jacobian evaluations. Even so, determining $\lambda$ to this precision is usually overkill, and the desired effect can be had by a much simpler method.

Instead of precisely tuning $\lambda$, we modify the traditional scheme to raise and lower the parameter by different amounts. Increasing $\lambda$ by very small amounts when a proposed step is uphill and then decreasing it by a large amount when a downhill step is finally found will mimic the desired behavior. We have found that increasing by a factor of $2$ and decreasing by a factor of $10$ works well, consistent with Lampton's results~\cite{Lampton1997}. We call this method, the \textit{delayed gratification} update scheme.

\begin{figure}

  \includegraphics[width=3.25in]{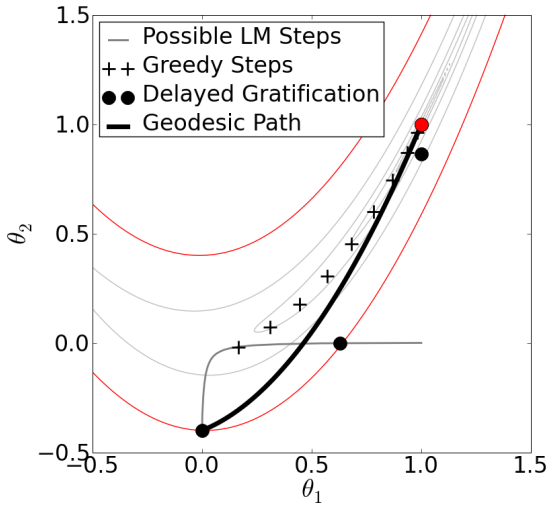}\caption{\label{fig:RosenbrockDelayedGratification}\textbf{Greedy       Step and Delayed Gratification Step Criterion.} In optimization     problems for which there is a long narrow canyon, such as for the     Rosenbrock function, choosing a delayed gratification step is     important to optimize convergence.  As the damping term is     increased, the Gauss-Newton direction is rotated into the gradient     direction, giving a number of possible steps an algorithm might     take. Choosing the step that lowers the cost the most will cause     an algorithm to descend quickly into the canyon, greatly reducing     the size of the future steps could take. This step choice is     excessively greedy, and can be improved upon. An algorithm that     takes the largest tolerable step size (in this case the largest     step that does not move uphill), will not decrease the cost     significantly in the first few steps, but will arrive at the best     fit in fewer steps and more closely approximate the true geodesic     path. What constitutes the largest tolerable step size should be     optimized for specific problems so as to guarantee convergence. }

\end{figure}

The reason that this update scheme is effective is due to the restriction that we do not allow uphill steps. If we move downhill as much as possible in the first few steps, we greatly restrict the steps that will be allowed as successive iterations, slowing down the convergence rate, as illustrated in Fig.~\ref{fig:RosenbrockDelayedGratification}.

By using the delayed gratification update scheme, we are using the smallest value of $\lambda$ that does not produce an uphill step.  If we choose a trust-region method, instead, each step will choose a much larger value of $\lambda$. The problem with using larger values of $\lambda$ at each step, is that they drive the algorithm downhill prematurely. Even if the trust region only cuts each possible step in half compared to the delayed gratification scheme, the cumulative effect will be much more damaging because of how this strategy reduces the possibility of future steps.

\subsection{Geodesic Acceleration\label{sub:Using-acceleration}}

We have seen that a geodesic is a natural path that an algorithm should follow in its search for the best fit. The application of geodesics to optimization algorithms is not new. It has been applied, for example to the problem of nonlinear programming with constraints~\cite{Luenberger1972,P'azman2002}, to neural network training~\cite{Igel2005}, and to the general problem of optimization on manifolds~\cite{Nishimori2005,Absil2008}. Here we apply it as a second order correction to the Levenberg-Marquardt step.

The geodesic equation is a second order differential equation, whose solution we have attempted to mimic by only calculating first derivatives of the residuals (Jacobians) and following a delayed gratification stepping scheme. From a single residual and Jacobian evaluation, an algorithm can calculate the gradient of the cost as well as the metric, which determines a direction. We would like to add a second order correction to the step, but one would expect its evaluation to require a knowledge of the second derivative matrix, which would be even more expensive to calculate than the Jacobian. We have already noted that most of the computer time is spent on Jacobian evaluations, so second order steps would have even more overhead. Fortunately, the second order correction to the geodesic path can be calculated relatively cheaply in comparison to a Jacobian evaluation.

The second order correction, or acceleration, to the geodesic path is given by \begin{equation}   a^{\mu}=-\Gamma_{\alpha\beta}^{\mu}v^{\alpha}v^{\beta},\label{eq:GeodesicAcceleration}\end{equation} as one can see by inspecting Eq.~\eqref{eq:GeodesicODE}. In the expression for the acceleration, the velocity contracts with the two lower indices of the connection. Recall from the definition, \[ \Gamma_{\alpha\beta}^{\mu}=g^{\mu\nu}\partial_{\nu}r_{m}\partial_{\alpha}\partial_{\beta}r_{m},\] that the lowered indices correspond to the second derivatives of the residuals. This means that the acceleration only requires a directional second derivative in the direction of the velocity. This directional derivative can be estimated with two residual evaluations in addition to the Jacobian evaluation. Since each step will always call at least one residual evaluation, we can estimate the acceleration with only one additional residuals call, which is very cheap computationally compared to a Jacobian evaluation.

\begin{figure}

  \includegraphics[width=3.25in]{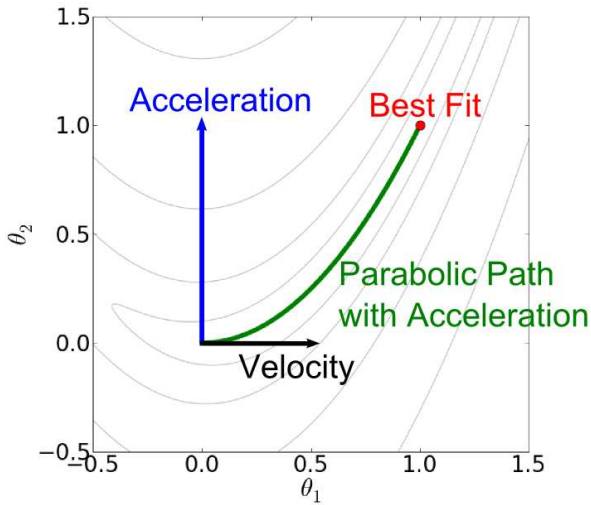}\caption{\label{fig:RosenbrockAcceleration}\textbf{Geodesic       Acceleration in the Rosenbrock Valley.} The Gauss-Newton     direction, or velocity vector, gives the correct direction that     one should move to approach the best fit while navigating a     canyon. However, that direction quickly rotates, requiring an     algorithm to take very small steps in order to avoid uphill     moves. The geodesic acceleration indicates the direction in which     the velocity rotates. The geodesic acceleration determines a     parabolic trajectory that can efficiently navigate the valley     without running up the wall. The linear trajectory quickly runs up     the side of the canyon wall. }

\end{figure}

With an easily evaluated approximation for the acceleration, we can then consider the trajectory given by \begin{equation}   \delta\theta^{\mu}=\dot{\theta}^{\mu}\delta\tau+\frac{1}{2}\ddot{\theta}^{\mu}\delta\tau^{2}.\label{eq:Acceleration-Step}\end{equation} By following the winding canyon with a parabolic path instead of a linear path, we expect to require fewer steps to arrive at the best fit. The parabola can more naturally curve around the corners of the canyon than the straight line path. This is illustrated for the Rosenbrock function in Fig.~\ref{fig:RosenbrockAcceleration}. Because the canyon of the Rosenbrock function is parabolic, it can be traversed exactly to the best fit by the acceleration in a single step.

\begin{figure*}

  \includegraphics[width=7in]{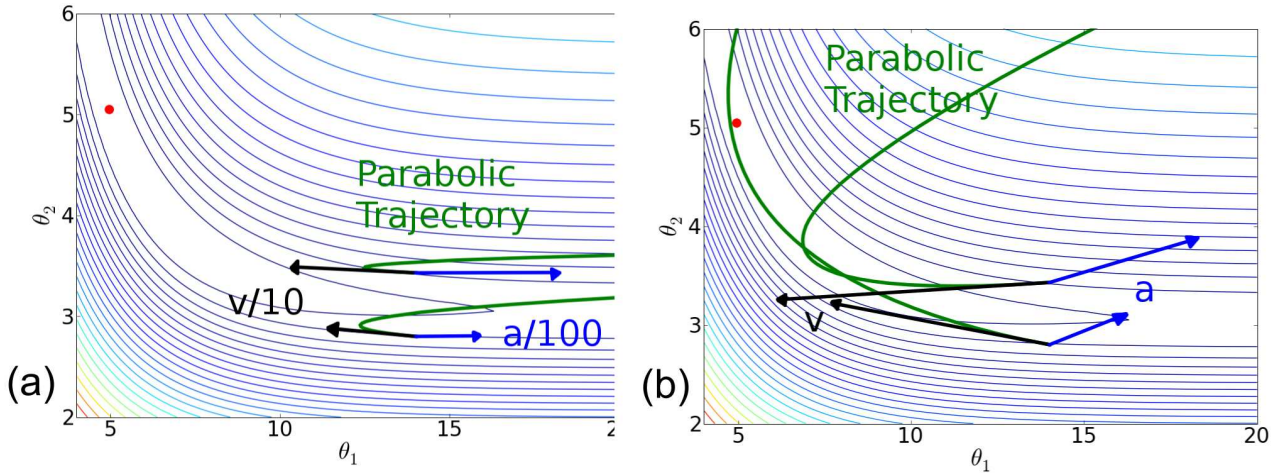}\caption{\label{fig:Acceleration-Rotation}a)     \textbf{De-acceleration when overstepping.} Typically the velocity     vector greatly overestimates the proper step size. (We have     rescaled both velocity and acceleration to fit in the figure.)     Algebraically, this is due to the factor of the inverse metric in     the velocity, which has very large eigenvalues.  The acceleration     compensates for this by pointing anti-parallel to the     velocity. However, the acceleration vector is also very large, as     it is multiplied twice by the velocity vector and once by the     inverse metric.To make effective use of the acceleration, it is     necessary to regularize the metric by including a damping term. b)     \textbf{Acceleration indicating the direction of the canyon. }As     the Levenberg-Marquardt parameter is raised, the velocity vector     shortens and rotates from the natural gradient into the downhill     direction. The acceleration vector also shortens, although much     more rapidly, and also rotates.  In this two dimensional cross     section, although the two velocity vectors rotate in opposite     directions, the accelerations both rotate to indicate the     direction that the canyon is turning. By considering the path that     one would optimally like to take (along the canyon), it is clear     that the acceleration vector is properly indicating the correction     to the desired trajectory.}

\end{figure*}

The relationship between the velocity and the acceleration depicted in Fig.~\ref{fig:RosenbrockAcceleration} for the Rosenbrock function is overly idealized. In general the velocity and the acceleration will not be perpendicular; in fact, it is much more common for them to be nearly parallel or anti-parallel. Notice that the expression for the connection coefficient involves a factor of the inverse metric, which will tend to bias the acceleration to align parallel to the sloppy directions, just as it does for the velocity. It is much more common for the acceleration to point in the direction opposite to the velocity, as for a summing exponentials model in Fig.~\ref{fig:Acceleration-Rotation}a.

Although an acceleration that is anti-parallel to the velocity may seem worthless, it is actually telling us something useful: our proposed step was too large. As we regulate the velocity by increasing the Levenberg-Marquardt parameter, we also regulate the acceleration. Once our velocity term is comparable to the distance over which the canyon begins to curve, the acceleration indicates into which direction the canyon is curving, as in Fig.~\ref{fig:Acceleration-Rotation}b.

If the damping term is too small, the acceleration points in the opposite direction to and is much larger than the velocity. This scenario is dangerous because it may cause the algorithm to move in precisely the opposite direction to the Gauss-Newton direction, causing parameter evaporation. To fix this problem, we add another criterion for an acceptable step. We want the contribution from the acceleration to be smaller than the contribution from the velocity; therefore, we typically reject proposed steps, increasing the Leveberg-Marquardt parameter until \begin{equation}   \frac{\sqrt{\sum\left(a^{\mu}\right)^{2}}}{\sqrt{\sum\left(v^{\mu}\right)^{2}}}<\alpha,\label{eq:avcriterion}\end{equation} where $\alpha$ is a chosen parameter, typically unity, although for some problems a smaller value is required.

The acceleration is likely to be most useful when the canyon is very narrow. As the canyon narrows, the allowed steps become smaller. In essence, the narrowness of the canyon is determining to what accuracy we are solving the geodesic equation. If the canyon requires a very high accuracy, then a second order algorithm is likely to converge much more quickly than a first order algorithm. We will see this explicitly in the next section when we compare algorithms.

We have argued repeatedly that for sloppy models whose parameter-effects curvature are dominant, a geodesic is the path that an algorithm should follow. One could object to this assertion on the grounds that, apart from choosing the initial direction of the geodesic to be the Gauss-Newton direction, there is no reference to the cost gradient in the geodesic equation. If a manifold is curved, then the geodesic will not lead directly to the best fit. In particular, the acceleration is independent of the data.

Instead of a geodesic, one could argue that the path that one should
follow is given by the first order differential equation \begin{equation}
  v^{\mu}=\frac{-g^{\mu\nu}\nabla_{\nu}C}{\sqrt{g^{\alpha\beta}\nabla_{\alpha}C\       \nabla_{\beta}C}},\label{eq:NaturalGradientODE}\end{equation}
where we have introduced the denominator to preserve the norm of
the tangent vector. Each Levenberg-Marquardt step chooses a direction
in the Gauss-Newton direction on the model graph, which seems to be
better described by Eq.~\eqref{eq:NaturalGradientODE} than by the geodesic
equation, Eq.~\eqref{eq:GeodesicODE}. In fact Eq.~\eqref{eq:NaturalGradientODE}
has been proposed as a Neural Network training algorithm by Amari
et al.~\cite{Amari2006}.

The second order differential equation corresponding to Eq.~\eqref{eq:NaturalGradientODE} which can be found by taking the second derivative of the parameters, is a very complicated expression. However, if one then applies the approximation that all non-linearities are parameter-effects curvature, the resulting differential equation is exactly the geodesic equation.  By comparing step sizes with inverse curvatures in Fig.~\ref{fig:Curvature-StepSize}, we can see that over a distance of several steps, the approximation that all non-linearities are parameter-effects curvature should be very good. In such a case, the deviation of Eq.~\eqref{eq:NaturalGradientODE} from Eq.~\eqref{eq:GeodesicODE} will not be significant over a few steps.

While the tensor analysis behind this result is long and tedious, the geometric meaning is simple and intuitive: if steps are much smaller than the extrinsic curvature on the surface, then the vector (in data space) corresponding to the Gauss-Newton direction can parallel transport itself to find the Gauss-Newton direction at the next point. That is to say the direction of the tangent vector of a geodesic does not change if the manifold is extrinsically flat.

Including second derivative information in an algorithm is not new. Newton's method, for example replaces the approximate Hessian of the Gauss-Newton method in Eq.~\eqref{eq:HessianApprox}, with the full Hessian in Eq.~\eqref{eq:Hessian}. Many standard algorithms seek to efficiently find the actual Hessian, either by calculating it directly or by estimation~\cite{Gill1978,Press2007}. One such algorithm, which we use for comparison in the next section, is a quasi-Newton method of Broyden, Fletcher, Goldfarb, and Shannon (BFGS)~\cite{Nocedal1999}, which estimates the second derivative from an accumulation of Jacobian evaluations at each step.

In contrast to these Newton-like algorithms, the geodesic acceleration is not an attempt to better approximate the Hessian. The results of section~\ref{sec:Extended-Geodesic-Coordinates} suggest that the approximate Hessian is very good. Instead of correcting the error in the size and direction of the ellipses around the best fit, it is more productive to account for how they are bent by non-linearities, which is the role of the geodesic acceleration. The geodesic acceleration is a \textit{cubic} correction to the Levenberg-Marquardt step.

There are certainly problems for which a quasi-Newton algorithm will make important corrections to the approximate Hessian. However, we have argued that sloppy models represent a large class of problems for which the Newton correction is negligible compared to that of the geodesic acceleration. We demonstrate this numerically with several examples in the next section.

\subsection{Algorithm Comparisons\label{sub:Algorithm-Comparisons}}

To demonstrate the effectiveness of an algorithm that uses delayed gratification and the geodesic acceleration, we apply it to a few test problems that highlight the typical difficulties associated with fitting by least squares.

\begin{figure}

  \includegraphics[width=3.25in]{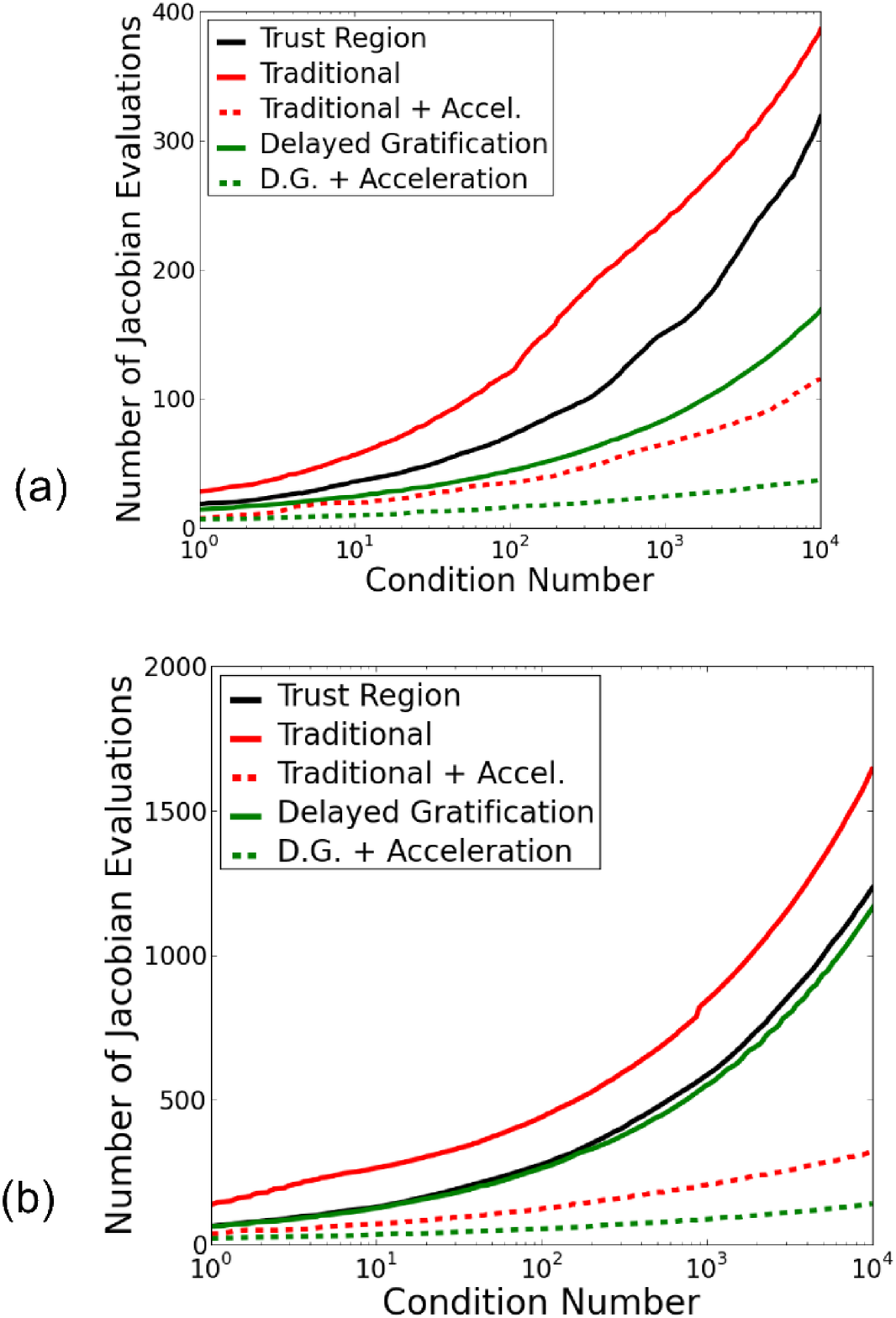}\caption{\label{fig:RosenbrockResults-LM}\textbf{Generalized       Rosenbrock results for Levenberg-Marquardt variants.} If the     canyon that an algorithm must follow is very narrow (measured by     the condition number of the metric at the best fit) or turns     sharply, the algorithm will require more steps to arrive at the     best fit. Those that use the geodesic acceleration term converge     more quickly as the canyon narrows. As the parameter-effects     curvature increases, the canyon becomes more curved and the     problem is more difficult. Notice that changing the canyon's path     from a cubic function in a) to a quartic function in b) slowed the     convergence rate by a factor of 5. We have omitted the quadratic     path since including the acceleration allows the algorithm to find     the best fit in one step, regardless of how narrow the canyon     becomes.}

\end{figure}

First, consider a generalized Rosenbrock function,\[ C=\frac{1}{2}\left(\theta_{1}^{2}+A^{2}\left(\theta_{2}-\frac{\theta_{1}^{n}}{n}\right)^{2}\right),\] where $A$ and $n$ are not optimizable parameters but set to control the difficulty of the problem. This problem has a global minimum of zero cost at the origin, with a canyon following the polynomial path $\theta_{1}^{n}/n$ whose width is determined by $A$. To compare algorithms we draw initial points from a Gaussian distribution centered at $(1,1/n)$ with standard deviation of unity, and compare the average number of Jacobian evaluations an algorithm requires in order to decrease the cost to $10^{-4}$. The results for the cubic and quartic versions of the problem are given in Fig.~\ref{fig:RosenbrockResults-LM} for several version of the the Levenberg-Marquardt algorithm.

We next consider a summing exponential problem; a summary of these results can be found in~\cite{Transtrum2010}. Here we expand it to include the delayed gratification algorithm outlined above in section~\ref{sub:Delayed-Gratification}.

\begin{table*}
  \begin{tabular}{|c||c|c|c|}
    \hline 
    Algorithm  & Success Rate  & Mean NJEV  & Mean NFEV\tabularnewline
    \hline
    \hline 
    Trust Region LM  & 12\%  & 1517  & 1649\tabularnewline
    \hline 
    Traditional LM  & 33\%  & 2002  & 4003\tabularnewline
    \hline 
    Traditional LM + accel  & 65\%  & 258  & 1494\tabularnewline
    \hline 
    Delayed Gratification  & 26\% & 1998 & 8625\tabularnewline
    \hline 
    Delayed Gratification + accel  & 65\% & 163 & 1913\tabularnewline
    \hline 
    BFGS  & 8\%  & 5363  & 5365\tabularnewline
    \hline
  \end{tabular}

  \caption{\label{tab:Exponential-Results}\textbf{ }The results of     several algorithms applied to a test problem of fitting a sum of four exponential terms (varying both rates and amplitudes -- 8 parameters) in log-parameters (to enforce positivity). Initial conditions are chosen near a manifold boundary with a best fit of zero cost near the center of the manifold. Among successful attempts, we further compare the average number of Jacobian and function evaluations needed to arrive at the fit. Success rate indicates an algorithm's ability to avoid the manifold boundaries (find the canyon from the plateau), while the number of Jacobian and function evaluations indicate how efficiently it can follow the canyon to the best fit. BFGS is a quasi newton scalar minimizer of Broyden, Fletcher, Goldfarb, and Shanno (BFGS)~\cite{Nocedal1999,Jones2001}. The traditional~\cite{Marquardt1963,Press2007} and trust region~\cite{More1977} implementations of Levenberg-Marquardt consistently outperform this and other general optimization routines on least squares problems, such as Powell, simplex, and conjugate gradient. Including the geodesic acceleration on a standard variant of Levenberg-Marquardt dramatically increases the success rate while decreasing the computation time. }

\end{table*}

A surprising result from Table~\ref{tab:Exponential-Results} is that including the geodesic acceleration not only improves the speed of convergence, but improves the likelihood of convergence, that is, the algorithm is less likely to evaporate parameters. This is a consequence of the modified acceptance criterion in Eq.~\eqref{eq:avcriterion}.  As an algorithm evaporates parameters, it approaches a singular point of the metric on the model manifold, causing the velocity vector in parameter space to diverge. The acceleration, however, also diverges, but much more rapidly than the velocity. By requiring the acceleration term to be smaller than the velocity, the algorithm is much more adept at avoiding boundaries.  Geodesic acceleration, therefore, helps to improve both the initial search for the canyon from the plateau, as well as the subsequent race along the canyon to the best fit.

Finally, we emphasize that the purpose of this section was to demonstrate that delayed gratification and geodesic acceleration are potentially helpful modifications to existing algorithms. The results presented in this section do not constitute a rigorous comparison, as such a study would require a much broader sampling of test problems. Instead, we have argued that ideas from differential geometry can be helpful to speed up the fitting process if existing algorithms are sluggish.  We are in the process of performing a more extensive comparison whose results will appear shortly~\cite{Transtrum2010c}.

\section{Conclusions\label{sec:Conclusions}}

A goal of this paper has been to use a geometric perspective to study nonlinear least squares models, deriving the relevant metric, connection, and measures of curvature, and to show that geometry provides useful insights into the difficulties associated with optimization.

We have presented the model manifold and noted that it typically has boundaries, which explain the phenomenon of parameter evaporation in the optimization process. As algorithms run into the manifold's boundaries, parameters are pushed to infinite or otherwise unphysical values. For sloppy models, the manifold is bounded by a hierarchy of progressively narrow boundaries, corresponding to the less responsive direction of parameter space. The model behavior spans a hyper-ribbon in data space.  This phenomenon of geometric sloppiness is one of the key reasons that sloppy models are difficult to optimize.  We provide a theoretical caricature of the model manifold characterizing their geometric series of widths, extrinsic curvatures, and parameter-effects curvatures.  Using this caricature, we estimate the number of evaporated parameters one might expect to find at the best fit for a given uncertainty in the data.

The model graph removes the boundaries and helps to keep the parameters at reasonable levels. This is not always sufficient, however, and we suggest that in many cases, the addition of thoughtful priors to the cost function can be a significant help to algorithms.

The second difficulty in optimizing sloppy models is that the model parameters are far removed from the model behavior. Because most sloppy models are dominated by parameter-effects curvature, if one could reparametrize the model with extended geodesic coordinates, the long narrow canyons would be transformed to one isotropic quadratic basin.  Optimizing a problem in extended geodesic coordinates would be a trivial task!

Inspired by the motion of geodesics in the curved valleys, we developed the delayed gratification update scheme for the traditional Levenberg-Marquardt algorithm and further suggest the addition of a geodesic acceleration term. We have seen that when algorithms must follow long narrow canyons, these can give significant improvement to the optimization algorithm.  We believe that the relative cheap computational cost of adding the geodesic acceleration to the Levenberg-Marquardt step gives it the potential to be a robust, general-purpose optimization algorithm, particularly for high dimensional problems. It is necessary to explore the behavior of geodesic acceleration on a larger problem set to justify this conjecture~\cite{Transtrum2010c}.

\section*{Acknowledgments}

We would like to thank Saul Teukolsky, Eric Siggia, John Guckenheimer, Cyrus Umrigar, Peter Nightingale, Stefanos Papanikolou, Bryan Daniels, and Yoav Kallus for helpful discussions, and acknowledge support from NSF grant number DMR-0705167.

\section*{Appendix A: Information Geometry }

The Fisher information matrix, or simply Fisher information, $I$, is a measure of the information contained in a probability distribution, $p$. Let $\xi$ be the random variable whose distribution is described by $p$, and further assume that $p$ depends on other parameters $\theta$ that are not random. This leads us to write \[ p=p(\xi;\theta),\] with the log likelihood function denoted by $l$:\[ l=\log\, p.\] The information matrix is defined to be the expectation value of the second derivatives of $l$, \begin{equation}   I_{\mu\nu}=\langle-\frac{\partial^{2}l}{\partial\theta^{\mu}\partial\theta^{\nu}}\rangle=-\int   d\xi\   p(\xi,\theta)\frac{\partial^{2}l}{\partial\theta^{\mu}\partial\theta^{\nu}}.\label{eq:Fisher1}\end{equation} It can be shown that the Fisher information can be written entirely in terms of first derivatives:\begin{equation}   I_{\mu\nu}=\langle\frac{\partial     l}{\partial\theta^{\mu}}\frac{\partial     l}{\partial\theta^{\nu}}\rangle=\int d\xi\   p(\xi,\theta)\frac{\partial l}{\partial\theta^{\mu}}\frac{\partial     l}{\partial\theta^{\nu}}.\label{eq:Fisher2}\end{equation}

Eq.~\eqref{eq:Fisher2} makes it clear that the Fisher information is a symmetric, positive definite matrix which transforms like a covariant rank-2 tensor. This means that it has all the properties of a metric in differential geometry. Information geometry considers the manifolds whose metric is the Fisher information matrix corresponding to various probability distributions. Under such an interpretation, the Fisher information matrix is known as the Fisher information metric.

As we saw in Section~\ref{sec:Introduction}, least squares problems arise by assuming a Gaussian distribution for the deviations from the model. Under this assumption, the cost function is the negative of the log likelihood (ignoring an irrelevant constant). Using these facts, it is straightforward to apply Eq.~\eqref{eq:Fisher1} or Eq.~\eqref{eq:Fisher2} to calculate the information metric for least squares problems. From Eq.~\eqref{eq:Fisher1}, we get \begin{equation}   g_{\mu\nu}=\langle\frac{\partial^{2}C}{\partial\theta^{\mu}\partial\theta^{\nu}}\rangle=\sum_{m}\langle\partial_{\mu}r_{m}\partial_{\nu}r_{m}+r_{m}\partial_{\mu}\partial_{\nu}r_{m}\rangle,\label{eq:Metric1}\end{equation} where we have replaced $I$ by $g$ to indicate that we are now interpreting it as a metric.

Eq.~\eqref{eq:Metric1}, being an expectation value, is really an integral over the random variable (i.e. the residuals) weighted by the probability.  However, since the integral is Gaussian, it can be evaluated easily using Wick's theorem (remembering that the residuals have unit variance).  The only subtlety is how to handle the derivatives of the residuals.  Inspecting Eq.~\eqref{eq:rdefinition}, reveals that the derivatives of the residuals have no random element, and can therefore be treated as constant. The net result is \begin{equation}   g_{\mu\nu}=\sum_{m}\partial_{\mu}r_{m}\partial_{\nu}r_{m}=(J^{T}J)_{\mu\nu},\label{eq:Metric}\end{equation} since $\langle r_{m}\rangle=0$. Note that we have used the Jacobian matrix, $J_{m\mu}=\partial_{\mu}r_{m}$ in the final expression.

We arrive at the same result using Eq.~\eqref{eq:Fisher2} albeit using different properties of the distribution: \[ g_{\mu\nu}=\sum_{m,n}\langle r_{m}\partial_{\mu}r_{m}r_{n}\partial_{\mu}r_{n}\rangle.\] Now we note that the residuals are independently distributed, $\langle r_{m}r_{n}\rangle=\delta_{mn}$, which immediately gives Eq.~\eqref{eq:Metric}, the same metric found in Section~\ref{sec:Introduction}.

There is a class of connections consistent with the Fisher metric, known as the $\alpha$-connections because they are parametrized by a real number, $\alpha$~\cite{Amari2007}. They are given by the formula \[ \Gamma_{\mu\nu,\epsilon}^{(\alpha)}=\langle\partial_{\epsilon}l\partial_{\mu}\partial_{\nu}l+\left(\frac{1-\alpha}{2}\right)\partial_{\epsilon}l\partial_{\mu}l\partial_{\nu}l\rangle.\] This expression is straightforward to evaluate. The result is independent of $\alpha$, \[ \Gamma_{\mu\nu}^{\epsilon}=g^{\epsilon\kappa}\sum_{m}\partial_{\kappa}r_{m}\partial_{\mu}\partial_{\nu}r_{m}.\] It has been shown elsewhere that the connection corresponding to $\alpha=0$ is in fact the Riemann connection. It is interesting to note that all the $\alpha$-connections, for the case of the nonlinear least squares problem, are the Riemann connection.

The field of information geometry is summarized nicely in several books~\cite{Amari2007,Murray1993}.

\section*{Appendix B: Algorithms}

Since we are optimizing functions with the form of sums of squares, we are primarily interested in algorithms that specialize in this form, specifically variants of the Levenberg-Marquardt algorithm.  The standard implementation of the Levenberg-Marquardt algorithm involves a trust region formulation. A FORTRAN implementation, which we use, is provided by MINPACK~\cite{More1980}.

The traditional formulation of Levenberg-Marquardt, however, does not employ a trust region, but adjusts the Levenberg-Marquardt term based on whether the cost has increased or decreased after a given step. An implementation of this algorithm is described in Numerical Recipes~\cite{Press2007} and summarized in Algorithm~\ref{alg:Traditional-Levenberg-Marquardt}.  Typical values of $\lambda_{up}$ and $\lambda_{down}$ are $10$.  We use this formulation as the basis for our modifications.

\begin{algorithm}
\begin{flushleft}
  1. Initialize values for the parameters, $x$, the   Levenberg-Marquardt parameter $\lambda$, as well as $\lambda_{up}$   and $\lambda_{down}$ to be used to adjust the damping term. Evaluate   the residuals $r$ and the Jacobian $J$ at the initial parameter   guess.

  2. Calculate the metric, $g=J^{T}J+\lambda I$ and the cost gradient   $\nabla C=J^{T}r$, $C=\frac{1}{2}r^{2}$.

  3. Evaluate the new residuals, $r_{new}$ at the point given by   $x_{new}=x-g^{-1}\nabla C$ , and calculate the cost at the new   point, $C_{new}=\frac{1}{2}r_{new}^{2}$.

  4. If $C_{new}<C$, accept the step, $x=x_{new}$ and set $r=r_{new}$   and $\lambda=\lambda/\lambda_{down}$. Otherwise, reject the step,   keep the old parameter guess $x$ and the old residuals $r$, and   adjust $\lambda=\lambda\times\lambda_{up}$.

  5. Check for convergence. If the method has converged, return $x$ as   the best fit parameters. If the method has not yet converged but the   step was accepted, evaluate the Jacobian $J$ at the new parameter   values. Go to step 2.

  \caption{\label{alg:Traditional-Levenberg-Marquardt}Traditional     Levenberg-Marquardt as described in~\cite{Levenberg1944,Marquardt1963,Press2007}}
\end{flushleft}
\end{algorithm}

The delayed gratification version of Levenberg-Marquardt that we describe in section~\ref{sub:Delayed-Gratification} modifies the traditional Levenberg-Marquardt algorithm to raise and lower the Levenberg-Marquardt term by differing amounts. The goal is to accept a step with the smallest value of the damping term that will produce a downhill step. This can typically be accomplished by choosing $\lambda_{up}=2$ and $\lambda_{down}=10$.

The geodesic acceleration algorithm can be added to any variant of Levenberg-Marquardt. We explicitly add it to the traditional version and the delayed gratification version, as described in Algorithm~\ref{alg:Geodesic-Acceleration}.  We do this by calculating the geodesic acceleration on the model graph at each iteration. If the step raises the cost or if the acceleration is larger than the velocity, then we reduce the Levenberg-Marquardt term and reject the step by default. If the step moves downhill and the velocity is larger than the acceleration, then we accept the step.  For accepted steps we raise the Levenberg-Marquardt term; otherwise, we decrease the Levenberg-Marquardt term.  In our experience the algorithm described in Algorithm~\ref{alg:Geodesic-Acceleration} is robust enough for most applications; however, we do not consider it to be a polished algorithm.  We will present elsewhere an algorithm utilizing geodesic acceleration that is further optimized and that we will make available as a FORTRAN routine~\cite{Transtrum2010c}.

\begin{algorithm}
\begin{flushleft}
  1. Initialize values for the parameters, $x$, the   Levenberg-Marquardt parameter $\lambda$, as well as $\lambda_{up}$   and $\lambda_{down}$ to be used to adjust the damping term, and   $\alpha$ to control the acceleration/velocity ratio. Evaluate the   residuals $r$ and the Jacobian $J$ at the initial parameter guess.

  2. Calculate the metric, $g=J^{T}J+\lambda I$ and the Cost gradient   $\nabla C=J^{T}r$, $C=\frac{1}{2}r^{2}$.

  3. Calculate the velocity $v=-g^{-1}\nabla C$, the geodesic   acceleration of the residuals in the direction of the velocity   $a=-g^{-1}J^{T}\left(v^{\mu}v^{\nu}\partial_{\mu}\partial_{\nu}r\right)$

  4. Evaluate the new residuals, $r_{new}$ at the point given by   $x_{new}=x+v+\frac{1}{2}a$ , and calculate the cost at the new   point, $C_{new}=\frac{1}{2}r_{new}^{2}$.

  5. If $C_{new}<C$ and $|a|/|v|<\alpha$, accept the step, $x=x_{new}$   and set $r=r_{new}$ and $\lambda=\lambda/\lambda_{down}$. Otherwise,   reject the step, keep the old parameter guess $x$ and the old   residuals $r$, and adjust $\lambda=\lambda\times\lambda_{up}$.

  6. Check for convergence. If the method has converged, return $x$ as   the best fit parameters. If the method has not yet converged but the   step was accepted evaluate the Jacobian $J$ at the new parameter   values. Go to step 2.   \caption{\label{alg:Geodesic-Acceleration}Geodesic Acceleration in     the traditional Levenberg-Marquardt algorithm}
\end{flushleft}
\end{algorithm}

In addition to the variations of the Levenberg-Marquardt algorithm, we also compare algorithms for minimization of arbitrary functions not necessarily of the least squares form. We take several such algorithms from the Scipy optimization package~\cite{Jones2001}. These fall into two categories, those that make use of gradient information and those that do not. Algorithms utilizing gradient information include a quasi-Newton of Broyden, Fletcher, Goldfarb, and Shannon (BFGS), described in~\cite{Nocedal1999}. We also employ a limited memory variation (L-BFGS-B) described in~\cite{Byrd1995} and a conjugate gradient (CG) method of Polak and Ribiere, also described in~\cite{Nocedal1999}.  We also explored the downhill simplex algorithm of Nelder and Mead and a modification of Powells' method~\cite{Jones2001}, neither of which make use of gradient information directly, and were not competitive with other algorithms.

\bibliographystyle{aps}
\bibliography{C:/Users/Mark/Documents/References}

\begin{thebibliography}{10}

\bibitem{Brown2003}
K.~Brown, J.~Sethna:
\newblock \emph{Physical Review E} \textbf{68}  (2003) 21904

\bibitem{Brown2004}
K.~Brown, C.~Hill, G.~Calero, C.~Myers, K.~Lee, J.~Sethna, R.~Cerione:
\newblock \emph{Physical biology} \textbf{1}  (2004) 184

\bibitem{Casey2007}
F.~Casey, D.~Baird, Q.~Feng, R.~Gutenkunst, J.~Waterfall, C.~Myers, K.~Brown,
  R.~Cerione, J.~Sethna:
\newblock \emph{Systems Biology, IET} \textbf{1}  (2007) 190

\bibitem{Daniels2008}
B.~Daniels, Y.~Chen, J.~Sethna, R.~Gutenkunst, C.~Myers:
\newblock \emph{Current Opinion in Biotechnology} \textbf{19}  (2008) 389

\bibitem{Gutenkunst2007}
R.~Gutenkunst, F.~Casey, J.~Waterfall, C.~Myers, J.~Sethna:
\newblock \emph{Annals of the New York Academy of Sciences} \textbf{1115}
  (2007) 203

\bibitem{Gutenkunst2007a}
R.~Gutenkunst, J.~Waterfall, F.~Casey, K.~Brown, C.~Myers, J.~Sethna:
\newblock \emph{PLoS Comput Biol} \textbf{3}  (2007) e189

\bibitem{Gutenkunst2008}
R.~Gutenkunst:
\newblock \emph{Sloppiness, modeling, and evolution in biochemical networks}:
\newblock Ph.D. thesis, Cornell University (2008)

\bibitem{Waterfall2006}
J.~Waterfall, F.~Casey, R.~Gutenkunst, K.~Brown, C.~Myers, P.~Brouwer,
  V.~Elser, J.~Sethna:
\newblock \emph{Physical Review Letters} \textbf{97}  (2006) 150601

\bibitem{Jeffreys1998}
H.~Jeffreys:
\newblock \emph{Theory of probability}:
\newblock Oxford University Press, USA (1998)

\bibitem{Rao1945}
C.~Rao:
\newblock \emph{Vull. Calcutta Math. Soc.} \textbf{37}  (1945) 81

\bibitem{Rao1949}
C.~Rao:
\newblock \emph{Sankhya} \textbf{9}  (1949) 246

\bibitem{Amari2007}
S.~Amari, H.~Nagaoka:
\newblock \emph{Methods of Information Geometry}:
\newblock Amer Mathematical Society (2007)

\bibitem{Murray1993}
M.~Murray, J.~Rice:
\newblock \emph{Differential geometry and statistics}:
\newblock Chapman \& Hall New York (1993)

\bibitem{Beale1960}
E.~Beale:
\newblock \emph{Journal of the Royal Statistical Society} \textbf{22}  (1960)
  41

\bibitem{Bates1980}
D.~Bates, D.~Watts:
\newblock \emph{J. Roy. Stat. Soc} \textbf{42}  (1980) 1

\bibitem{Bates1981}
D.~Bates, D.~Watts:
\newblock \emph{Ann. Statist} \textbf{9}  (1981) 1152

\bibitem{Bates1983}
D.~Bates, D.~Hamilton, D.~Watts:
\newblock \emph{Communications in Statistics-Simulation and Computation}
  \textbf{12}  (1983) 469

\bibitem{Bates1988}
D.~Bates, D.~Watts:
\newblock \emph{Nonlinear Regression Analysis and Its Applications}:
\newblock John Wiley (1988)

\bibitem{Cook1985}
R.~Cook, J.~Witmer:
\newblock \emph{American Statistical Association} \textbf{80}  (1985) 872

\bibitem{Cook1986}
R.~Cook, M.~Goldberg:
\newblock \emph{The Annals of Statistics}  (1986) 1399

\bibitem{Clarke1987}
G.~Clarke:
\newblock \emph{Journal of the American Statistical Association}  (1987) 844

\bibitem{Transtrum2010}
M.~K. Transtrum, B.~B. Machta, J.~P. Sethna:
\newblock \emph{Physical Review Letters} \textbf{104}  (2010) 1060201

\bibitem{Barndorff-Nielsen1986}
O.~Barndorff-Nielsen, D.~Cox, N.~Reid:
\newblock \emph{International statistical review} \textbf{54}  (1986) 83

\bibitem{Gabay1982}
D.~Gabay:
\newblock \emph{Journal of Optimization Theory and Applications} \textbf{37}
  (1982) 177

\bibitem{Mahony1994}
R.~Mahony:
\newblock \emph{Optimization algorithms on homogeneous spaces}:
\newblock Ph.D. thesis, Australian National University (1994)

\bibitem{Mahony2002}
R.~Mahony, J.~Manton:
\newblock \emph{Journal of Global Optimization} \textbf{23}  (2002) 309

\bibitem{Manton2004}
J.~Manton:
\newblock In: \emph{Proceedings of the 16th International Symposium on
  Mathematical Theory of Networks and Systems, Leuven, Belgium} (2004)

\bibitem{Peeters1993}
R.~Peeters:
\newblock \emph{Research-Memorandum}  (1993)

\bibitem{Smith1993}
S.~Smith:
\newblock \emph{Harvard University, Cambridge, MA}  (1993)

\bibitem{Smith1994}
S.~Smith:
\newblock \emph{Hamiltonian and gradient flows, algorithms and control}
  \textbf{3}  (1994) 113

\bibitem{Udriste1994}
C.~Udriste:
\newblock \emph{Convex functions and optimization methods on Riemannian
  manifolds}:
\newblock Kluwer Academic Pub (1994)

\bibitem{Yang2007}
Y.~Yang:
\newblock \emph{Journal of Optimization Theory and Applications} \textbf{132}
  (2007) 245

\bibitem{Absil2008}
P.~Absil, R.~Mahony, R.~Sepulchre:
\newblock \emph{Optimization Algorithms on Matrix Manifolds}:
\newblock Princeton University Press (2008)

\bibitem{Press2007}
W.~Press, S.~A. Teukolsky, W.~T. Vetterling, B.~P. Flannery:
\newblock \emph{Numerical recipes: the art of scientific computing,}:
\newblock Cambridge University Press (2007)

\bibitem{Misner1973}
C.~Misner, K.~Thorne, J.~Wheeler:
\newblock \emph{Gravitation}:
\newblock WH Freeman and Company (1973)

\bibitem{Spivak1979}
M.~Spivak:
\newblock \emph{PublishorPerish, California}  (1979)

\bibitem{Eisenhart1997}
L.~Eisenhart:
\newblock \emph{Riemannian geometry}:
\newblock Princeton Univ Pr (1997)

\bibitem{Ivancevic2007}
T.~Ivancevic:
\newblock \emph{Applied differential geometry: a modern introduction}:
\newblock World Scientific Pub Co Inc (2007)

\bibitem{Stoer2002}
J.~Stoer, R.~Bulirsch, W.~Gautschi, C.~Witzgall:
\newblock \emph{Introduction to numerical analysis}:
\newblock Springer Verlag (2002)

\bibitem{Hertz1991}
J.~Hertz, A.~Krogh, R.~Palmer:
\newblock \emph{Introduction to the theory of neural computation}:
\newblock Westview Press (1991)

\bibitem{Frederiksen2004}
S.~Frederiksen, K.~Jacobsen, K.~Brown, J.~Sethna:
\newblock \emph{Physical Review Letters} \textbf{93}  (2004) 165501

\bibitem{Amari2006}
S.~Amari, H.~Park, T.~Ozeki:
\newblock \emph{Neural Computation} \textbf{18}  (2006) 1007

\bibitem{Levenberg1944}
K.~Levenberg:
\newblock \emph{Quart. Appl. Math} \textbf{2}  (1944) 164

\bibitem{Marquardt1963}
D.~Marquardt:
\newblock \emph{Journal of the Society for Industrial and Applied Mathematics}
  \textbf{11}  (1963) 431

\bibitem{More1977}
J.~More:
\newblock \emph{Lecture notes in mathematics} \textbf{630}  (1977) 105

\bibitem{Kass1984}
R.~Kass:
\newblock \emph{Journal of the Royal Statistical Society. Series B
  (Methodological)}  (1984) 86

\bibitem{Hamilton1982}
D.~Hamilton, D.~Watts, D.~Bates:
\newblock \emph{Ann. Statist} \textbf{10}  (1982) 393

\bibitem{Donaldson1987}
J.~Donaldson, R.~Schnabel:
\newblock \emph{Technometrics} \textbf{29}  (1987) 67

\bibitem{Wei1994}
B.~Wei:
\newblock \emph{Australian \& New Zealand Journal of Statistics} \textbf{36}
  (1994) 327

\bibitem{Haines2004}
L.~Haines, T.~O~Brien, G.~Clarke:
\newblock \emph{Statistica Sinica} \textbf{14}  (2004) 547

\bibitem{Demidenko2006}
E.~Demidenko:
\newblock \emph{Computational Statistics and Data Analysis} \textbf{51}  (2006)
  1739

\bibitem{Hilbert1999}
D.~Hilbert, S.~Cohn-Vossen:
\newblock \emph{{Geometry and the Imagination}}:
\newblock American Mathematical Society (1999)

\bibitem{Hartley1961}
H.~Hartley:
\newblock \emph{Technometrics}  (1961) 269

\bibitem{Transtrum2010c}
M.~K. Transtrum, B.~B. Machta, C.~Umrigar, P.~Nightingale, J.~P. Sethna:
\newblock \emph{Development and comparison of algorithms for nonlinear least
  squares fitting}:
\newblock In preparation

\bibitem{Bakushinskii1992}
A.~Bakushinskii:
\newblock \emph{Computational Mathematics and Mathematical Physics} \textbf{32}
   (1992) 1353

\bibitem{Lampton1997}
M.~Lampton:
\newblock \emph{Computers in Physics} \textbf{11}  (1997) 110

\bibitem{Luenberger1972}
D.~Luenberger:
\newblock \emph{Management Science}  (1972) 620

\bibitem{P'azman2002}
A.~P{\'a}zman:
\newblock \emph{Journal of Statistical Planning and Inference} \textbf{103}
  (2002) 401

\bibitem{Igel2005}
C.~Igel, M.~Toussaint, W.~Weishui:
\newblock \emph{Trends and Applications in Constructive Approximation,
  International Series of Numerical Mathematics} \textbf{151}  (2005)

\bibitem{Nishimori2005}
Y.~Nishimori, S.~Akaho:
\newblock \emph{Neurocomputing} \textbf{67}  (2005) 106

\bibitem{Gill1978}
P.~Gill, W.~Murray:
\newblock \emph{SIAM Journal on Numerical Analysis}  (1978) 977

\bibitem{Nocedal1999}
J.~Nocedal, S.~Wright:
\newblock \emph{Numerical optimization}:
\newblock Springer (1999)

\bibitem{Jones2001}
E.~Jones, T.~Oliphant, P.~Peterson, \emph{et~al.}:
\newblock \emph{URL http://www. scipy. org}  (2001)

\bibitem{More1980}
J.~Mor{\'e}, B.~Garbow, K.~Hillstrom:
\newblock \emph{User guide for MINPACK-1} (1980)

\bibitem{Byrd1995}
R.~Byrd, P.~Lu, J.~Nocedal, C.~Zhu:
\newblock \emph{SIAM Journal on Scientific Computing} \textbf{16}  (1995) 1190

\bibitem{Brown2003a}
K.~S. Brown:
\newblock \emph{Signal Transduction, Sloppy Models, and Statistical Mechanics}:
\newblock Ph.D. thesis, Cornell University (2003)

\bibitem{Golub1973}
G.~Golub, V.~Pereyra:
\newblock \emph{SIAM Journal on Numerical Analysis} \textbf{10}  (1973) 413

\bibitem{Kaufman1975}
L.~Kaufman:
\newblock \emph{BIT Numerical Mathematics} \textbf{15}  (1975) 49

\bibitem{Golub2003}
G.~Golub, V.~Pereyra:
\newblock \emph{Inverse Problems} \textbf{19}  (2003) R1

\end{thebibliography}

\end{document}